\RequirePackage{fix-cm}
\documentclass{svjour3}  
\usepackage{graphicx}
\usepackage{epsfig,amsmath,amsfonts,bm,empheq}
\usepackage{amssymb}
\usepackage[inner=2cm,outer=2cm]{geometry}
\usepackage{cuted}
\usepackage{standalone}
\usepackage{listings}
\usepackage{courier}
\usepackage{subfiles}
\usepackage{hyperref}
\usepackage{tikz}
\usepackage{booktabs}
\usepackage{multirow}
\usepackage{subcaption}
\usepackage{import}
\usepackage{pgf}
\usetikzlibrary{patterns}
\usetikzlibrary{calc}
\usetikzlibrary{decorations.pathmorphing,decorations.markings}

\definecolor{or}{rgb}{0.9,0.3,0}
\newcommand{\vonkar}{von K\'arm\'an }
\newcommand{\vect}[1]{{\bm #1}}
\newcommand{\mat}[1]{{\bm #1}}
\newcommand{\tp}[1]{{#1^{\operatorname{T}}}}
\newcommand{\dd}{{\textrm{d}}}

\newcommand{\vdiver}{\operatorname{div}}
\newcommand{\tgrad}{\bm{\nabla}}
\newcommand{\tptgrad}{\tp{\tgrad}}
\newcommand{\trace}{\operatorname{tr}}

\graphicspath{{BEAM_ModeShapes/}{PLATE_ModeShapes/}{Other_Figures/}}
\begin{document}
\title{Non-intrusive reduced order modelling for the dynamics of geometrically nonlinear flat structures using three-dimensional finite elements}
\titlerunning{Non-intrusive ROM for the dynamics of geometrically nonlinear flat structures using 3D FEM}

\author{
  Alessandra Vizzaccaro
  \and
  Arthur Givois
  \and
  Pierluigi Longobardi
  \and
  Yichang Shen
  \and
  Jean-Fran\c{c}ois De\"u
  \and
  Lo\"ic Salles
  \and
  Cyril Touz\'e
  \and 
  Olivier Thomas}
   
\authorrunning{A. Vizzaccaro, A. Givois, P. Longobardi, Y. Shen, J.-F. De\"u, L. Salles, C. Touz\'e, O. Thomas}

\institute{
A. Vizzaccaro
\at
Vibration University Technology Centre,
Imperial College London, SW7 2AZ London, UK\\
\email{a.vizzaccaro17@imperial.ac.uk}
\and
A. Givois, O. Thomas
\at
Arts et Metiers Institute of Technology, LISPEN, HESAM Universit\'e,
 F-59000 Lille, France
\and
A. Givois, J.-F. De\"u
\at
Structural Mechanics and Coupled Systems Laboratory, Conservatoire National des Arts et M\'etiers, 2 Rue Cont\'e, 75003 Paris, France
\and
P.Longobardi, L. Salles
\at
Vibration University Technology Centre,
Imperial College London, SW7 2AZ London, UK
\and
Y. Shen, C. Touz\'e
\at
IMSIA, ENSTA Paris, CNRS, EDF, CEA, Institut Polytechnique de Paris, 828 Boulevard des Mar\`echaux, 91762 Palaiseau Cedex, France}


\maketitle
\begin{abstract}
Non-intrusive methods have been used since two decades to derive reduced-order models for geometrically nonlinear structures, with a particular emphasis on the so-called STiffness Evaluation Procedure (STEP), relying on the static application of prescribed displacements in a finite-element context. We show that a particularly slow convergence of the modal expansion is observed when applying the method with 3D elements, because of nonlinear couplings occurring with very high frequency modes involving 3D thickness deformations. Focusing on the case of flat structures, we first show by computing all the modes of the structure that a converged solution can be exhibited by using either static condensation or normal form theory. We then show that static modal derivatives provide the same solution with fewer calculations. Finally, we propose a modified STEP, where the prescribed displacements are imposed solely on specific degrees of freedom of the structure, and show that this adjustment also provides efficiently a converged solution.
\keywords{Reduced order modeling \and Geometric nonlinearities \and Three-dimensional effect \and Thickness modes \and Modified STiffness Evaluation Procedure \and Nonlinear modes \and Modal derivatives}
\end{abstract}

\section{Introduction}

Geometrically nonlinear effects appear generally in thin structures such as beams, plates and shells, when the amplitude of the vibration is of the order of the thickness \cite{NayfehPai,ThomasBilbao08}. The \vonkar family of models for beams, plates and shells allows one to derive explicit partial differential equations (PDE)~\cite{Landau1986,Bazant,ThomasBilbao08,thomas04,givois2019}, showing clearly that a coupling between bending and longitudinal motions causes a nonlinear restoring force of polynomial type in the equations of motion. 
This geometric nonlinearity is then at the root of complex behaviours, that also need dedicated computational strategies in order to derive quantitative predictions. On the phenomenological point of view, structural nonlinearities give rise to numerous nonlinear phenomena that have been analysed in a number of studies: frequency dependence on amplitude \cite{Nayfeh79,Lewandowki97a,KerschenNNM09}, hardening/softening behaviour \cite{touze03-NNM,touze-shelltypeNL}, hysteresis and jump phenomena \cite{Nayfeh00,Wanda1}, mode coupling through internal resonances \cite{Nayfeh00,thomas04,givois2020}, bifurcations and loss of stability \cite{Touze:JSV:2012,guillot2020}, chaotic and turbulent vibrations \cite{Duc:PhysD:2014,WTbook}. On the computational point of view, nonlinear couplings break the invariance property of the linear eigenmodes. Consequently, deriving reduced-order models (ROM) is no longer a straightforward problem and care has to be taken in order to find out a ROM that is capable of describing the dynamics of the whole system without losing accuracy. 

When the structure under study is discretized with the finite element (FE) method, the problem of deriving accurate ROM is more stringent since the user cannot rely on a PDE in order to unfold an ad hoc mathematical method for building the ROM. Moreover, because of the intrinsic nature of the geometrical nonlinearities, all degrees of freedom of the FE model are nonlinearly coupled and substructuring techniques are not suitable, on the contrary to localized nonlinearities occurring frequently in contact and friction problems \cite{deklerk2008,yuan2019}. 
Consequently, for geometric nonlinearity, the computation of the nonlinear coupling coefficients is as important as finding out a correct reduced basis, and sometimes the two problems are interwoven. Moreover, a number of recent studies highlighted the importance of using so-called {\em indirect} or {\em non-intrusive} methods, where the idea is to use the standard operations that any FE code is used to perform in order to build the ROM, hence avoiding the need to enter in the code and write new lines  computing the needed quantities. On the other hand, direct methods also exist, for which there is a need to implement the computations at the level of the element \cite{senechal11,Touze:compmech:2014}.
The STEP  (STiffness Evaluation Procedure) is a non-intrusive method and has been first introduced by Muravyov and Rizzi in 2003 \cite{muravyov}. In its first version as described in~\cite{muravyov}, it  allows computation of the nonlinear coupling coefficients of the discretized problem in the modal basis from a series of static computations with prescribed modal displacements. It has then been used in a number of contexts~\cite{Mignolet08,LazarusThomas2012,mignolet13,Perez2014,givois2019}, and is generally connected to the modal basis. However one has to understand that {\em per se},  STEP is just an evaluation technique, a computational non-intrusive method, that can be used with other inputs than those from the modal basis.


The STEP, although being largely applied in numerous cases, is known to suffer from a number of problems, making it not as so simple as its formulation could let one think. A first one lies in the amplitude of the prescribed displacement one has to impose in order to excite sufficiently the nonlinearity. As shown in \cite{givois2019}, there is a clear range of amplitude for which the method works properly, between a minimal value where nonlinearity is not sufficiently excited and a maximum value for which other nonlinear effects are appearing. Another problem is connected to the use of the modal basis with a STEP computation, and must be interpreted as a drawback of using the modal basis for nonlinear computation, but is not directly linked with the STEP calculation. The problem is that of the slow convergence linked to the loss of invariance of eigenspaces, and the numerous couplings between low frequency bending modes and high frequency longitudinal modes, as underlined in a number of papers, see {\em e.g.}~\cite{mignolet13,givois2019} and references therein.  Consequently, numerous methods have been proposed in order to overcome this limitation: dual modes \cite{KIM2013}, POD modes \cite{RIZZI2008}, discrete empirical interpolation \cite{TisoDEIM}, modal derivatives \cite{IDELSOHN1985,Weeger2016}, quadratic manifold \cite{Jain2017,Rutzmoser}, just to name a few.


Incidentally, the majority of papers where the STEP has been applied to thin structures, report results obtained with structures discretized with beam, plate or shell finite elements, as being mostly interested in slender  structures. A few examples with block elements can be found in the literature, see {\em e.g.}~\cite{PEREZ2014notch,WANG20181}, where dual modes have been used in order to achieve convergence. Using 3D finite elements can be interesting in practice, sometimes mandatory. In engineering applications, structures are often defined with a 3D geometrical model, for which a 3D FE discretization is straightforward. In other cases, some particular physical effects (piezoelectricity for instance) are sometimes implemented in existing FE codes only with 3D elements. Preliminary studies of the authors reveal a number of difficulties when blindly applying the STEP with modal basis to 3D finite elements, with more stringent inaccuracies and problems than those encountered with plate or shell elements. In particular, an unexpected slow convergence was observed when using the modal basis, and very high frequency modes appear to be involved.



The objective of this paper is to diagnose properly the issues one can encounter when applying the STEP with a modal basis to a structure discretized with 3D finite elements and present methods to overcome the problems. In the course of the paper, we will show that the problem is intrinsically related to the use of the modal basis, and that the STEP can be used with other inputs than the modal basis, in order to get better results. We restrict our attention to thin structures that are symmetric in the thickness direction, such as straight beams or plates, for which there is a bending / longitudinal uncoupling at the linear level that greatly simplifies the understanding of the phenomena and enables to obtain reference results. 

The paper is organized as follows. Section \ref{sec:2} is dedicated to the framework of the study, the equations of motion and a brief recall of the STEP. Section~3 addresses on test examples the issues of the STEP applied to 3D FE. It is shown that a nonlinear coupling of bending modes with very high frequency modes involving deformations in the thickness of the structures occurs, and thus called thickness modes. Those modes are the result of 3D deformation effects that are not present when using beam or plate models. This unexpected coupling is different from the traditional bending-longitudinal coupling and the numerical examples show that they are of prime importance to achieve a converged solution. If one can compute all the coupled modes, two strategies are given to overcome the large dimension of the reduced basis: static condensation and the reduction to a single nonlinear normal modes. Both method shows that when a single master mode drives the dynamics, the ROM can still be composed of a single nonlinear oscillator. However, finding out all the coupled high-frequency modes is generally out of reach for complex structures. In section~4, we then show that using a static condensation of a single modal derivative allows to retrieve the same converged result, in a more direct and efficient way. In addition, a modified version of the STEP is proposed, to directly embed the coupling with thickness modes. It consists in applying the prescribed displacement only on selected degrees of freedom and let the other free. In section~5, the physical mechanisms of those nonlinear couplings are explained and section~6 presents numerical results to validate the proposed numerical methods, able to overcome the convergence issues of the classical STEP.

\section{Modelling}\label{sec:2}

\subsection{Reduced order model and STEP}

An elastic mechanical structure discretized by the finite element method and having geometrical nonlinearities is considered. The time-dependent displacement vector $\vect{x}$ gathers all the degrees of freedom of the model (displacements/rotations at each nodes) and is  $N$-dimensional. The equation can be written:
\begin{equation}
\mat{M}\ddot{\vect{x}} + \mat{C}\dot{\vect{x}} + \vect{f}(\vect{x}) = \vect{f}_e,
\label{eq:Nddl}
\end{equation}
where $\mat{M}$ and $\mat{C}$ are the $N\times N$ dimensional mass and damping matrices, $\vect{f}(\vect{x})$ is the internal force vector, $\vect{f}_e(t)$ is the external force vector and the overdot represents the classical differentiation with respect to time $t$: $\dot{\bullet}=\dd\bullet/\dd t$. Note that since we are more interested in the computation of the nonlinear restoring force,  a linear viscous damping model has been used. In the present case of geometrically nonlinear structures, the internal force vector encompasses only polynomial terms up to order three and involves the displacement vector $\vect{x}$ only~\cite{LazarusThomas2012,mignolet13,Touze:compmech:2014,holzapfel2000nonlinear}. For the present study, we restrict our attention to this case, but extensions to more complex cases involving for examples velocity terms can also be handled.

It is convenient to split the internal force vector into a linear part and a purely nonlinear part. Assuming that the equilibrium point, the structure at rest, is given by  $\vect{x} = \vect{0}$, the tangent stiffness matrix $\mat{K}$ classically writes:
\begin{equation}
\bm{K}=\left.\frac{\partial \bm{f}}{\partial\bm{x}}\right|_{\bm{x}=\bm{0}},
\end{equation}
so that the nonlinear internal force vector is defined as 
\begin{equation}
\label{eq:fnl}
\bm{f}_\text{nl}(\bm{x}) = \bm{f}(\bm{x}) - \bm{K}\bm{x},
\end{equation}
and the equations of motion reads:
\begin{equation}
\bm{M}\ddot{\bm{x}} + \mat{C}\dot{\vect{x}} +\bm{K}\bm{x} + \bm{f}_\text{nl}(\bm{x}) = \bm{f}_\text{e} \label{eq:MvtEFGal},
\end{equation}	
where the geometrically nonlinear part of the problem is concentrated in the nonlinear internal force vector $\bm{f}_\text{nl}(\bm{x})$. 

The modal basis can be used as a first step in order to make the linear terms diagonal. The eigenmodes are the family of couples eigenfrequency-eigenvectors $(\omega_k,\vect{\phi}_k)$, $k=1,\ldots, N$, solutions of the undamped, free and linearized Eq.~(\ref{eq:MvtEFGal}):
\begin{equation}
\label{eq:linmode}
(\mat{K}-\omega_k^2\mat{M})\vect{\phi}_k=\vect{0}.
\end{equation}
Assuming the modal expansion for the displacement vector:
\begin{equation}
\label{eq:projmod}
\vect{x}(t)=\sum_{k=1}^{N} \vect{\phi}_k X_k(t),
\end{equation}
where $X_k(t)$ is the modal amplitude, and using Galerkin projection allows one to rewrite the equations of motion in the modal space, for all $k=1,\ldots, N$, as:
\begin{equation}
\label{eq:eqq}
\ddot{X}_k+
2\zeta_k\omega_k\dot{X}_k+
\omega_k^2X_k+
\sum_{i=1}^{N}\sum_{j=i}^{N} \alpha_{ij}^k X_iX_j+
\sum_{i=1}^{N}\sum_{j=i}^{N}\sum_{l=j}^{N} \beta_{ijl}^k X_i X_j X_l = Q_k,
\end{equation}
with
\begin{equation}
\label{eq:Qk}
Q_k=\tp{\vect{\phi}_k}\vect{f}_e/m_k,\qquad m_k=\tp{\vect{\phi}_k}\mat{M}\vect{\phi}_k,
\end{equation}
where $m_k$ is the modal mass, and where a modal damping (of factor $\zeta_k$) has been assumed uncoupled (an assumption valid for small damping, even with non-proportional $\mat{C}$ matrix \cite{Geradin2015}). The nonlinear part of this reduced order is written with quadratic and cubic polynomial terms with coefficients $\alpha_{ij}^k$ and $\beta_{ijl}^k$. It is an exact expansion in the case of 3D FE \cite{Touze:compmech:2014} or beam/plate/shell FE based on \vonkar strain/displacement law \cite{LazarusThomas2012}, whereas it is truncated in the case of geometrically exact theories \cite{thomas16-ND}.



In Eqs.~(\ref{eq:eqq}), the linear parameters $\omega_k$, $\bm{\phi}_k$ and $Q_k$ are obtained by the modal analysis of Eq.~(\ref{eq:linmode}), available in any finite element code. The main issue is thus to compute the nonlinear coupling coefficients $\alpha_{ij}^k$ and $\beta_{ijl}^k$. 
The STEP (STiffness Evaluation Procedure) when used with the modal basis as first introduced by Muravyov and Rizzi~\cite{muravyov}, is a non-intrusive (or indirect) method, allowing one to get these coupling coefficients from standard computations available in any FE code.  It relies on imposing prescribed static displacements having the shapes of selected eigenmodes, with a given amplitude. From a clever choice of the modes and the amplitudes, a simple algebra allows one to retrieve all the coefficients from the internal force vector given by the FE code, the key idea being to impose plus/minus the displacement with selected combinations of modes. The reader can find detailed explanation on this calculation in a number of papers, including the original one~\cite{muravyov}, as well as the improvement proposed in \cite{Perez2014} using the tangent stiffness matrix. 

To illustrate the method, we only show the computation of coefficients $\alpha_{pp}^k$ and $\beta_{ppp}^k$; for the general case the interested reader is referred to~\cite{muravyov,Perez2014}. The following static displacements are prescribed to the structure:
\begin{equation}
\label{eq:x1}
\bm{x}_p =\pm \lambda \vect{\phi}_{p}\quad\Rightarrow\quad 
\left\{\begin{array}{l}
q_p=\lambda, \\
q_j=0\quad\forall j\neq p,
\end{array}\right.
\end{equation}
where $\lambda$ refers to an amplitude coefficient of the eigenvector $\vect{\phi}_{p}$ whose value has to be chosen so as to activate the geometrical nonlinearities. A value of $h/20$, where $h$ is the thickness of the plate, is recommended in~\cite{givois2019}.  Since the modes are orthogonal, imposing a displacement along mode $p$ is equivalent to consider that only the modal coordinate $q_p$ is not vanishing, as detailed in the second part of Eq.~\eqref{eq:x1}. Since $\bm{x}_p$ is time independent, introducing Eq.~\eqref{eq:x1} into  Eqs.~\eqref{eq:MvtEFGal} and \eqref{eq:eqq} leads to, for all $k=1,\ldots, N$:
\begin{subequations}
\label{eq:SysLinMur1Mode}
\begin{align}
\lambda^2 \alpha_{pp}^k  + \lambda^3  \beta_{ppp}^k & = \tp{\vect{\phi}_k} \bm{f}_\text{nl}(\lambda\bm{\phi}_p)/m_k, \\ 
\lambda^2 \alpha_{pp}^k  - \lambda^3  \beta_{ppp}^k & = \tp{\vect{\phi}_k} \bm{f}_\text{nl}(-\lambda\bm{\phi}_p)/m_k.
\end{align}
\end{subequations}
Hence the unknown quadratic and cubic coefficients are found easily as
\begin{subequations}
\label{eq:zicoefSTEP}
\begin{align}
\alpha_{pp}^k = \frac{1}{2\lambda^2}\frac{\vect{\phi}_k}{m_k}\left( \bm{f}_\text{nl}(\lambda\bm{\phi}_p) + \bm{f}_\text{nl}(-\lambda\bm{\phi}_p) \right), \label{eq:zicoefSTEPq}\\
\beta_{ppp}^k = \frac{1}{2\lambda^3}\frac{\vect{\phi}_k}{m_k}\left( \bm{f}_\text{nl}(\lambda\bm{\phi}_p) - \bm{f}_\text{nl}(-\lambda\bm{\phi}_p) \right). \label{eq:zicoefSTEPq}
\end{align}
\end{subequations}
Similar algebraic manipulations with more modes involved in the prescribed displacement then allows one to get the full family of quadratic and cubic coefficients. The non intrusive nature of the method appears clearly: in the FE software, one prescribes the displacement field $\bm{x}_p$ and computes the corresponding external force vector $\bm{f}_\text{e}=\bm{f}(\bm{x}_p)$ with successive linear and nonlinear static computations. Then, $\bm{f}_\text{nl}(\bm{x}_p)$ is obtained with Eq.~(\ref{eq:fnl}).


\subsection{The case of flat structures}

In this article, we restrict the analysis to flat structures, such as straight beams or plates with a boundary of arbitrary shape. The thickness of the plate can be non-constant, but the geometrical and material distribution must be symmetric with respect to the middle line / plane of the structure.

If a plate theory, with a Kirchhoff-Love kinematics for instance, is applied to this structure, the displacement field of any point of the structure is described by the displacement field of the middle plane. There is a membrane / bending decoupling in linear elasticity and two families of modes are obtained: bending modes, for which only the transverse component of the middle plane displacement are non-zero, and membrane modes, for which the transverse component of the middle plane displacement field is zero. Analogous properties are valid for a beam theory, with a longitudinal / transverse decoupling on the middle line. 

In the present case of a 3D structure which has the shape of a beam / plate, it is also possible to split the eigenmodes into two families in the same manner. The first family includes the bending modes, analogous to the ones obtained in the plate theory. Their frequencies are in the lower part of the spectrum, while their deformed shapes are dominated by transverse displacements. They, up to the accuracy of the plate theory, have the same displacement field in the middle plane / line, with no longitudinal displacement. The second family gathers all the other modes, denoted by non-bending (NB) modes, that appear at higher frequencies in the spectrum. Some of them are analogous to the longitudinal  modes of the plate theory, with the same transverse displacement field in the middle plane / line. Other modes are also present, linked to 3D effects and thus with no counterpart in the beam / plate theory, with mode shapes dominated by thickness deformations. For any mode of this second family, the displacement field in the middle plane / line has only longitudinal components, and thus no transverse component. Examples of NB modes will be shown throughout the paper, especially in Tab.~\ref{tab:BeamModes}.

Let us decompose the displacement vector $\vect{X}$ by denoting as  $q_r$, $r=1, ... , N_B$ the bending coordinates, and $p_s$, $s=N_B + 1, ... , N$ the membrane coordinates: $\vect{X} = \tp{[q_1, ... , q_{N_B}, p_{N_B+1}, ..., p_{N}]}$. Then, Eq.~\eqref{eq:eqq} can be rewritten for each coordinates~\cite{Jain2017,givois2019} and involves quadratic and cubic coupling terms between the $q_r$ and the $p_s$. We restrict ourselves to the case of a transverse low frequency excitation, for which the external forces remain normal to the middle plane of the plate. As a consequence, the dynamics is dominated by the bending modes, which are the only ones that receive external excitation. In this case, Eq.~\eqref{eq:eqq} can be simplified. First, all quadratic $\alpha_{ij}^r$ coefficients involving two bending coordinates $i,j$ vanishes, in order to fulfil the symmetry of the restoring force~\cite{givois2019,thomas04}. In addition, one can assume that the bending coordinates, which are directly excited by the external forcing, are considered of the order magnitude of a small parameter $\varepsilon$: $q_r=O(\varepsilon)$, for all $r\in\{1,\ldots, N_B\}$. On the other hand, NB coordinates, since they are not directly excited and shall vibrate at a lower order of magnitude, are assumed to scale as  $\varepsilon^2$: $p_s=O(\varepsilon^2)$ for all $s\in\{N_B + 1,\ldots, N\}$. Plugging these two scaling in Eq.~\eqref{eq:eqq} and keeping only the leading order, one arrives to, for the bending coordinates, $\forall \; r=1, ... , N_B$:
\begin{equation}
\label{eq:qsimple}
\ddot{q}_r+ 2\zeta_r\omega_r\dot{q}_r+ \omega_r^2q_r+\sum_{i=1}^{N_B}\sum_{l=N_B+1}^{N} \alpha_{il}^r q_ip_l+  \sum_{i=1}^{N_B}\sum_{j=i}^{N_B}\sum_{k=j}^{N_B} \beta_{ijk}^r q_i q_j q_k +O(\varepsilon^4) = Q_r,
\end{equation}
and for the membrane coordinate, $\forall \; s=N_B+1, ... , N$:
\begin{equation}
\label{eq:psimple}
\ddot{p}_s+ 2\zeta_s\omega_s\dot{p}_s+ \omega_s^2p_s+\sum_{i=1}^{N_B}\sum_{j=i}^{N_B} \alpha_{ij}^s q_iq_j +O(\varepsilon^3) = 0.
\end{equation}
In the above equation, one can notice that because of the transverse excitation, the second member of Eq.~(\ref{eq:psimple}) is zero.

Eqs.~(\ref{eq:qsimple},\ref{eq:psimple}) approximate the dynamics of the structure in the present case of a 3D FE model. If an analytic \vonkar model was used, those equations would be exact, with the terms $O(\varepsilon^3)$ and $O(\varepsilon^4)$ identically vanishing \cite{givois2019}. Those equations also show that the in-plane vibrations are quadratically coupled to the bending coordinates, while in the equations of motion of the transverse modes, only two nonlinear terms have to be taken into account: a quadratic coupling involving a product between a transverse and an in-plane coordinate, and a cubic term involving three transverse modes. This very specific form of equations renders the case of flat structures easier to solve than general shell problems that encompass all the possible nonlinear couplings as stated in Eq.~\eqref{eq:eqq}.

\subsection{Static condensation and nonlinear normal modes}
\label{sec:stat_cond}

Since the non-bending (NB) modes have natural frequencies very large as compared to those of the directly excited bending modes, $\omega_s\gg\omega_r$, the dynamical part of Eqs.~\eqref{eq:psimple} can be neglected. The nonlinearities being more simple in this case, one can directly express the non-bending coordinate as function of the bending ones as:
\begin{equation}
\label{eq:pstatic}
p_s = -\sum_{i=1}^{N_B}\sum_{j=i}^{N_B} \frac{\alpha_{ij}^s}{\omega_s^2} q_iq_j.
\end{equation}
Substituting Eq.~\eqref{eq:pstatic} into Eq.~\eqref{eq:qsimple}, one can rewrite the dynamics of the structure as a closed system involving only bending coordinates as
\begin{equation}
\label{eq:qcubic}
\ddot{q}_r+ 2\zeta_r\omega_r\dot{q}_r+ \omega_r^2q_r+ \sum_{i=1}^{N_B}\sum_{j=i}^{N_B}\sum_{k=j}^{N_B} \Gamma_{ijk}^r q_i q_j q_k = Q_r,
\end{equation}
where the cubic $\Gamma_{ijk}^r$ coefficients appear. Their general expression is derived in Appendix~\ref{app:A}.

If one is interested in deriving a reduced-order model for a single bending mode (say the master mode with  label $p$), taking into account all the other non-bending mode, then Eq.~\eqref{eq:qcubic} can be used and the leading nonlinear cubic term simply reduces to:
\begin{equation}
\label{eq:Gammappp}
\Gamma_{ppp}^p = \beta_{ppp}^p - \sum_{s=N_B+1}^{N} \mathcal{C}^{ps}_{ppp}
\end{equation}
where the correction factors have been introduced and read:
\begin{equation}
\label{eq:Cpsppp}
\mathcal{C}^{ps}_{ppp}= \frac{\alpha_{ps}^p\alpha_{pp}^s}{\omega_s^2}=\frac{2(\alpha_{pp}^s)^2}{\omega_s^2}.
\end{equation}
These expressions show that the cubic term $\beta_{ppp}^p$ of the standard modal expansion must be corrected by the summation of all NB modes quadratically coupled to the master one. 

The quadratic coefficients $\alpha^p_{ij}$ have some symmetry relationships, provided that the nonlinear stiffness derives from a potential~\cite{muravyov}. In particular, the following relationship holds: $\alpha_{ps}^p = 2 \alpha_{pp}^s$, which leads to the second equation~\eqref{eq:Cpsppp}. Recalling Eq.~\eqref{eq:SysLinMur1Mode}, the evaluation of $\alpha_{pp}^s$ requires only the computation of the nonlinear force $\bm{f}_\text{nl}$ when the displacement along the $p$-th linear mode is prescribed, whereas the calculation of the $\alpha_{ps}^p$ coefficients for each $s$-th mode would require as many evaluation of $\bm{f}_\text{nl}$ as the number of non-bending modes. Nevertheless, it requires the projection of nonlinear force onto each membrane eigenvector $\phi_s$, and thus their computation, which is costly in practice since a large number of them is required to reach convergence (see section \ref{subsec:condens}).

From a physical perspective, following Eq.~(\ref{eq:SysLinMur1Mode}), coefficient $\alpha_{pp}^s$ can be seen as the projection onto the $s$-th membrane mode of the quadratic stiffness forces arising in the structure when a displacement along the linear $p$-th mode is imposed. Interestingly, this quadratic coefficient is related to a monomial $q_p^2$ on the $s$-th oscillator equation for $p_s$. These terms are recognized as invariant-breaking terms (see {\em e.g.} \cite{touze03-NNM,TouzeCISM}), in the sense that as soon as energy is given to the master mode $p$, all $s$ modes having these important invariant-breaking terms will no longer be vanishing. These invariant-breaking terms are responsible for the loss of invariance of the linear eigenspaces, and they are found back naturally as correction factors when applying static condensation. They are also key in the formulation of invariant manifolds in order to define NNMs in phase space~\cite{touze03-NNM,ShawPierre91}.

In parallel to the static condensation emphasised here, one can use the reduction formulae given by the normal form approach, restricting the motion to a single Nonlinear Normal Mode (NNM) \cite{touze03-NNM,TouzeCISM,Touze:compmech:2014}. In this case, the reduced order model is directly constructed from Eqs.~\eqref{eq:eqq}. The main advantage as compared to the above described static condensation is that there is no need to assume the particular structure of the equations obtained for flat structures (Eqs.~(\ref{eq:qsimple},\ref{eq:psimple})),  thus generalizing the results to arches and shells. Considering only the NNM label $p$, the reduced-order model reads:
\begin{equation} 
\ddot{q_p} + \omega_p^2 q_p + \left(\sum_{s=N_B+1}^{N} -\dfrac{\alpha^p_{ps}\alpha^s_{pp}}{\omega_s^2}\left(\dfrac{\omega_s^2-2\omega_p^2}{\omega_s^2-4\omega_p^2}\right) + \beta_{ppp}^p\right) q_p^3 + \left(\sum_{s=N_B+1}^{N} \dfrac{\alpha^p_{ps}\alpha^s_{pp}}{\omega_s^2}\left(\dfrac{2}{\omega_s^2-4\omega_p^2}\right)\right) q_p \dot{q_p}^2 \; = \; 0 \; . 
\label{eq:dynsingledof}
\end{equation}
Once again, one can observe that the correction brought to the cubic term $\beta^p_{ppp}$ is solely  given by the quadratic invariant-breaking terms. If one has been able to compute all the quadratic $\alpha^p_{ij}$ coefficients appearing in Eq.~\eqref{eq:dynsingledof}, then the model can be used to simulate the dynamics. Also, it is worth mentioning that since the NB modes have high frequencies, we can assume that $\omega_s \gg \omega_p$ (which is equivalent to neglect the membrane inertia). Then the term in factor of $q_p^3$ in Eq.~\eqref{eq:dynsingledof} exactly reduces to the one obtained with static condensation in Eq.~\eqref{eq:qcubic}. On the other hand, the term in factor of $q_p \dot{q_p}^2$ has no counterpart in static condensation, but is an order of magnitude smaller since it scales as $1/\omega_s^4$, so that both models are almost equivalent when a slow/fast decomposition can be assumed. This extends the results of \cite{denis18-MSSP}, in which the term $q_p^3$ is chosen as the leading term for experimental identification purposes. A complete comparison of static condensation and nonlinear normal modes is also provided in~\cite{ICEvsNNMpre} in the context of clarifying the implicit condensation and expansion method. Finally, one can note that the formula used in Eq.~\eqref{eq:dynsingledof} have been obtained thanks to a normal form approach on the conservative system~\cite{touze03-NNM}, but they can be extended in order to take into account the damping of the slave modes in the master coordinate ROM, hence accounting for a finer prediction of the losses~\cite{TOUZE:JSV:2006}, a feature that once again is not possible with static condensation.

\section{STEP convergence with 3D elements}\label{sec:appl_STEP3D}

\subsection{Test examples and direct computation of coefficients with the STEP}\label{subsec:test}

\begin{figure*}[ht]\centering
\begin{subfigure}{.5\textwidth}\centering
\includegraphics[width=6cm]{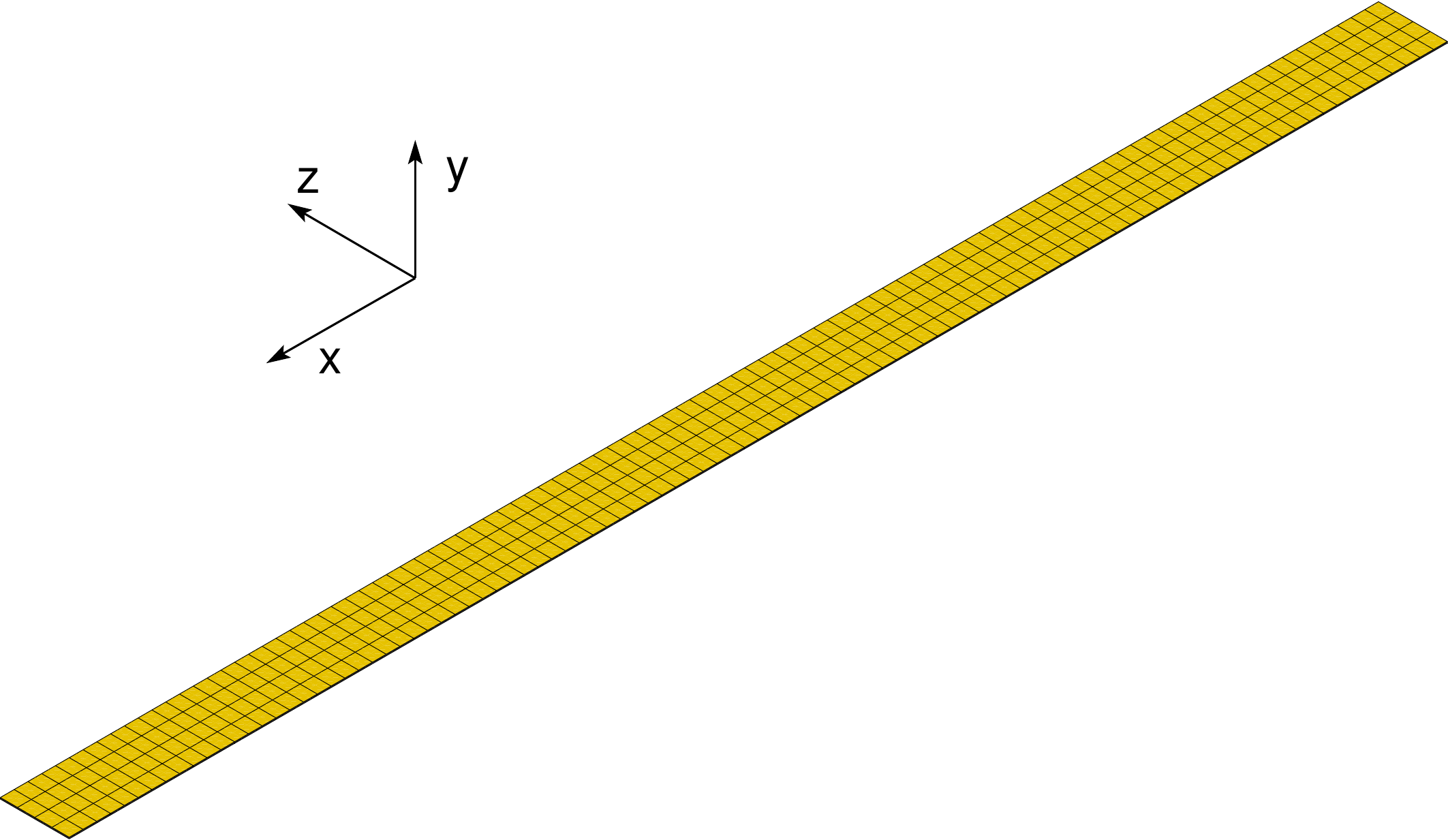}
\caption{Thin beam mesh}\label{fig:meshbeama}
\end{subfigure}
\begin{subfigure}{.49\textwidth}\centering
\includegraphics[width=6cm]{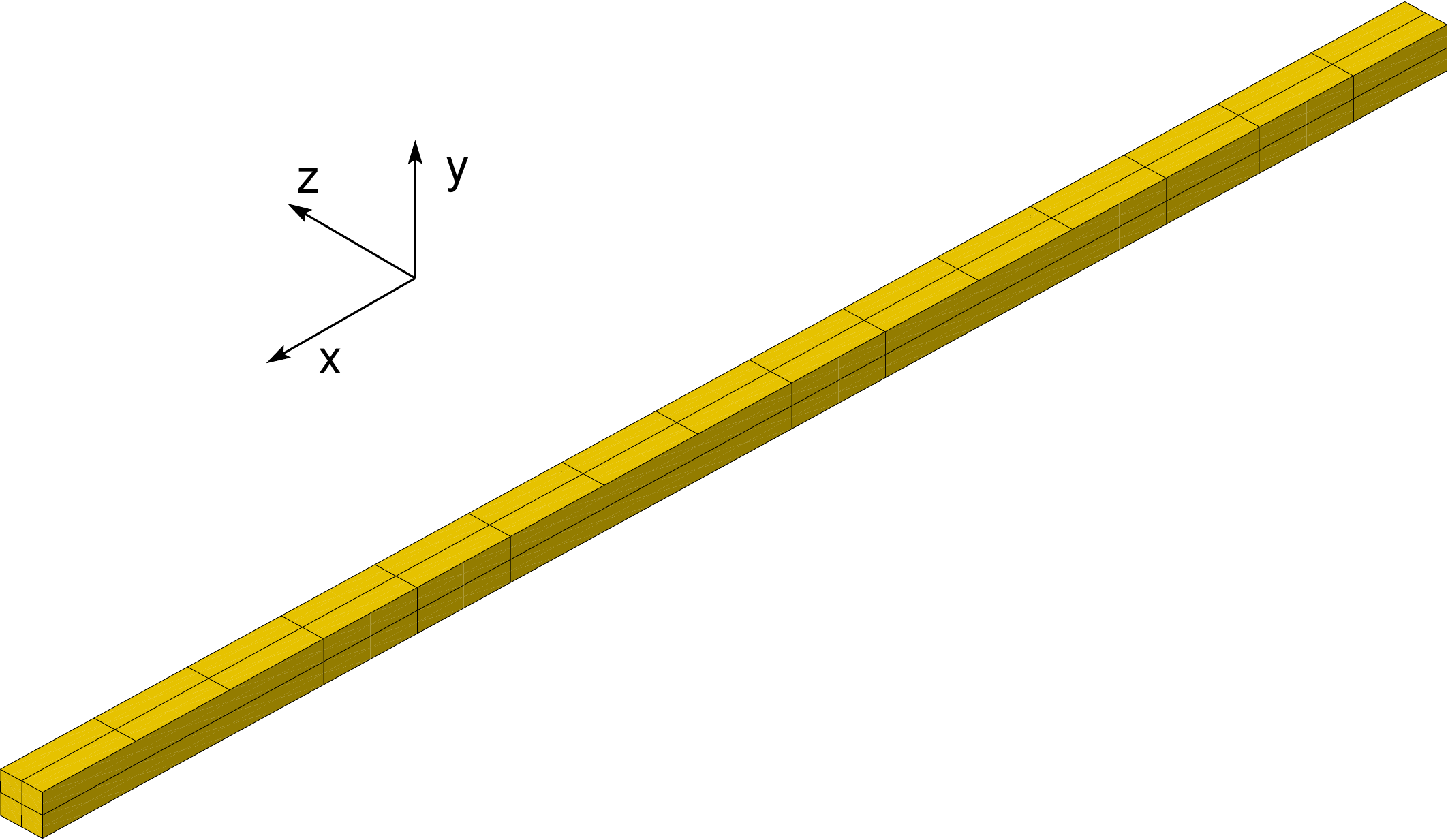}
\caption{Thick beam mesh}\label{fig:meshbeamb}
\end{subfigure}
\caption{Mesh used in the FE computations for the first two test cases on thin and thick beams.}\label{fig:meshbeam}
\end{figure*}

\begin{table*}[ht]
\begin{center}
\begin{tabular}{| c | c | c | c | c |}\hline
\rule{0pt}{0.4cm}
\multirow{2}{*}
 & $\qquad \beta_{iii}^r \qquad $ & $ \qquad \beta_{iii}^r \qquad  $ & $\qquad \alpha_{ii}^s \qquad  $ &  $\qquad  \alpha_{il}^r \qquad $ \rule[-1.4ex]{0pt}{0pt}  \\
\multirow{2}{*}
& \small $i = r = 1$  & \small $i = r =2$ &  \small$i \hspace{-0.06cm} = \hspace{-0.06cm} 2, \hspace{-0.06cm} \enskip s \hspace{-0.06cm} = \hspace{-0.06cm} N_B \hspace{-0.08cm} + \hspace{-0.06cm} 2$ & \small $i \hspace{-0.08cm} = \hspace{-0.08cm} r \hspace{-0.08cm} = \hspace{-0.08cm} 2, \hspace{-0.06cm} \enskip l \hspace{-0.07cm} = \hspace{-0.09cm} N_B \hspace{-0.08cm} + \hspace{-0.06cm} 2 \hspace{-0.08cm}$ \normalsize
  \rule[-1.4ex]{0pt}{0pt}  \\ \hline 
\rule{0pt}{0.7cm}    
 Analytic coefficient (Ac)  & 1.334e+03  \rule[-2.4ex]{0pt}{0pt} & 2.128e+04 & -110.0 & -660.24 \\ \hline
\rule{0pt}{0.5cm}    
 STEP, shell elements, $\nu = 0$ & 1.334e+03 & 2.128e+04 & -110.4 & -660.96 \\ \rule[-1.4ex]{0pt}{0pt} 
\rule{0pt}{0.35cm}    
Relative error with Ac (\%) & 0.02 \% & 0.005 \%  & 0.36 \% & 0.11 \% \rule[-1.4ex]{0pt}{0pt} \\ \hline 
\rule{0pt}{0.5cm}    
 STEP, shell elements, $\nu = 0.3$ & 1.458e+03 & 2.343e+04 & -110.4 & -663.60  \\ \rule[-1.4ex]{0pt}{0pt} 
\rule{0pt}{0.2cm}    
Relative error with Ac (\%) & 9.3 \% & 10.1 \% & 0.36 \% & 0.51 \% \rule[-1.4ex]{0pt}{0pt}
\\ \hline 
\rule{0pt}{0.5cm}    
 STEP, 3D elements, $\nu = 0$ & 2.668e+03 & 4.257e+04 & -109.9 & -660.26 
\\ \rule[-1.4ex]{0pt}{0pt} 
\rule{0pt}{0.2cm}    
Relative error with Ac (\%) & 100 \% & 100\% & 0.09 \% & 0.003 \% \rule[-1.4ex]{0pt}{0pt}
  \\ \hline 
\rule{0pt}{0.5cm}    
 STEP, 3D elements, $\nu = 0.3$ & 5.185e+03 & 8.229e+04 & -111.69 & -664.74  
\\ \rule[-1.4ex]{0pt}{0pt} 
\rule{0pt}{0.3cm}    
Relative error with Ac (\%) & 288.7 \% & 286.7 \%  & 1.54 \% & 0.68 \% \rule[-1.4ex]{0pt}{0pt} 
 \\ \hline 
 \end{tabular}\caption{Nonlinear dimensionless coefficients $\alpha_{il}^r$, $\alpha_{ij}^s$ and $\beta_{ijk}^r$ of the clamped beam, with non-zero and zero Poisson's ratios ($\nu=0$ and $\nu = 0.3$). The modes considered in the coefficient indexes are the first two bending modes ($i=1,2$) and the second axial mode ($l=N_B+2$). The maximum of displacement amplitudes has been fixed at $h/20$ for these computations.\label{tab:ErrorsCoeffsSTEP}}\end{center}
 \end{table*}

In order to properly point out the convergence issues faced by using the modal basis as input prescribed displacements for the STEP,   we consider the simple case of a clamped-clamped thin beam, shown in Fig. \ref{fig:meshbeama}, with length, thickness and width equal to $L=1~\text{m}$, $h=1~\text{mm}$, $b=50~\text{mm}$. The Young's modulus is chosen as $E=210~\text{GPa}$. This particular geometry has been chosen to be thin (the thickness to length ratio is $10^{-3}$) to compare the results of the STEP computation to analytic values, obtained from a beam model with Euler-Bernoulli kinematics and \vonkar assumptions, see \cite{givois2019} where these comparisons have been more fully addressed.  Table~\ref{tab:ErrorsCoeffsSTEP} presents the computations of nonlinear modal coupling coefficients obtained by the classical STEP with 3D and shell elements, compared to the analytical values. The two meshes used here consist of four node DKQ shell elements and twenty-node brick elements (HEX20), respectively. The computations are realized with the open software Code\_Aster \cite{ASTER}. 100 elements in length, 4 elements in the width have been used for both meshes, with 2 elements in the thickness for the 3D mesh. This sufficiently refined mesh ensures that there is no convergence issue for the computations of the nonlinear coefficients.

The results clearly highlights the fact that using blindly 3D elements in a STEP computation with the modal basis leads to individual values of coupling coefficients that are far from their reference, analytical values.  In particular, the cubic coefficients are largely overestimated and a strong dependence to the Poisson's ratio is found with the 3D elements: the $\beta_{iii}^r$ are exactly twice the expected result with 3D elements and a zero Poisson's ratio, but they become almost  three times overestimated with $\nu=0.3$. On the other hand, using shell elements allows recovering the exact analytical result if selecting $\nu=0$, whereas a $10$~\% error is found for the same STEP computation with $\nu = 0.3$. These results clearly demonstrate that the modal basis as input for the STEP can be used safely with 2D elements but its extension to 3D elements is very problematic and should lead to large errors. As already noticed in the introduction, the problem comes from the fact the one uses eigenmodeshape functions as projection basis, but not from the calculation procedure itself.


\subsection{Condensation of the cubic coefficient and frequency-response curves}\label{subsec:condens}

\begin{figure*}[h]\centering
\begin{subfigure}{.49\textwidth}\centering
\includegraphics[width=7.5cm]{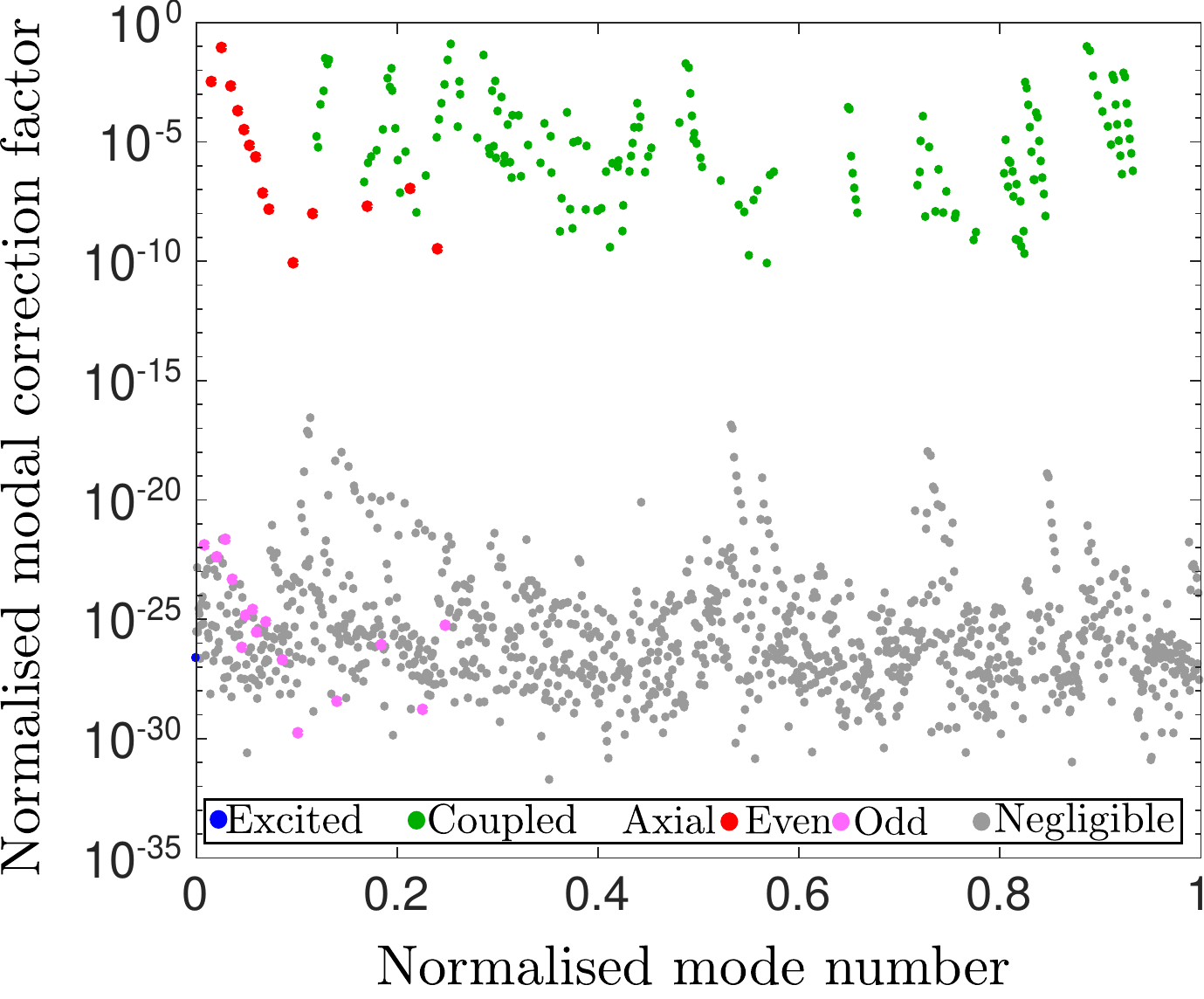}
\caption{Beam discretised with 1287 dofs}
\end{subfigure}
\begin{subfigure}{.49\textwidth}\centering
\includegraphics[width=7.5cm]{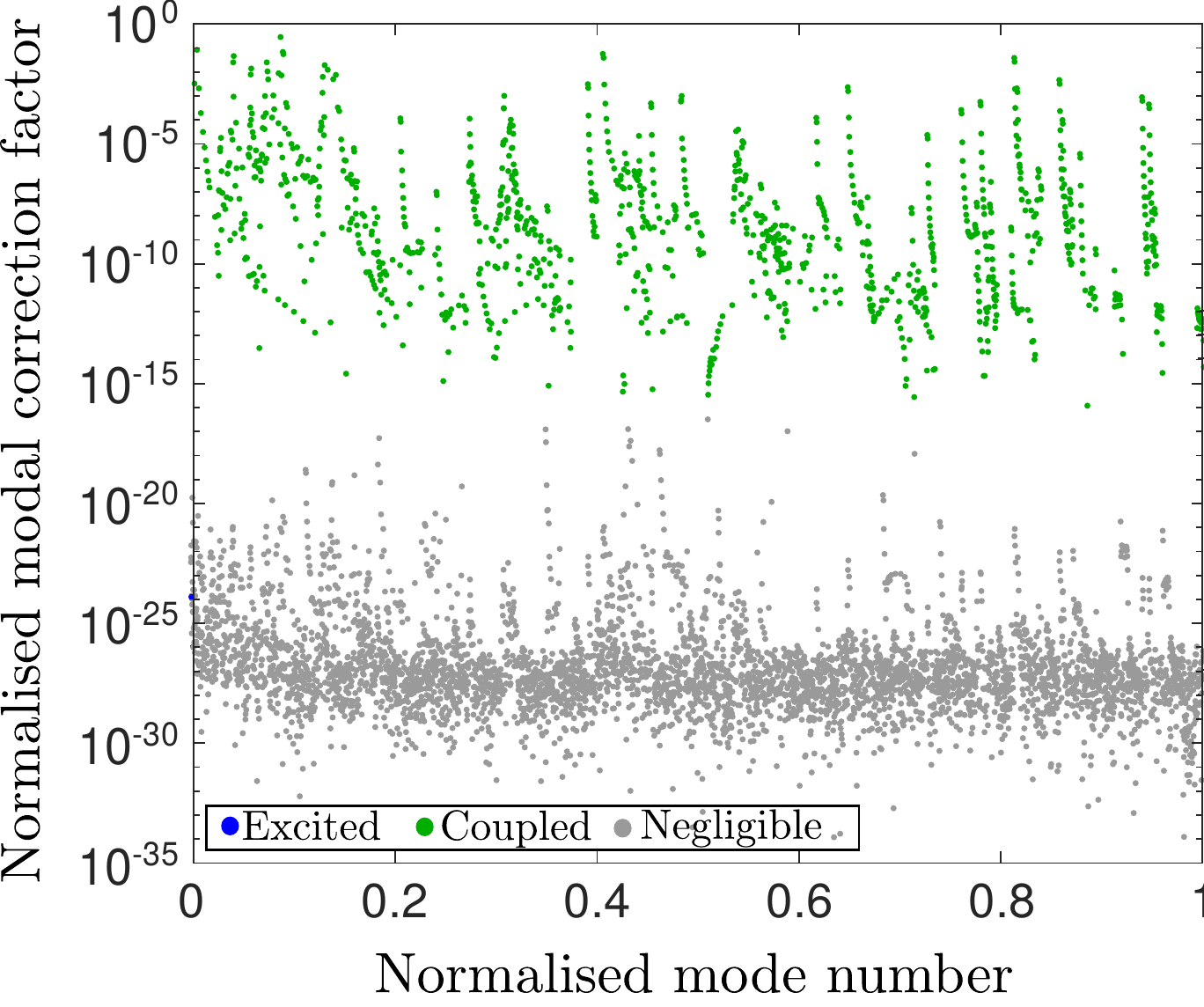}
\caption{Beam discretised with 5733 dofs}
\end{subfigure}
\caption{Nondimensional correction factor $\mathcal{C}^{1s}_{111}/\beta^1_{111}$ (Eq.~(\ref{eq:Cpsppp})), associated to the first bending mode ($p=1$) and to all the other modes ($s=2,\ldots N$) of the thick beam testcase, over the nondimensional mode number $s/N$. The Poisson ratio is $\nu=0.3$.}\label{fig:unsorted}
\end{figure*}

In the previous section, we showed that a direct calculation of individual coefficients leads to different values as compared to analytical results. However, of main importance is the prediction of the global behaviour of the structure, in a dynamical regime where modes are nonlinearly coupled and interacting together. In this section, we show how the static condensation presented in Section \ref{sec:stat_cond} can help to understand how the modes are coupled in order to define the hardening/softening behaviour of bending modes.

In order to shed light on the couplings arising between the modes, another test case is chosen. It is a thick beam, with the same length $L=1~\text{m}$ but with a square cross section with $h=b=30~\text{mm}$, as shown in Fig. \ref{fig:meshbeamb}. The cross section was chosen square to be able to easily observe the 3D deformations of the cross section. The material properties are $E=210~\text{GPa}$ for the Young's modulus and $\rho=7800~\text{kg/m$^3$}$ for the density. A coarse mesh of 15 HEX20 elements along the axis and $2\times2$ in the cross section is chosen, to obtain full model with a reduced number of degrees of freedom (1287). The analyses on this beam are run in the software CodeAster \cite{ASTER}. For the sake of simplicity, we restrict attention to the convergence of the effective cubic coefficient of the first bending mode, $\Gamma_{111}^1$.

Fig.~\ref{fig:unsorted}(a) shows the behaviour of the correction factor $\mathcal{C}^{1s}_{111}$ (defined in Eq.~(\ref{eq:Cpsppp})), normalized by the cubic coefficient $\beta_{111}^1$, as a function of the mode number $s$, for the thick beam having 1287 dofs. This plot shows that the number of modes that are coupled to the first bending mode by invariant-breaking terms is very large, and uniformly distributed along the frequency spectrum. In order to facilitate the readings, the modes for which the correction factor is below 10$^{-15}$ have been sorted as negligible. In this family of modes that are not important, one find backs all the odd axial modes, for symmetry reason. On the other hand, all even axial modes are strongly coupled to the first one. The most surprising result is that if one does considers only axial modes, then only a few portion of the couplings will be revealed and taken into account. Indeed, Fig.~\ref{fig:unsorted}(a)  shows that there is a very large number of modes having very large frequencies, and still being strongly coupled to the master bending mode.

In order to check the independence of this behaviour from the mesh refinement, a second mesh of 20 elements on the axis and 3 x 3 elements on the section has been defined on the same beam geometry. Fig.~\ref{fig:unsorted}(b) reports a very similar behaviour for this second test case, where the distribution of coupled modes is uniform along the whole set of modes.

\begin{figure*}[h!]\centering
\begin{subfigure}{.49\textwidth}\centering
\includegraphics[width=7.5cm]{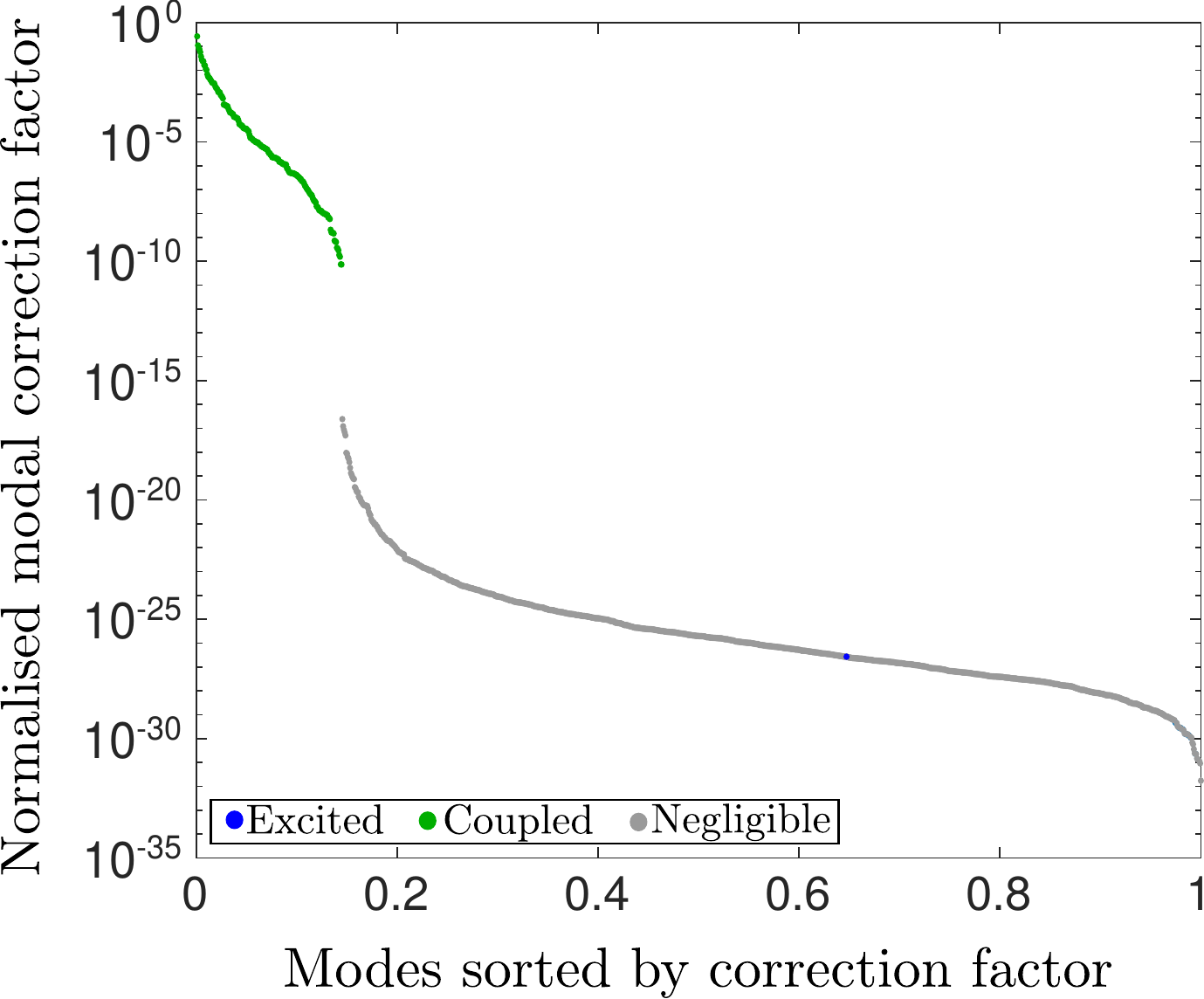}
\caption{Beam discretised with 1287 dofs}
\end{subfigure}
\begin{subfigure}{.49\textwidth}\centering
\includegraphics[width=7.5cm]{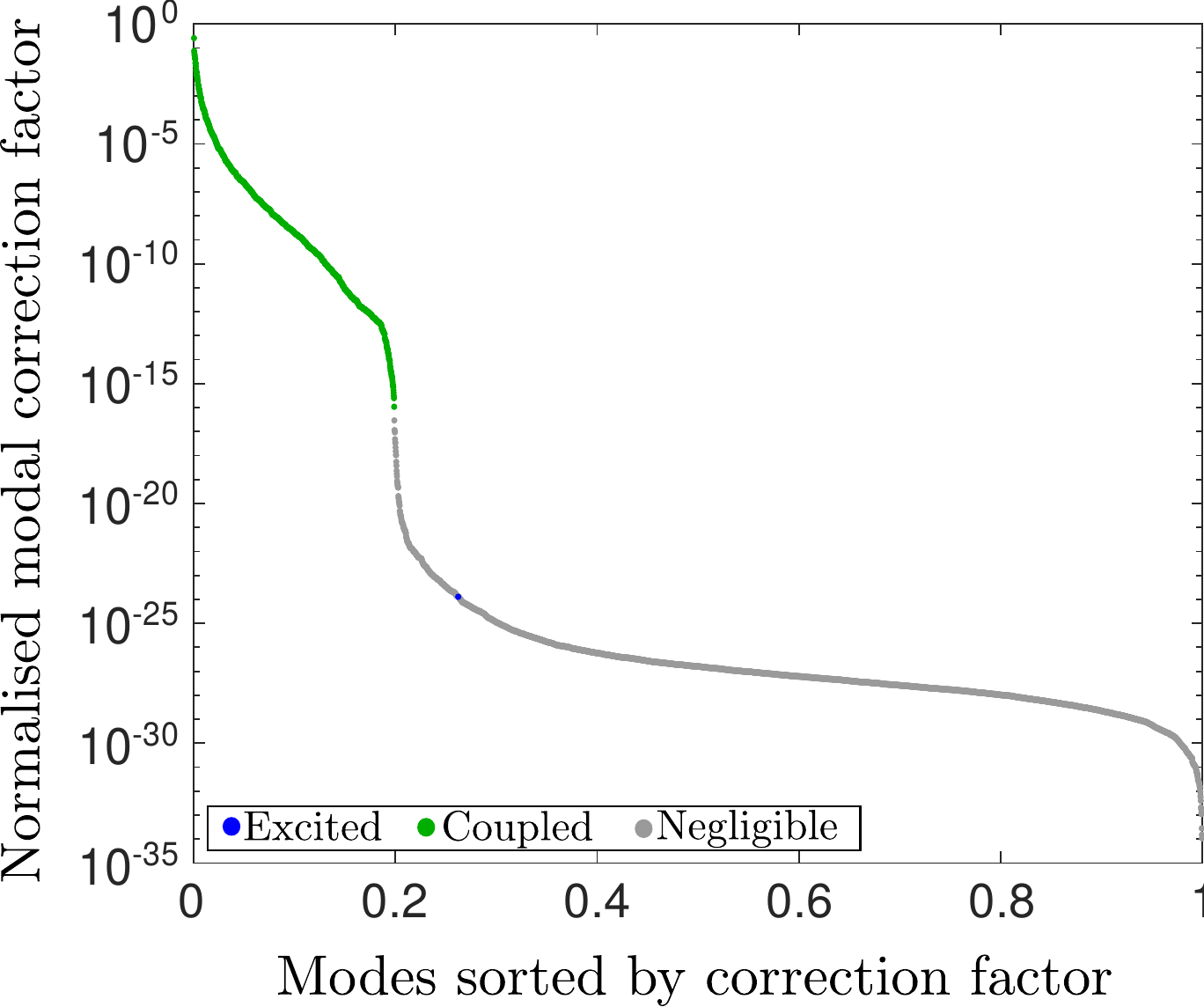}
\caption{Beam discretised with 5733 dofs}
\end{subfigure}
\caption{Nondimensional correction factor $\mathcal{C}^{1s}_{111}/\beta^1_{111}$ (Eq.~(\ref{eq:Cpsppp})), associated to the first bending mode ($p=1$) and to all the other modes ($s=2,\ldots N$) of the thick beam testcase, over the sorted nondimensional mode number $s/N$, where the imposed order is by decreasing correction factor. The Poisson ratio is $\nu=0.3$.}\label{fig:sorted}
\end{figure*}

Fig.~\ref{fig:sorted} shows the same data than Fig.~\ref{fig:unsorted}, but with now the modes sorted by decreasing correction factor. It is possible to observe that, by choosing 10$^{-15}$ as a threshold for the significance of each contribution, a small percentage of modes, around 20\%, is actually relevant. Consequently, the number of relevant modes depends on the mesh: the more refined it is, the more relevent modes are needed to reach convergence. Moreover, as seen on Fig.~\ref{fig:unsorted}, these modes are spread over the entire spectrum, which would need the computation of all the eigenmodes of the structure, an operation impossible in practice for a complex structure with a larger number of dofs. 

\begin{table*}[p]
\begin{center}
\begin{tabular}{ccccc}
\hline
\multicolumn{5}{c}{Coupled modes }\\
  \# 
& $\omega_l/(2\pi)[Hz]$ 
& $\mathcal{C}^{1l}_{111}/\Gamma^1_{111}$ 
& Shape 
& Section 
 \\
\hline
  330 
& 1.137e5 
& 3.22e-1
& \includegraphics[height=2.15cm]{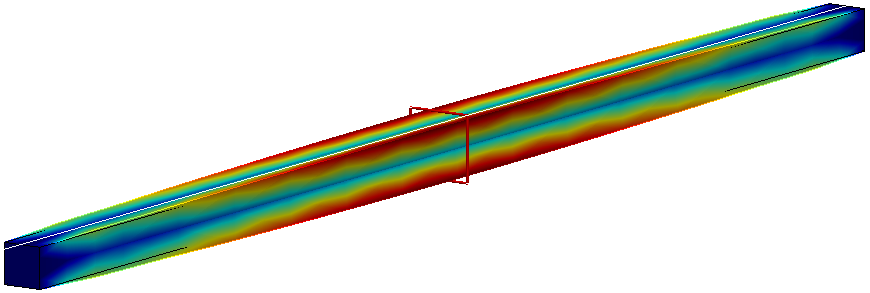} \rule{0pt}{2.25cm}
& \includegraphics[height=1.5cm]{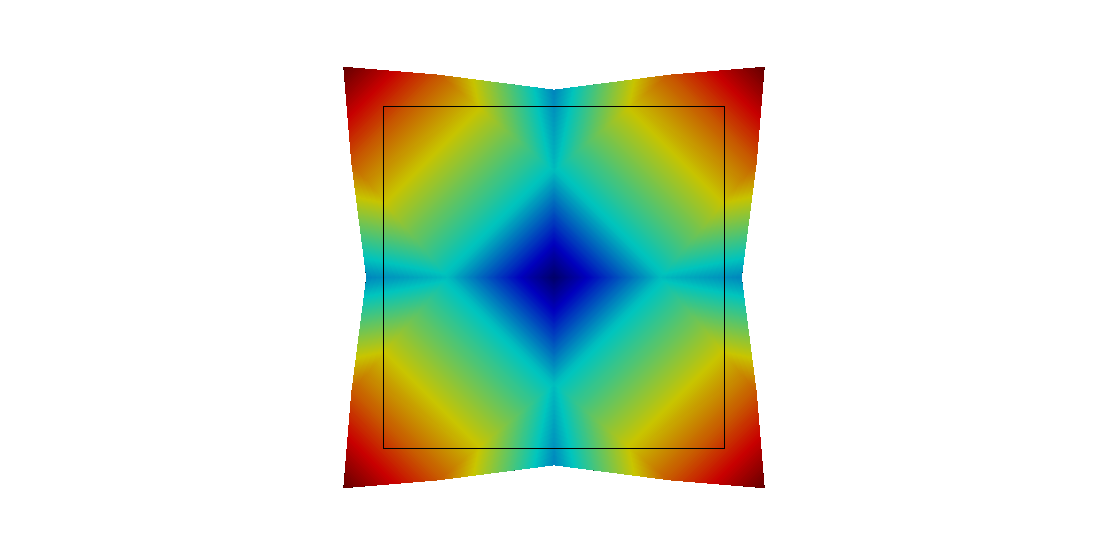} 
 \\
\hline
  328 
& 1.125e5 
& 1.34e-1
& \includegraphics[height=2.15cm]{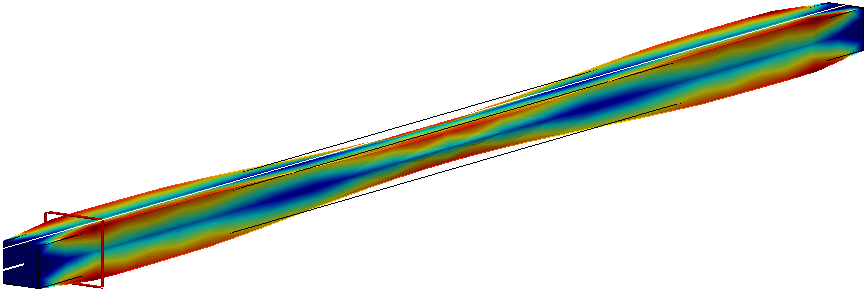}\rule{0pt}{2.25cm}
& \includegraphics[height=1.5cm]{s_324_328_330_370.png} 
 \\
\hline
  1143 
& 3.698e5 
& 1.05e-1
& \includegraphics[height=2.15cm]{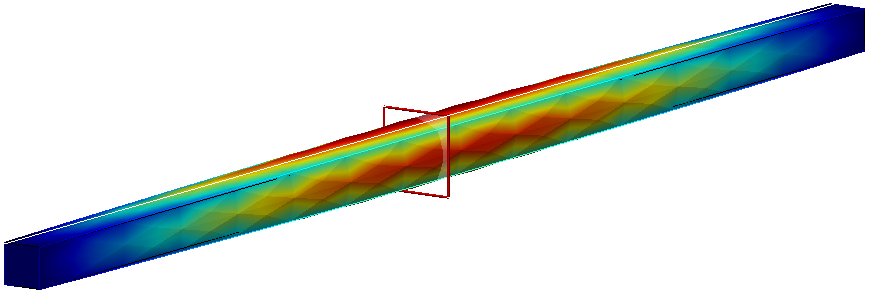} \rule{0pt}{2.25cm}
& \includegraphics[height=1.5cm]{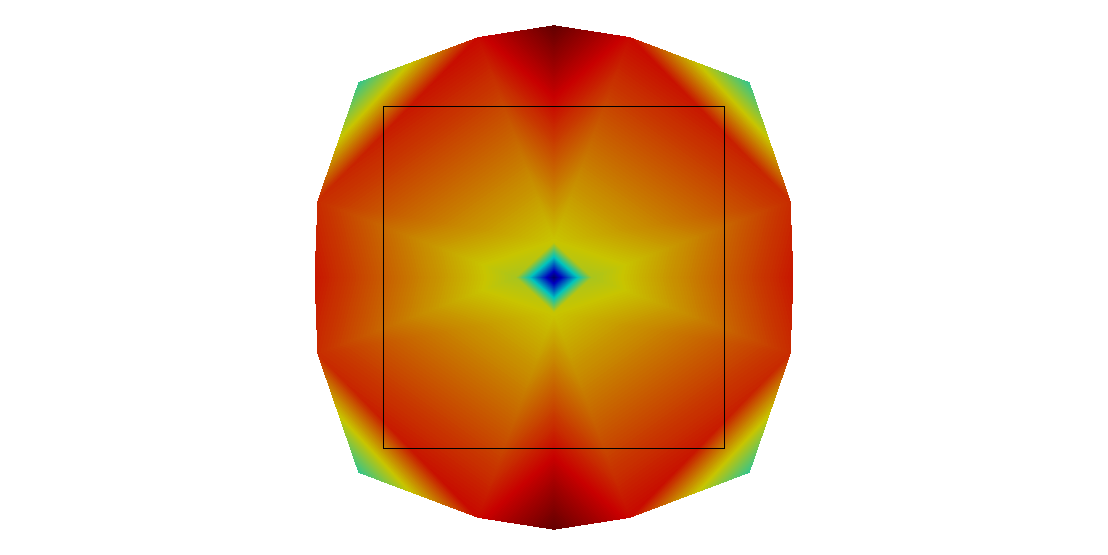} 
 \\
\hline
  34 
& 1.043e4 
& 9.26e-2
& \includegraphics[height=2.15cm]{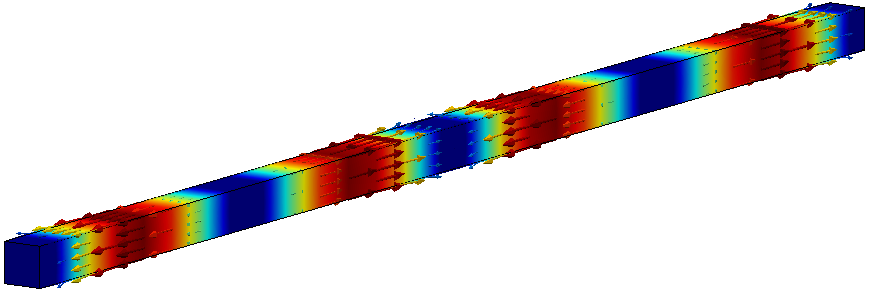} \rule{0pt}{2.25cm}
& \includegraphics[height=1.5cm]{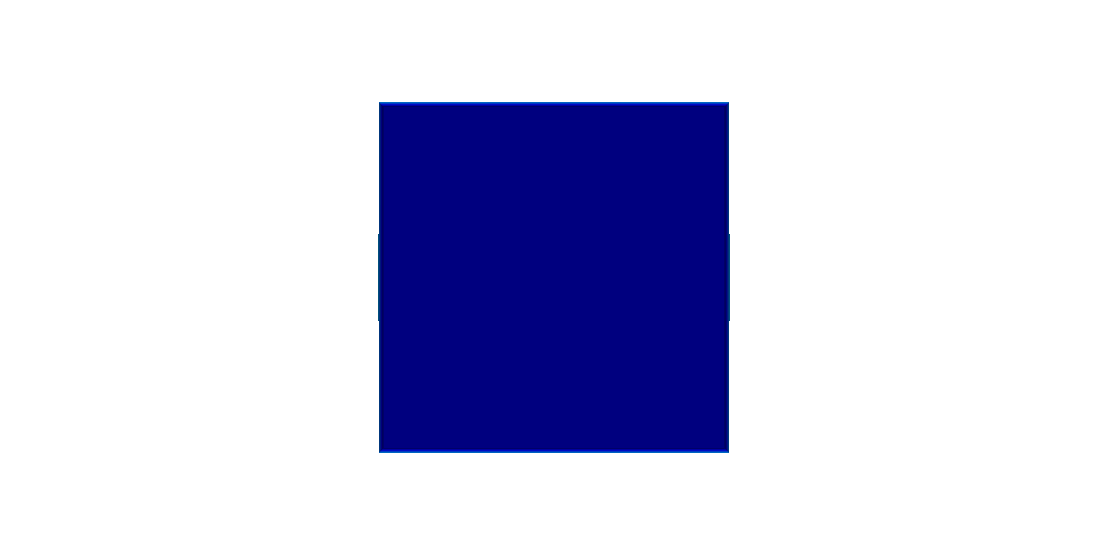} 
 \\
\hline
1147 
& 3.699e5 
& 7.01e-2
& \includegraphics[height=2.15cm]{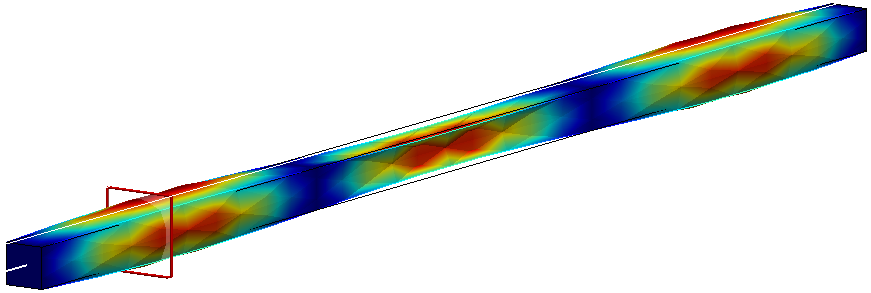}\rule{0pt}{2.25cm}
& \includegraphics[height=1.5cm]{s_1143_1147.png} 
 \\
\hline
370
& 1.202e5
& 4.60e-2
& \includegraphics[height=2.15cm]{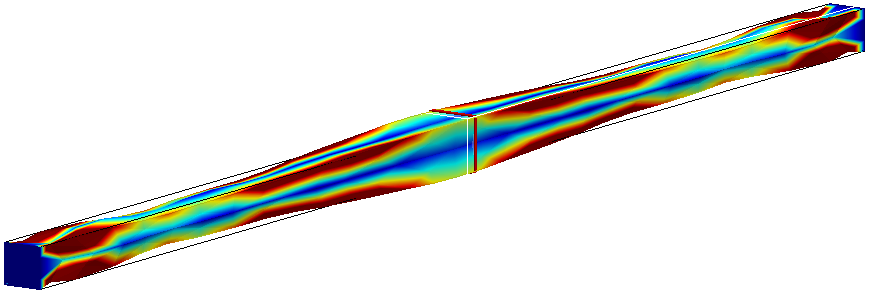}\rule{0pt}{2.25cm}
& \includegraphics[height=1.5cm]{s_324_328_330_370.png} 
 \\
\hline
167
& 7.644e4
& 3.35e-2
& \includegraphics[height=2.15cm]{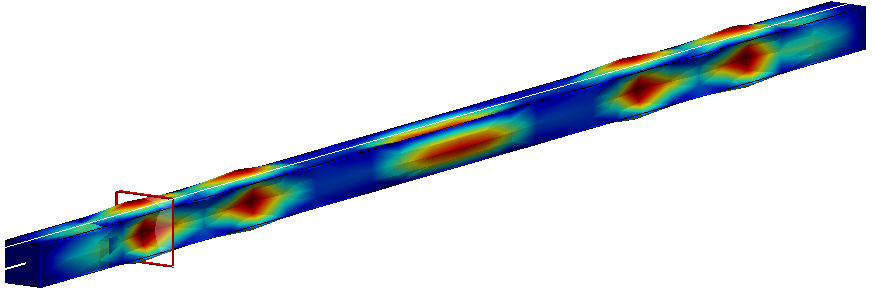}\rule{0pt}{2.25cm}
& \includegraphics[height=1.5cm]{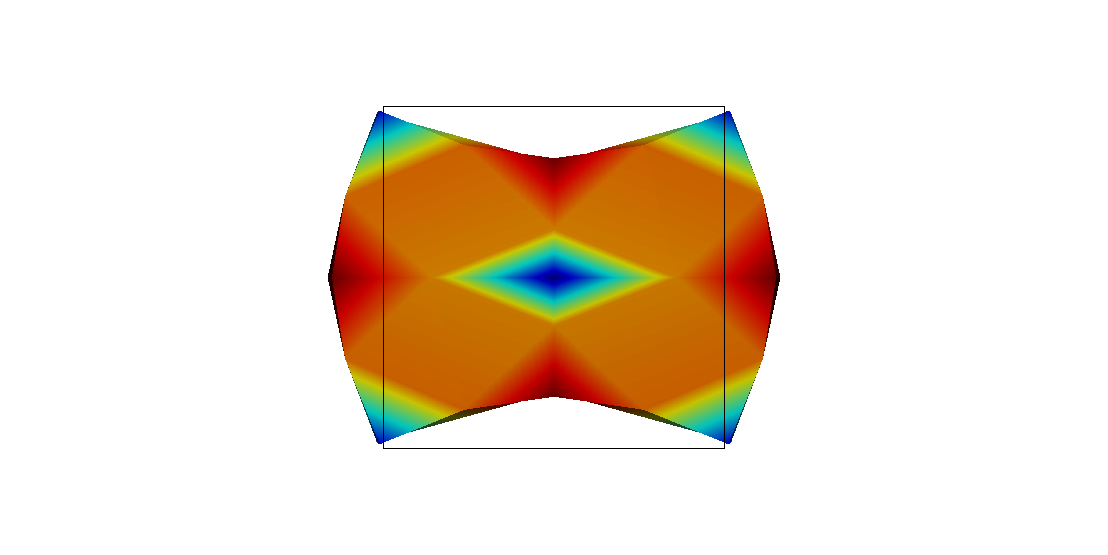} 
 \\
\hline
172
& 7.653e4
& 2.93e-2
& \includegraphics[height=2.15cm]{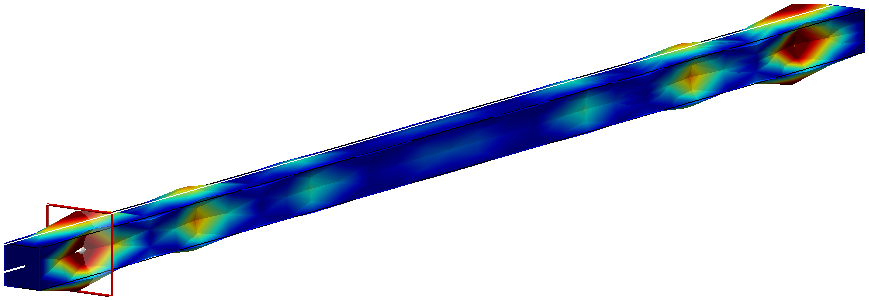}\rule{0pt}{2.25cm}
& \includegraphics[height=1.5cm]{s_167_172.png} 
 \\
\hline
324
& 1.112e5
& 2.89e-2
& \includegraphics[height=2.15cm]{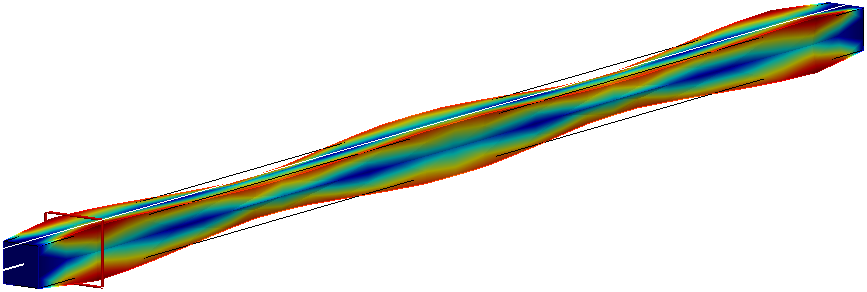}\rule{0pt}{2.25cm}
& \includegraphics[height=1.5cm]{s_324_328_330_370.png} 
 \\
\hline
\end{tabular}\end{center}
\hfill
Colormap\includegraphics[width=5cm]{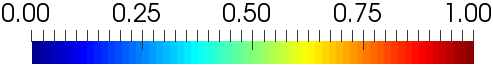}
\hspace{1.5cm}
\caption{Order of appearance in the basis (\#), 
eigenfrequencies, correction factors and shapes of the most relevant modes coupled with the first bending mode in $y$-direction for the thick clamped-clamped beam. The colors scale the modulus of the displacement field and arrows display the axial displacement for axial mode~34. The Poisson ratio is $\nu=0.3$.}
\label{tab:BeamModes}
\end{table*}

In order to gain insight into these coupled modes, Table \ref{tab:BeamModes} shows the associated mode shapes, sorted according to the importance of their contribution in the correction factor, thus following Fig.~\ref{fig:sorted}(a). The table shows the first nine eigenmode shapes, recalling in the first column their number of appearance when the modes are sorted according to the eigenfrequencies. One can observe that only one of these modes is a pure axial mode: the fourth one appearing in table \ref{tab:BeamModes}, also being the 34$^{th}$ by order of increasing frequencies. All the other ones involve important deformation in the thickness of the beam. They are thus called thickness modes, their presence being the direct consequence of 3D effects. The second column of Table \ref{tab:BeamModes} displays the eigenfrequencies, showing that they all are high-frequency modes. The last column shows the deformation of the section, showing the importance of thickness deformation.

\begin{figure*}[h!]
\centering
\begin{subfigure}{.49\textwidth}\centering
\includegraphics[scale=.5]{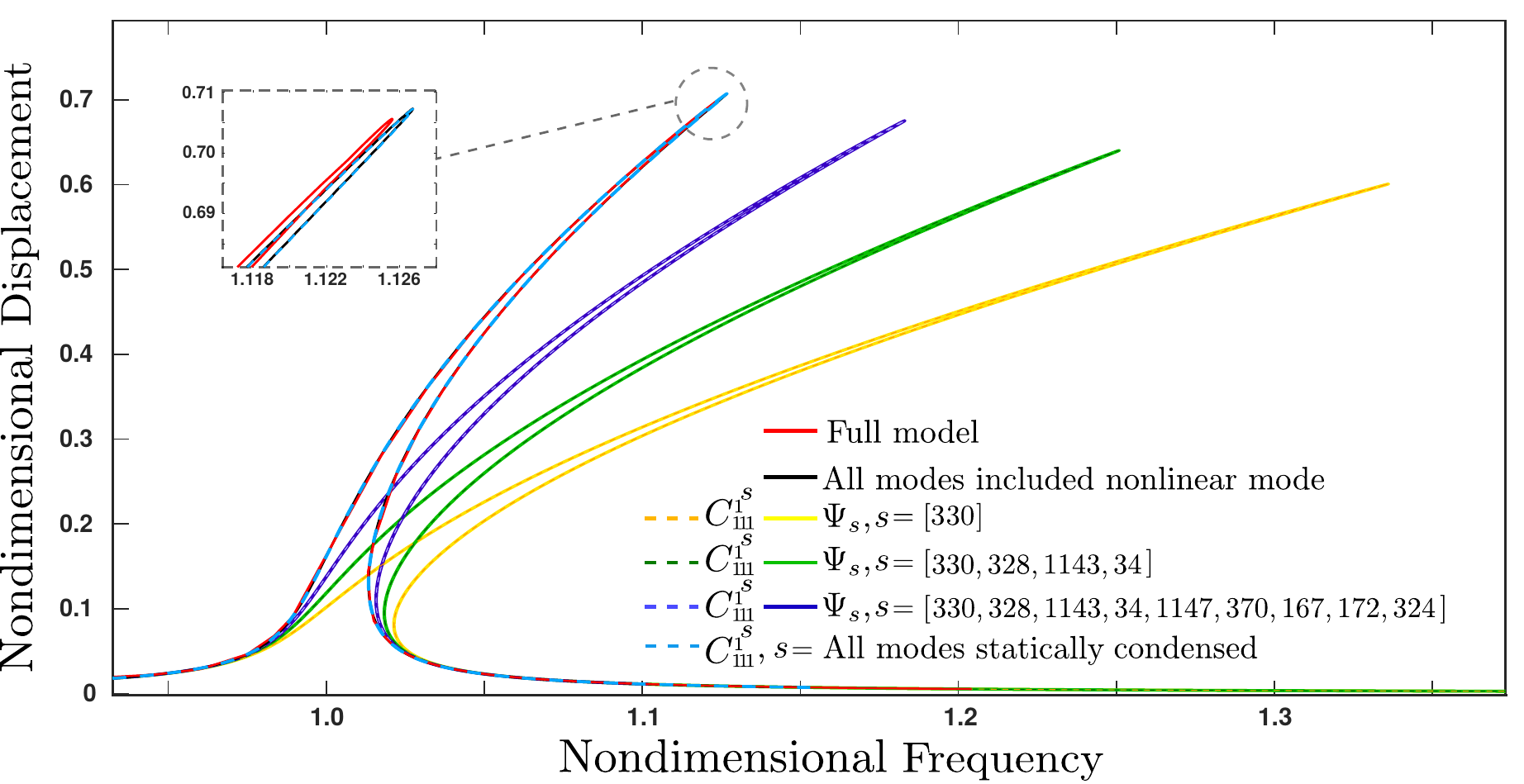}
\caption{Forced response functions at beam centre node}\label{fig:FRF}
\end{subfigure}
\begin{subfigure}{.49\textwidth}\centering
\includegraphics[scale=.5]{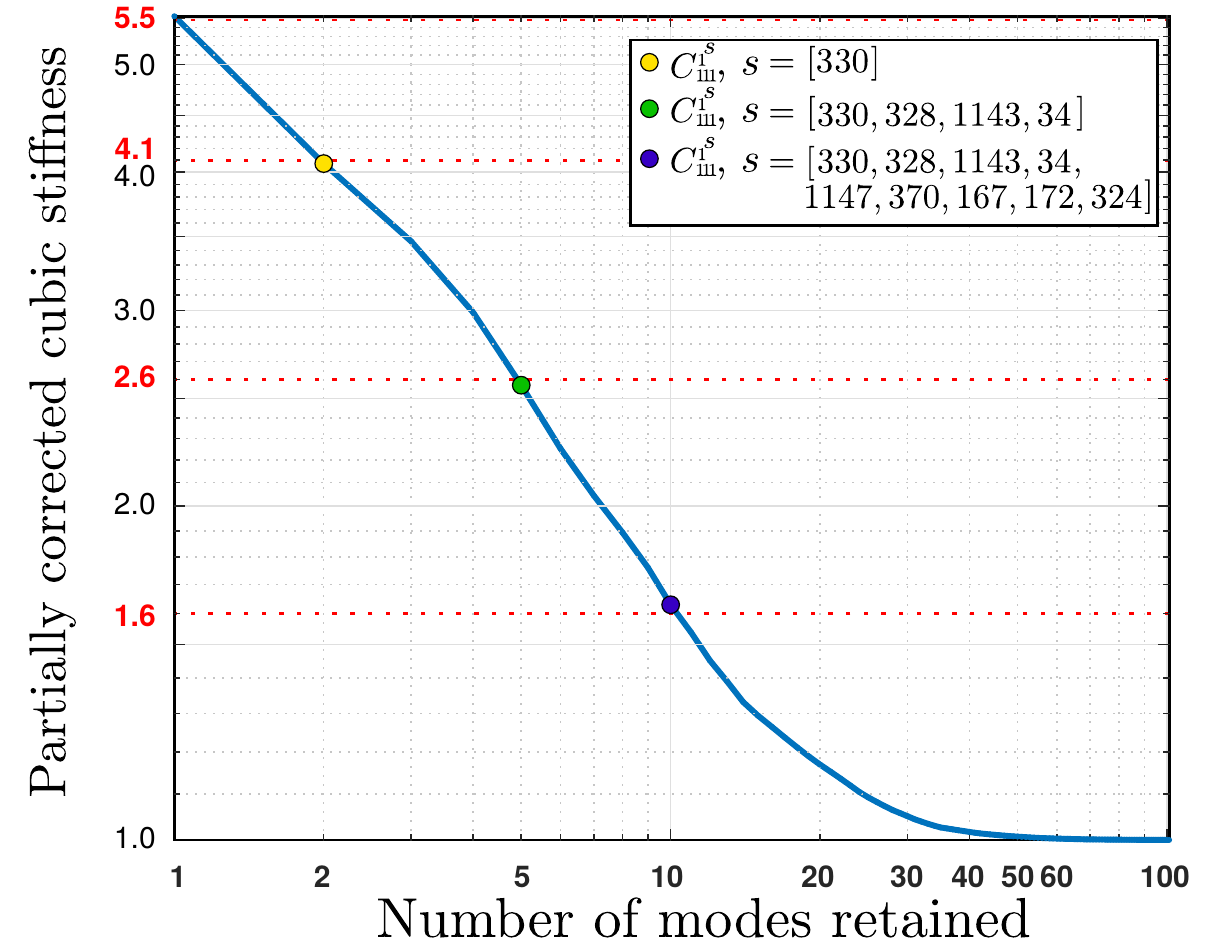}
\caption{Convergence of partially corrected cubic coefficients}\label{fig:CONV}
\end{subfigure}
\caption{Convergence to the full model solution with the increase of coupled modes taken into account in the reduced model. (a) Frequency response curve of the beam at center, in the vicinity of the first bending mode eigenfrequency; case of the thick beam with 1287 dofs. Red: full model, and convergence curves with increasing number of modes retained in the truncation: 1 mode (yellow), 4 modes (green), 9 modes (blue), all modes statically condensed (dashed light blue). Solution with one NNM in black. (b) Convergence of the evaluated corrected cubic stiffness coefficient $\Gamma^1_{111}$, defined in Eq.~\eqref{eq:Gammappp}, with increasing number of linear modes kept in the truncation.}\label{fig:FRF1}
\end{figure*}

The frequency-response curves of the thick beam in the vicinity of its first eigenfrequency is investigated in order to illustrate how the static condensation and the NNM approach are able to retrieve the correct nonlinear behaviour. Fig.~\ref{fig:FRF1}(a) shows the comparison of the solutions obtained by continuation, for different reduced-order models and the full model solution. The latter has been obtained by solving all the degrees of freedom of the system with a parallel implementation of harmonic balance method and pseudo-arc length continuation \cite{Blahos2020}; the computation of the full forced response with 3 harmonics lasted approximately 36 hours. The convergence of the solution using static condensation with an increasing number of modes to compute the correction is also shown. Despite only few modes have a very high correction factor $\mathcal{C}^{s1}_{111}$, i.e. play a major role in the decrease of $\beta^1_{111}$ (see Eq. \eqref{eq:Gammappp}), it is the sum of the contributions from all the coupled modes that makes the reduced model converge to the solution of the full one. In Fig. \ref{fig:FRF1}(b), the strong stiffening effect coming from not having included enough coupled modes, is slowly reduced by their inclusion in the basis; however, only the response obtained by static condensation of all coupled modes (cyan dashed) approximates the solution correctly (almost overlapped with the full model solution in red). On the other hand, the NNM solution with all the modes taken into account show also a direct convergence to the frequency-response curve.

Fig. \ref{fig:FRF1}(b) shows the convergence of the corrected cubic coefficient $\Gamma^1_{111}$ defined in Eq.~\eqref{eq:Gammappp} with the number of modes retained, i.e. the first mode plus the number of coupled modes condensed. When only one coupled mode is taken into account, the cubic coefficient $\beta^1_{111}$ overestimates largely $\Gamma^1_{111}$ (5.5 times): it is first explained by the classical bending-membrane coupling effect, and secondly to Poisson effect relating to the results given in Table \ref{tab:ErrorsCoeffsSTEP}. With 9 coupled modes the error on $\Gamma^1_{111}$ is still significant (more than 60\%). This strong overestimation of the cubic stiffness value results in the unrealistic stiffening effect observed in the forced response. The number of coupled modes that must be taken into account to ensure an acceptable accuracy makes the use of STEP in its first classic formulation (i.e. with the  eigenmodeshape functions as projection basis and without condensation) quite impractical: 44 modes give a 1\% error and 68 an error of 0.1\%. The condensation of these modes onto the excited one becomes then a viable option to drastically reduce the computational burden without affecting the accuracy of the solution.

\section{Alternative computational methods}

In the previous section, we have shown that in the case of 3D elements, a strong coupling with thickness modes occurred, thus rendering the convergence of the modal ROM particularly stringent. When one is able to compute all the linear modes and associated coefficients, then static condensation and normal form approach can be used to finally produce accurate ROMs. However in most of the cases, the computation of all the linear modes, including the thickness modes appearing at very high frequencies, is out of reach. In this section, we investigate two alternative methods, for which there is no need to compute all the linear modes: static modal derivative, and a modified STEP.

\subsection{Static modal derivatives}\label{sec:SMD}

Sections \ref{sec:2} and \ref{sec:appl_STEP3D} were devoted to the derivation of a reduced order model from a modal point of view. In fact, a modal projection of the quadratic nonlinear forces onto each mode $\phi_s$ is required to obtain the coefficients $\alpha^s_{pp}$. Here we want to introduce the concept of static modal derivatives (see \cite{IDELSOHN1985}) because its application provides the same results as the static condensation of all non-bending modes, but without requiring the computation of their associated  eigenvectors.

The definition of modal derivatives have arisen from the recognition of the fact that in the nonlinear range, mode shapes and frequencies depend on amplitude~\cite{IDELSOHN1985}. Introducing this dependency in the eigenproblem defining the modes, one arrives at a quantity defined as the modal derivative~\cite{IDELSOHN1985,Weeger2016}. Following the definition of static modal derivatives $\vect{\theta}_{pr}$ (SMD) from \cite{Jain2017}, it reads:
\begin{equation}
\vect{\theta}_{pr}=-\vect{K}^{-1}\left.\left(\dfrac{\partial}{\partial q_p}
\dfrac{\partial\vect{f}_\text{nl}}{\partial\vect{x}}(\boldsymbol{\phi}_p q_p) 
\right)
\right|_{q_p=0}
\cdot
\boldsymbol{\phi}_r.\label{eq:Jain}
\end{equation}
When $p=r$ a more convenient expression (in the point of view of its direct computation from a FE code) for the modal derivative $\vect{\theta}_{pp}$ writes:
\begin{equation}\label{eq:SMD_fnl}
\vect{\theta}_{pp}=-\vect{K}^{-1}\left(\dfrac{\vect{f}_\text{nl}(\lambda\vect{\phi}_p)+\vect{f}_\text{nl}(-\lambda\vect{\phi}_p)}{\lambda^2}\right).
\end{equation}
The equivalent general expression for $\vect{\theta}_{pr}$, with $p\neq r$ is provided in Appendix \ref{app:B}. 
Eq.~\eqref{eq:SMD_fnl} shows how the SMD can be easily computed from a set of applied static displacement, in a manner having analogies with the STEP.  The term in parenthesis in Eq.~\eqref{eq:SMD_fnl} can be seen as the numerical second order derivatives with respect to $\lambda$ of the nonlinear force along mode $p$, evaluated at the equilibrium position. This can be easily evaluated in any  FE software by imposing, on a nonlinear structure, a displacement proportional to the linear $p$-th mode with first positive and then negative sign in order to isolate the quadratic part of the nonlinear forces. 

The SMD can thus be seen as an added displacement vector that enriches the basis constituted by the linear mode $p$, in a way that takes into account the nonlinear deformation of the structure. The use of SMDs as added vectors in the reduced order model basis is extensively documented in \cite{Tiso2011,Weeger2016,SOMBROEK2018,Rutzmoser,Jain2017}, and a complete comparison of quadratic manifolds derived from modal derivatives with normal for theory is given in~\cite{VizzaMDNNM}. Here the focus is on the relationship between modal derivatives and non-bending modes and on the equivalence between static condensation of all non-bending modes and static condensation of modal derivatives.

\begin{figure*}[h!]\centering
\begin{subfigure}{.49\textwidth}\centering
\includegraphics[scale=.3]{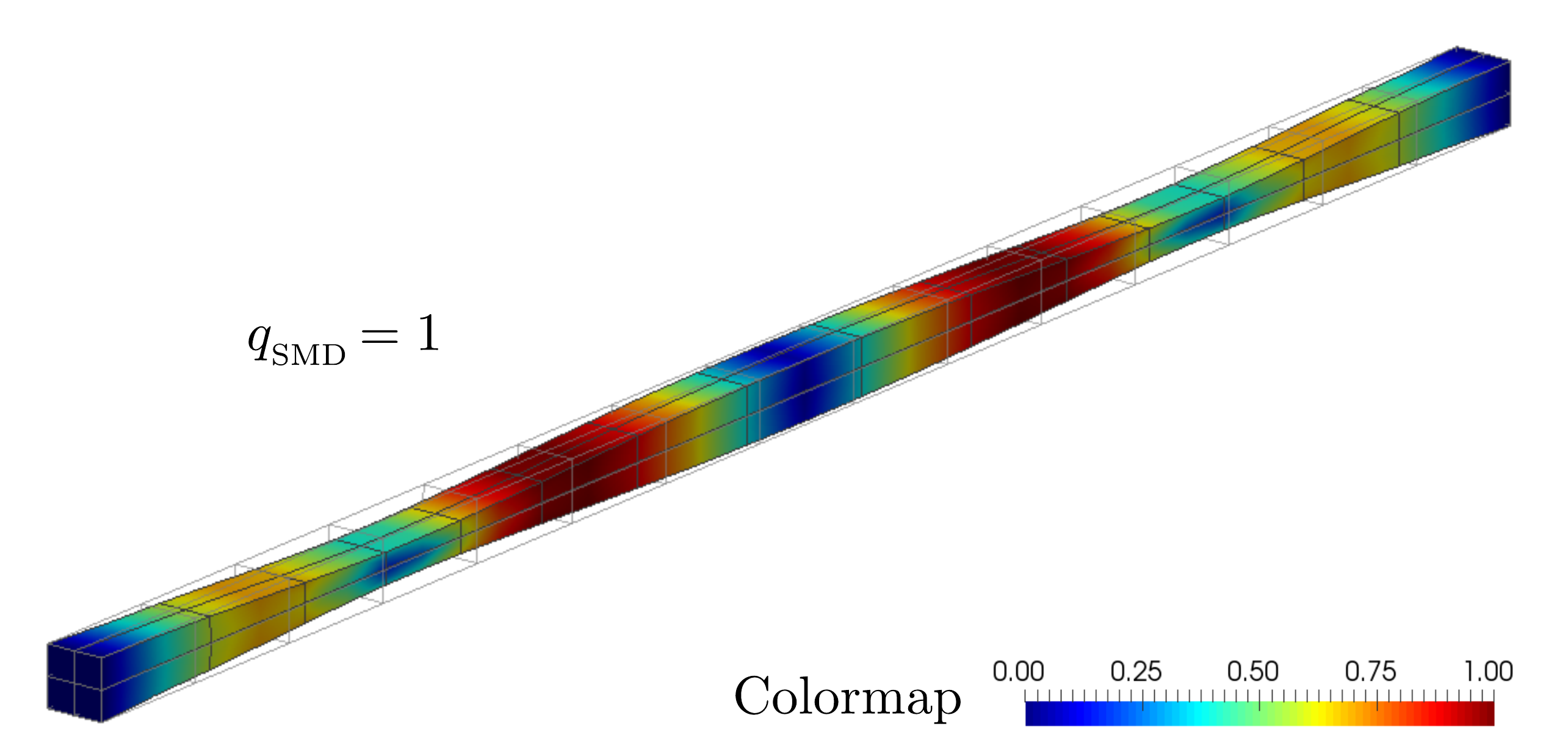}
\caption{Static modal derivatives}\label{fig:SMD_a}
\end{subfigure}
\begin{subfigure}{.49\textwidth}\centering
\includegraphics[scale=.3]{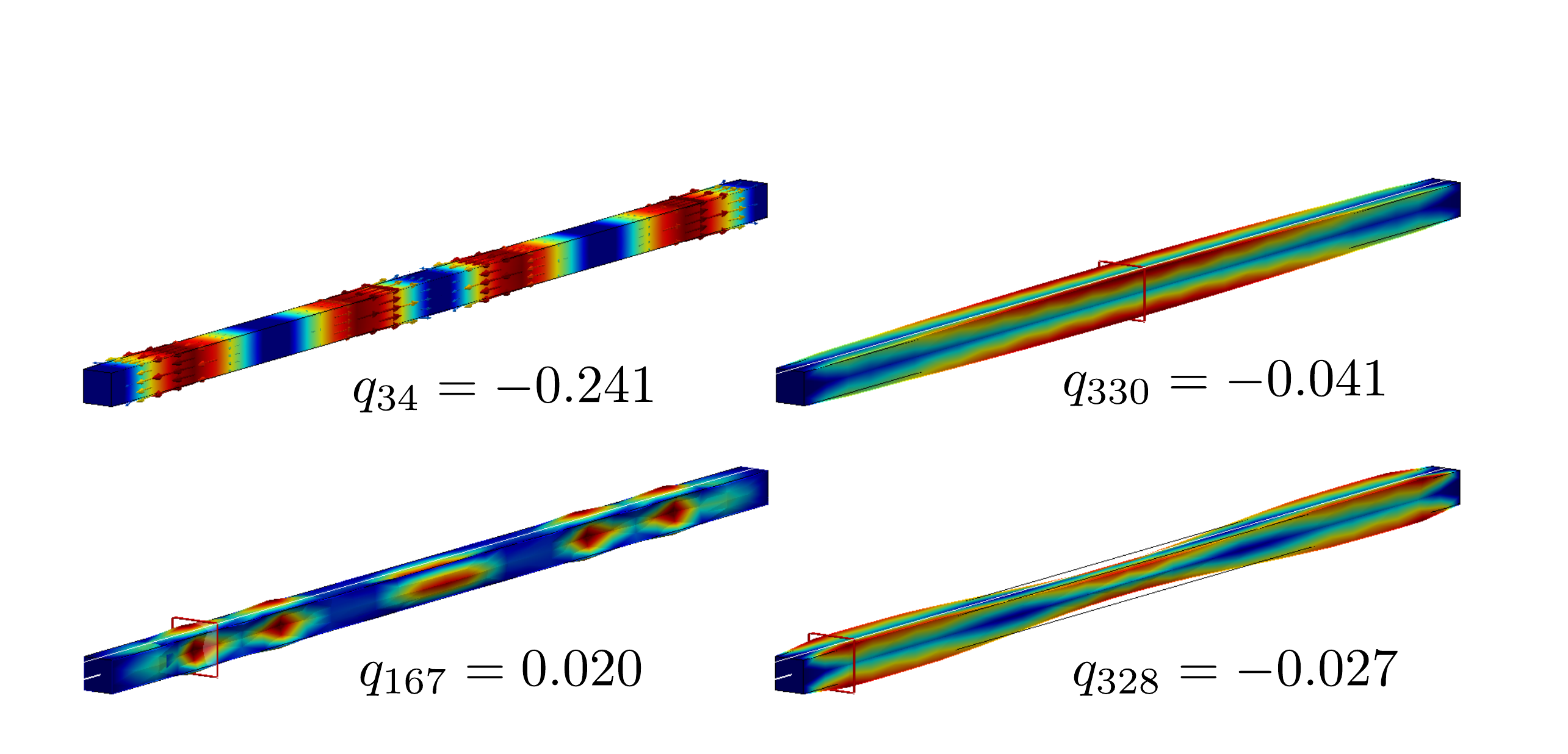}
\caption{Non-bending modes}\label{fig:SMD_b}
\end{subfigure}
\caption{(a) Static modal derivative $\bm{\theta}_{11}$ associated to the first bending mode. (b) Four first non-bending modes contained in the SMD $\bm{\theta}_{11}$, found equivalently by projecting $\bm{\theta}_{11}$ on the linear mode basis, or by application of Eq.~\eqref{eq:SMD_pp}. The relative modal participation factors $q_s$ of each of these modes numbered 34, 167, 328 and 330 (see also Table~1) is also numerically given, normalized by the total amplitude of the SMD $\bm{\theta}_{11}$ ($q_{SMD}=1$), and exactly recovers the factors exhibited in Eq.~\eqref{eq:SMD_pp}.}\label{fig:SMD}
\end{figure*}

Given a system with geometric nonlinearities up to cubic order, the static modal derivatives related to the $p$-th and $r$-th linear modes can then be expressed in terms of the vector of quadratic coefficients $\vect{\alpha}_{pr}$ as:
\begin{equation}
\label{eq:link_pr}
\bm{\theta}_{pr}=-\vect{V}\vect{\Omega}^{-2}\;\bm{\alpha}_{pr}\, ,
\end{equation}
and the one relative to the $p$-th linear mode as:
\begin{equation}\label{eq:link_pp}
\bm{\theta}_{pp}=-2\;\vect{V}\vect{\Omega}^{-2}\;\bm{\alpha}_{pp}\, ,
\end{equation}
where the full matrix of eigenvectors $\vect{V}$ has been introduced. The detailed derivation of these two equations is given in Appendix \ref{app:B} together with the classical orthonormality properties of the matrix of eigenvectors $\vect{V}$. By expanding Eq.~\eqref{eq:link_pp} over all the modes and by noticing that, for a flat structure, $\alpha^s_{pp}$ is non-zero only when $s$ is a non-bending mode, one obtains this important relationship (see Appendix \ref{app:B} and~\cite{VizzaMDNNM}):
\begin{equation}\label{eq:SMD_pp}
\vect{\theta}_{pp}=-\sum^{N}_{s=N_B+1}2\vect{\phi}_s \dfrac{\alpha^s_{pp}}{\omega^2_s}.
\end{equation}
The SMD thus appears as a linear combination of coupled modes with factor $-2\alpha^s_{pp}/\omega^2_s$; therefore, it can be seen as a displacement field that takes into account the contribution of all non-bending modes into one equivalent vector. To show this property, the SMD relative to the first bending mode of the thick beam test case is depicted in Fig.~\ref{fig:SMD_a}. The projection of the SMD onto the linear modes of the system recovers Eq.~\eqref{eq:SMD_pp}.  The contributions from the non-bending modes that have been identified in previous calculations, the fourth axial mode as well as various thickness modes, appear in the SMD. For each mode $s$, The modal amplitudes $q_s$ obtained from the projection coincides with those from Eq.~\eqref{eq:SMD_pp} i.e. they are equal to $-2{\alpha^s_{pp}}/{\omega^2_s}$.
These results recover and elaborate on those obtained in~\cite{Jain2017,SOMBROEK2018}, where it was shown that axial modes are contained in the SMDs of bending modes. The result is here extended to thickness modes and specified since the exact participation factor of each mode is made explicit.

Once understood that the SMD allows gathering in a single vector the participation of all coupled modes, we want to show how to retrieve directly, from the calculation of the SMD, the correct nonlinear behaviour of the structure, when the motion is restricted to a single master mode. In the specific case of a flat structure, Eq.~\eqref{eq:pstatic} shows that the amplitudes of the NB modes can be explicitely related to the squared amplitude of the single master mode labeled $p$, thanks to the static condensation, as:
\begin{equation}\label{eq:p_s}
p_s = - \dfrac{\alpha^s_{pp}}{\omega^2_s}q_p^2.
\end{equation}
The physical displacement $\bm{x}(q_p)$ that corresponds to the solution gathering together the master bending mode $p$ and all its coupled NB modes can be written as:
\begin{equation}
\bm{x}(q_p)=q_p\bm{\phi}_p+\sum^{N}_{s=N_B+1}p_s\;\vect{\phi}_s.
\end{equation}
In this last equation, replacing $p_s$ by its value obtained from static condensation, Eq.~\eqref{eq:p_s}, and then using Eq.~\eqref{eq:SMD_pp} defining $\vect{\theta}_{pp}$ as a summation on the NB modes, one arrives easily at the fact that this physical displacement can be expressed as a function of the modal coordinate plus the participation of the SMD as:
\begin{equation}\label{eq:qm}
\bm{x}(q_p)=q_p\bm{\phi}_p+\frac{1}{2}q_p^2\bm{\theta}_{pp}\;.
\end{equation}

If one wants now to derive a reduced-order model composed of a single master coordinate (say $q_p$ here) and that contains the correct nonlinear behaviour, then the equation of motion would simply read:
\begin{equation}\label{eq:romSMDsc}
\ddot{q}_p+ 2\zeta_p\omega_p\dot{q}_p+ \omega_p^2q_p+ \tilde{\Gamma}_{ppp}^p q_p^3 = Q_p,
\end{equation}
with $\tilde{\Gamma}_{ppp}^p$ a corrected cubic coefficient. Thanks to Eq.~\eqref{eq:zicoefSTEPq}, one knows that a cubic coefficient can be found from this computation, provided the imposed displacement is selected correctly. If the imposed displacement is along a linear mode, as in Eq.~\eqref{eq:zicoefSTEPq}, then one will retrieve the modal nonlinear coupling coefficient, but other choice of imposed displacement can be made. In particular, if one selects the one given by Eq.~\eqref{eq:qm}, then the cubic coefficient $\tilde{\Gamma}_{ppp}^p$ will contain the contribution of the master mode plus that of the SMD. Since Eq.~\eqref{eq:SMD_pp} shows that in our particular case (flat structure, one master mode), the SMD is completely equivalent to the static condensation of all coupled NB modes, then one will easily understand that $\tilde{\Gamma}_{ppp}^p = \Gamma_{ppp}^p$, the corrected cubic coefficient given in Eq.~\eqref{eq:Gammappp}. The complete proof of this result is provided in Appendix \ref{app:C}. 

The main result, from the SMD perspective, is that if one restricts to a single master mode $p$, then the SMD can be easily computed thanks to Eq.~\eqref{eq:SMD_fnl}. Then the corrected cubic coefficient can be directly computed from:
\begin{equation}\label{eq:Gamma_smd}
\Gamma^p_{ppp}=\tp{\bm{\phi}_p}\left(
\vect{f}_\text{nl}(\bm{x}(\lambda))-\vect{f}_\text{nl}(\bm{x}(-\lambda))\right)/2\lambda^3,
\end{equation}
where the imposed displacement is selected as in Eq.~\eqref{eq:qm}. This procedure allows then to find exactly the same corrected cubic coefficient of the master mode, but without resorting to the computation of all eigenvectors, as needed in the static condensation. It is thus a much more computationally efficient to use this methodology. Numerical examples are provided in Section \ref{sec:num}.

\subsection{A modified STEP for 3D elements}\label{sec:mstep}

\begin{figure*}[h]
\begin{center}
\includegraphics[scale=.6]{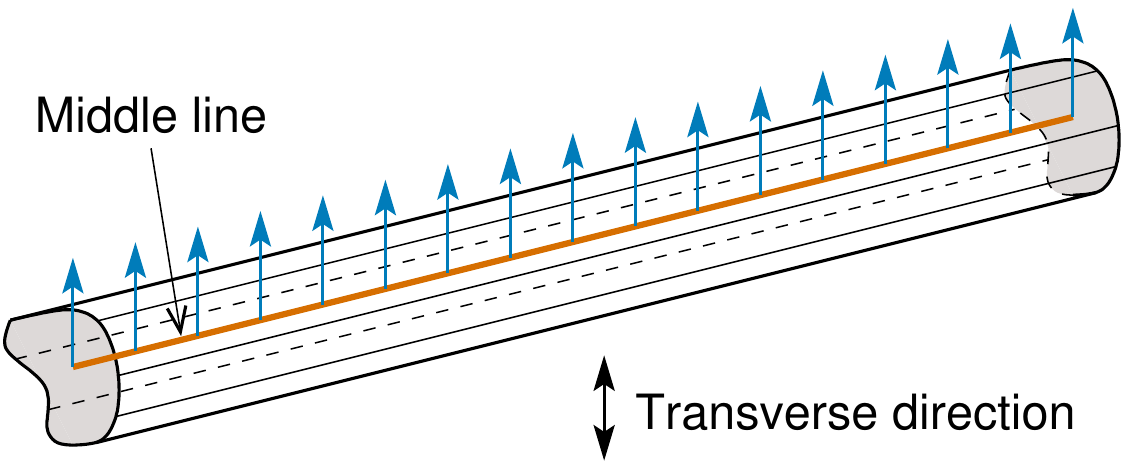}\hspace{4em}
\includegraphics[scale=.6]{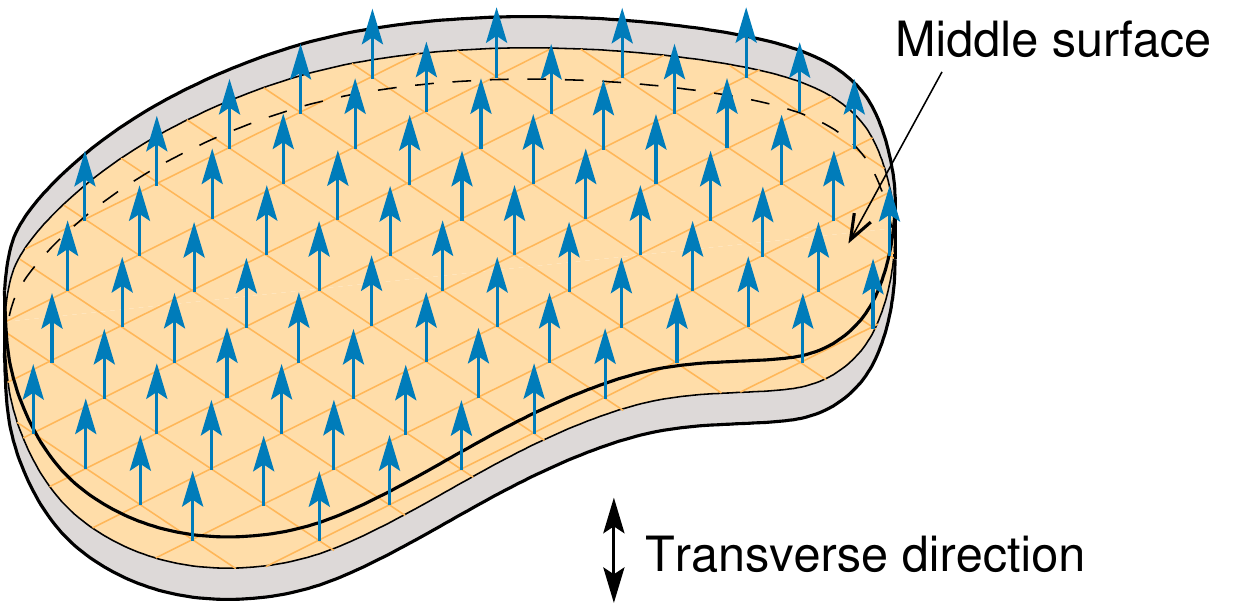}
\caption{Examples of prescribed displacement field  of the M-STEP, in the transverse direction and in the middle surface / line of a plate / beam}
\label{fig:MSTEP}
\end{center}
\end{figure*}

As observed in the previous sections, the modal ROM associated to 3D FE discretization shows a slow convergence because of the couplings with very high frequency modes involving thickness deformations. Since these thickness modes are a peculiarity of the 3D model, they have no counterpart in plate or beam theory, which concentrate the kinematical description on the middle plane / line. Also, using the STEP with plate / beam elements show a faster convergence since one has to recover only the well-known coupling between bending and in-plane motions. In order to circumvent these difficulties, we propose here to modify the STEP by prescribing the displacements only on the middle line / plane  of the structure, and to let free the other degrees of freedom (Fig.~\ref{fig:MSTEP}). The idea is to include automatically the effects of NB modes, by a kind of implicit condensation of their motion, embedded into the prescribed displacement on the middle line / plane. The obtained method is called the M-STEP, for Modified-STEP. Note that a comparable idea has also been introduced in~\cite{KimCantilever,wangohara}, but for 2D flat structures only, where only transverse motions were prescribed, leaving the other degrees of freedom free and thus building directly a condensed model.

\subsubsection{Formulation}
\label{subsec:MStepFormulation}

We show in this section that it is possible to compute directly the cubic coefficients $\Gamma_{ijk}^r$  of Eq.~(\ref{eq:qcubic}) with a modified STEP, without having to compute all the coefficients $\alpha_{ij}^k$ and $\beta_{ijl}^k$ beforehand. Note that in the classical STEP, the three components of the displacement field are prescribed to all the nodes of the FE mesh: the whole vector of unknown $\vect{x}$ is imposed and the FE code computation is just an evaluation of the internal force vector $\vect{f}_e=\vect{f}(\vect{x})$. 

%
Here, we choose to apply the STEP by prescribing the displacement field only to selected nodes and for selected components. Precisely, we denote by $\mathcal{S}$ the middle plane / line of the structure and $\vect{n}$ the bending direction. To compute $\Gamma_{ppp}^k$ for a given $p\in\{1,\ldots N_B\}$, we choose to perform a FE computation by prescribing (i) only the transverse component of the bending mode $\vect{\phi}_p$ (ii) only on the nodes of the middle plane / line of the structure (Fig.~\ref{fig:MSTEP}). We then prescribe the following time independent displacement to the structure:
\begin{equation}
\vect{x}|_{\mathcal{S},\vect{n}} = \lambda\, \vect{\phi}_p|_{\mathcal{S},\vect{n}},\quad \lambda\in\mathbb{R}.
\end{equation}
where $\vect{x}|_{\mathcal{S},\vect{n}}$ corresponds to displacement $\vect{x}$ restricted to (i) the nodes of the FE mesh belonging to $\mathcal{S}$ and to (ii) its components along the direction defined by $\vect{n}$. In all the other nodes and directions, a zero forcing is prescribed. We then use the finite element code to solve this problem and obtain $\vect{x}$ as well as $\vect{f}(\vect{x})=\vect{f}_e$ everywhere. Since some components of $\vect{x}$ are not prescribed, a Newton-Raphson procedure is necessary to solve this nonlinear algebraic problem.

To precise the method, we call master dofs the ones for which the displacement is prescribed, and slave dofs the other ones. The full displacement vector $\vect{x}$ and the internal forcing $\vect{f}$ can thus be decomposed as:
\begin{equation}
\vect{x}=
\begin{bmatrix}
\vect{x}_{\text{M}}\\
\vect{x}_{\text{S}}
\end{bmatrix},\qquad 
\vect{f}=
\begin{bmatrix}
\vect{f}_{\text{M}}\\
\vect{f}_{\text{S}}
\end{bmatrix},
\end{equation}
where the index $\text{M}$ and $\text{S}$ are associated to the master and slave dofs, respectively. The M-STEP consists in prescribing:
\begin{equation}
\label{eq:MSTEPdef}
\begin{cases}
\vect{x}_{\text{M}}=\lambda \vect{\phi}_p^\text{M}, \\
\vect{f}_{\text{S}}=\vect{0}.
\end{cases}
\end{equation}
Then, solving the static problem $\vect{f}(\vect{x})=\vect{f}_e$ with the FE code leads to compute the internal force vector $\vect{f}_\text{M}(\vect{x})$ on the master nodes and the displacement  $\vect{x}_S$ on the slave nodes. The solution of the problem then reads:
\begin{equation}
\label{eq:MSTEP}
\vect{x}=
\begin{bmatrix}
\lambda\vect{\phi}_p^\text{M}\\
\vect{x}_{\text{S}}
\end{bmatrix},\qquad 
\vect{f}=\vect{f}_e=
\begin{bmatrix}
\vect{f}_{\text{M}}\\
\vect{0}
\end{bmatrix}.
\end{equation}

Translated in the modal space, the above computations are close to the following situation. Prescribing via $\vect{x}$ only the transverse motion in the form of $\vect{\phi}_p$, and because $\vect{\phi}_p$ is orthogonal to the other bending modes $\vect{\phi}_k$, $k\neq p$, the modal coordinate are $q_p\simeq\lambda$ and $q_k\simeq0$. Considering the orthogonality relations associated to the stiffness matrix, this leads to assume that, for all $s\in\{1,\ldots N_B\}, \; s\neq p$:
\begin{equation}
\label{eq:quasiortho}
\tp{\vect{\phi}_p}\bm{K}\vect{x}\simeq\tp{\vect{\phi}_p}\bm{K}(\lambda \vect{\phi}_p)=\lambda\omega_p^2 m_p,\quad \tp{\vect{\phi}_s}\bm{K}\vect{x}\simeq\tp{\vect{\phi}_s}\bm{K}(\lambda \vect{\phi}_p)=0.
\end{equation}
In other words, it is assumed that the nonzero slave part $\vect{x}_\text{S}$ of $\vect{x}$ is not involved in the orthogonality relations of the bending modes. Moreover, since the part of the displacement associated to longitudinal and thickness displacements is not prescribed by $\vect{x}$, the associated NB modal coordinates are not zero and their value depend on the nonlinear coupling and the geometric nonlinearities. Finally, because any NB eigenmode $\vect{\phi}_s$ has zero displacements on the middle plane / line in the transverse direction, the modal forcing of the NB modes is exactly zero:
\begin{equation}
\vect{\phi}_s=
\begin{bmatrix}
\vect{0} \\
\vect{\phi}_s^{\text{S}}
\end{bmatrix}\quad\Rightarrow\quad
Q_s=\frac{\tp{\vect{\phi}_s}\vect{f}_e}{m_s}=0.
\end{equation}
To summarize, one has:
\begin{subequations}
\label{eq:assume1}
\begin{empheq}[left=\empheqlbrace]{align}
& q_p\simeq\lambda,\\
& q_k\simeq0\qquad \forall k=1,\ldots N,\;k\neq p\\
& Q_s=0 \qquad  \forall s=N_B,\dots,N
\end{empheq}
\end{subequations}
Error estimates of those assumptions will be introduced in section~\ref{sec:quality}.

Using the assumptions (\ref{eq:assume1}) in Eqs.~(\ref{eq:qsimple},\ref{eq:psimple}) leads to:
\begin{empheq}[left=\empheqlbrace]{align}
\lambda \omega^2_p +\sum_{s=N_B+1}^{N} \lambda\alpha^p_{ps}  p_s +\lambda^3\beta^p_{ppp}&=\tp{\vect{\phi}_p}\vect{f}(\vect{x})/m_p,
\label{p_mode}\\
\sum_{s=N_B+1}^{N} \lambda \alpha^k_{ps} p_s +\lambda^3 \beta^k_{ppp}&=\tp{\vect{\phi}_k}\vect{f}(\vect{x})/m_k, && \forall k=1,\dots,N_B,\,k\neq p
\label{k_mode}\\
\omega^2_s p_s + \lambda^2\alpha^s_{pp}& =  0, &&\forall s=N_B+1,\dots,N
\label{s_mode}
\end{empheq}
The above Eq.~(\ref{s_mode}) leads to:
\begin{equation}
p_s =-\dfrac{ \alpha^s_{pp} \lambda^2}{\omega^2_s },
\end{equation}
that can be condensed into~(\ref{p_mode}),(\ref{k_mode}) to give:
\begin{align}
\left(\beta^p_{ppp}-\sum_{l=N_B+1}^{N} \frac{\alpha^p_{ps}  \alpha^s_{pp}}{\omega^2_s } \right)\lambda^3 & = \tp{\vect{\phi}_p}\vect{f}(\vect{x}) /m_p - \lambda\omega_p^2,\label{p_mode_cond} \\
\left(\beta^k_{ppp}-\sum_{l=N_B+1}^{N} \frac{\alpha^k_{ps}  \alpha^s_{pp}}{\omega^2_s } \right)\lambda^3 & = \tp{\vect{\phi}_k}\vect{f}(\vect{x}) /m_k,\quad \forall k\neq p.
\label{k_mode_cond}
\end{align}
One then recognizes the terms in parenthesis as the sought corrected cubic coefficients $\Gamma^i_{ppp}$, defined by Eqs.~(\ref{eq:Gammappp}) and (\ref{eq:Gammadef}), so that: 
\begin{equation}
\Gamma^p_{ppp}=\frac{\tp{\vect{\phi}_p} \vect{f}(\vect{x})}{m_p \lambda^3}-\frac{\omega_p^2}{\lambda^2},\qquad \Gamma^k_{ppp}=\frac{\tp{\vect{\phi}_k} \vect{f}(\vect{x})}{m_k \lambda^3}.
\end{equation}
Consequently, all $\Gamma_{ppp}^i$, $i=1,\ldots N_B$ relies on a single nonlinear static finite elements computation, defined by Eq.~(\ref{eq:MSTEP}). Other coefficients $\Gamma_{ijl}^k$ can be obtained in the same manner, by mixing different $x_\text{M}$ on several bending modes, following the classical STEP.

This above described  M-STEP method is a way of automatically embed in the computation the effect of all the NB modes nonlinearly coupled to the bending modes associated to $\Gamma_{ijl}^k$. The essence of the method is to select the prescribed displacement $\vect{x}$ so that (i) it leaves free the degrees of freedom associated to the NB modes, so that the forcing $Q_s$ of the longitudinal modal coordinates in Eq.~(\ref{s_mode}) is exactly zero and (ii) it is as orthogonal as possible to the other bending modes than the $p$-th.

\subsubsection{Quality indicator for the convergence of the method}
\label{sec:quality}

In order to be able to quantify {\em a priori} the quality of the computation, a main idea is to check the validity of  the assumptions used in the two first equations~(\ref{eq:assume1}), which are true at first order but might deteriorate in case of an incorrect selection of master dofs. Equivalently, one can  verify   the orthogonality of the displacement vector $\vect{x}$ to the bending modes written in Eqs.~(\ref{eq:quasiortho}). To that purpose, let us define the following errors:
\begin{equation}
e_1^{pp} = \frac{\tp{\vect{\phi}_p}\bm{K}\vect{x}}{\lambda\omega_p^2 m_p}-1,\qquad e_1^{pk} = \frac{\tp{\vect{\phi}_k}\bm{K}\vect{x}}{\lambda\omega_p^2 m_p},\quad \forall k\neq p,
\end{equation}
that should be small as compared to 1 because of Eqs.~(\ref{eq:quasiortho}). If one wants to compute those errors with a FE in a non intrusive way, $\bm{K}\vect{x}=\vect{f}_1$ can be computed as the reaction force vector $\vect{f}_1$ of a linear static computation where $\vect{x}$ is prescribed to all the nodes of the FE mesh. 

Another check can also be performed by prescribing Eq.~(\ref{eq:MSTEPdef}) into a linear static computation:
\begin{equation}
\label{eq:Kxl=f}
\mat{K}\vect{x}_\text{l} = \vect{f}_e, 
\end{equation}
which gives:
\begin{equation}
\label{eq:MSTEP}
\vect{x}_\text{l}=
\begin{bmatrix}
\lambda\vect{\phi}_p^\text{M}\\
\vect{x}_{\text{lS}}
\end{bmatrix}.
\end{equation}
Since there are no geometrical nonlinearities, imposing $\vect{\phi}_p$ on the middle line / surface in the transverse direction should result in a vector almost collinear to $\vect{\phi}_p$, that is $\vect{x}_\text{l}\simeq\lambda\vect{\phi}_p$. In particular, the slave part $\vect{x}_{\text{lS}}$ of $\vect{x}_{\text{l}}$ should be very close to $\vect{\phi}_p^\text{S}$, the slave part of $\vect{\phi}_p$. We then define the following error:
\begin{equation}
e(\hat{\vect{y}},\vect{y})=\frac{||\hat{\vect{y}}-\vect{y}||}{||\vect{y}||},
\end{equation}
where $||\cdot||$ is the norm of vector $\cdot$, and we check that $e_2^p = e(\vect{x}_l,\lambda\vect{\phi}_p)$ is very small as compared to 1.

\section{Physical mechanisms of the nonlinear couplings}

\subsection{Poisson effect}

\begin{figure}[h]
\begin{center}\def\svgwidth{9cm}
\begingroup%
  \makeatletter%
  \providecommand\color[2][]{%
    \errmessage{(Inkscape) Color is used for the text in Inkscape, but the package 'color.sty' is not loaded}%
    \renewcommand\color[2][]{}%
  }%
  \providecommand\transparent[1]{%
    \errmessage{(Inkscape) Transparency is used (non-zero) for the text in Inkscape, but the package 'transparent.sty' is not loaded}%
    \renewcommand\transparent[1]{}%
  }%
  \providecommand\rotatebox[2]{#2}%
  \ifx\svgwidth\undefined%
    \setlength{\unitlength}{608bp}%
    \ifx\svgscale\undefined%
      \relax%
    \else%
      \setlength{\unitlength}{\unitlength * \real{\svgscale}}%
    \fi%
  \else%
    \setlength{\unitlength}{\svgwidth}%
  \fi%
  \global\let\svgwidth\undefined%
  \global\let\svgscale\undefined%
  \makeatother%
  \begin{picture}(1,0.75)%
    \put(0,0){\includegraphics[width=\unitlength,page=1]{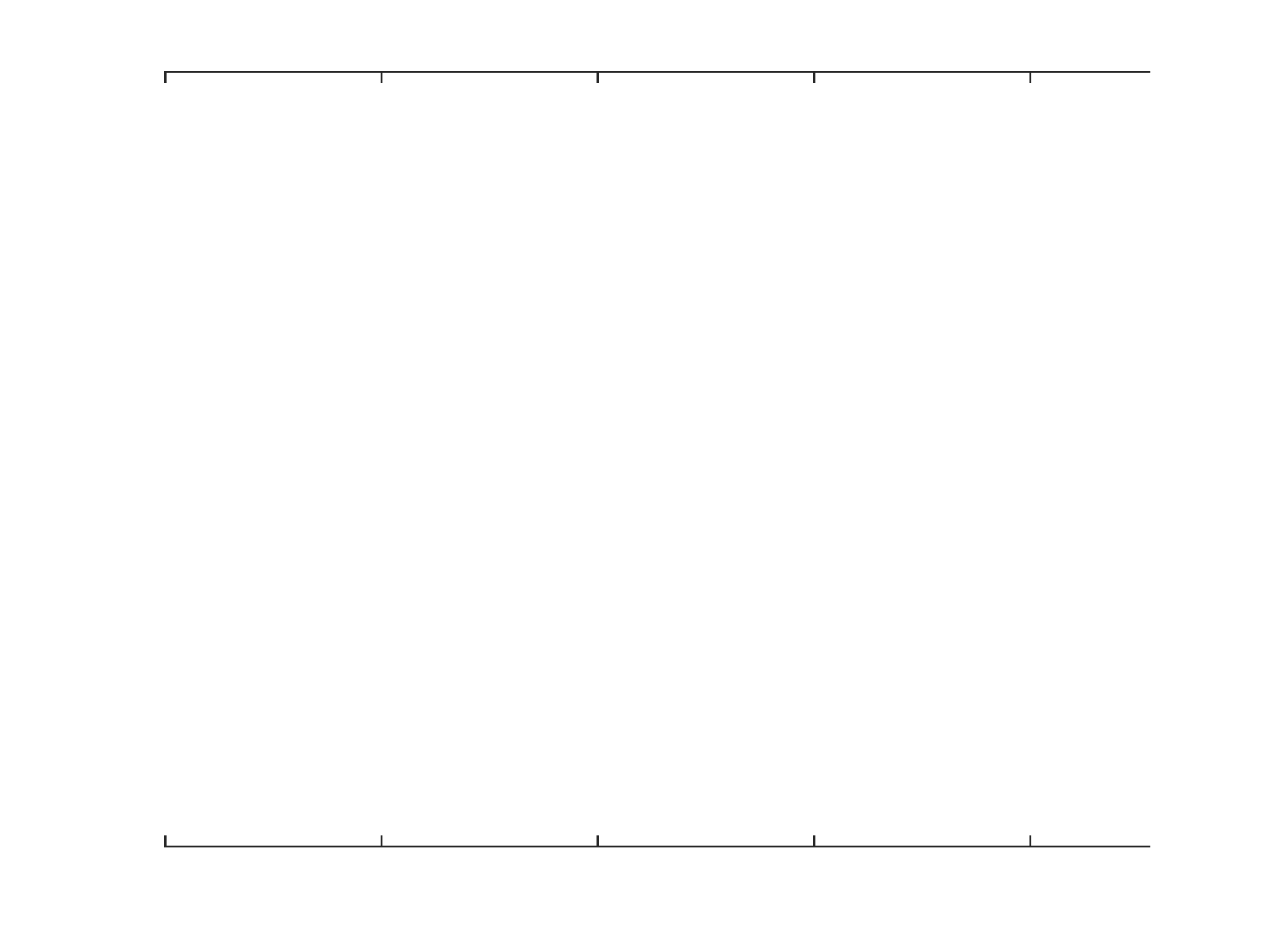}}%
\small
    \put(0.12236842,0.04707868){\makebox(0,0)[lb]{\smash{0}}}%
    \put(0.28151382,0.04444711){\makebox(0,0)[lb]{\smash{0.1}}}%
    \put(0.45184355,0.04444711){\makebox(0,0)[lb]{\smash{0.2}}}%
    \put(0.62217329,0.04444711){\makebox(0,0)[lb]{\smash{0.3}}}%
    \put(0.79250289,0.04444711){\makebox(0,0)[lb]{\smash{0.4}}}%
    \put(0.40131618,0.00323724){\makebox(0,0)[lb]{\smash{Poisson's ratio $\nu$}}}%
    \put(0,0){\includegraphics[width=\unitlength,page=2]{adj_coef_poiss_heuristic.pdf}}%
    \put(0.09313132,0.09814276){\makebox(0,0)[lb]{\smash{1}}}%
    \put(0.09313132,0.25039434){\makebox(0,0)[lb]{\smash{2}}}%
    \put(0.09313132,0.45165987){\makebox(0,0)[lb]{\smash{5}}}%
    \put(0.08128921,0.60391145){\makebox(0,0)[lb]{\smash{10}}}%
    \put(0.05371237,0.22828974){\rotatebox{90}{\makebox(0,0)[lb]{\smash{$\beta_{iii}^{i}/\beta_{iii}^{i,AN}$, $\enskip i \in [1,4]$}}}}%
    \put(0,0){\includegraphics[width=\unitlength,page=3]{adj_coef_poiss_heuristic.pdf}}%
    \put(0.21184211,0.63806513){\makebox(0,0)[lb]{\smash{$\beta_{111}^1$}}}%
    \put(0,0){\includegraphics[width=\unitlength,page=4]{adj_coef_poiss_heuristic.pdf}}%
    \put(0.21184211,0.59119711){\makebox(0,0)[lb]{\smash{$\beta_{333}^3$}}}%
    \put(0,0){\includegraphics[width=\unitlength,page=5]{adj_coef_poiss_heuristic.pdf}}%
    \put(0.34342089,0.63806505){\makebox(0,0)[lb]{\smash{$\beta_{222}^2$}}}%
    \put(0,0){\includegraphics[width=\unitlength,page=6]{adj_coef_poiss_heuristic.pdf}}%
    \put(0.34342105,0.59119663){\makebox(0,0)[lb]{\smash{$\beta_{444}^4$}}}%
    \put(0,0){\includegraphics[width=\unitlength,page=7]{adj_coef_poiss_heuristic.pdf}}%
    \put(0.55758101,0.55879151){\color[rgb]{0,0,0}\makebox(0,0)[lb]{\smash{$\frac{2}{1-\nu-2\nu^2}$}}}%
    \put(0,0){\includegraphics[width=\unitlength,page=8]{adj_coef_poiss_heuristic.pdf}}%
    \put(0.66062871,0.27544624){\color[rgb]{0,0,0}\makebox(0,0)[lb]{\smash{$\frac{1}{1-\nu^2}$}}}%
    \put(0,0){\includegraphics[width=\unitlength,page=9]{adj_coef_poiss_heuristic.pdf}}%
    \put(0.16151091,0.37330555){\color[rgb]{0,0,0}\makebox(0,0)[lb]{\smash{3D elements}}}%
    \put(0,0){\includegraphics[width=\unitlength,page=10]{adj_coef_poiss_heuristic.pdf}}%
    \put(0.32019192,0.20770521){\color[rgb]{0,0,0}\makebox(0,0)[lb]{\smash{shell elements}}}%
  \end{picture}%
\endgroup%
\caption{Evolution of the ratio of the cubic coefficients $\beta_{iii}^{i}$ compared with the analytic values $\beta_{iii}^{i,AN}$, with regard to the Poisson ratio, for the first four ($i=1,\ldots 4$) bending modes of the thin beam (Fig.~\ref{fig:meshbeama}) meshed with shell or 3D elements, as specified on the plot. The heuristic dependences on the Poisson's ratio are plotted with black dashed lines. All symbols are merged.}
\label{fig:SensPoisson}
\end{center}
\end{figure}

The previous sections show that in the case of a 3D model, a given bending mode is nonlinearly coupled to numerous high frequency modes, most of them involving thickness deformations. To understand this effect, a numerical study of the sensitivity of the coefficients computed with the STEP on the Poisson ratio is here given. It is found that a precise dependence of the cubic coefficient $\beta^r_{iii}$ can be established: in Fig. \ref{fig:SensPoisson}, the values obtained by the direct application of the STEP are fitted to an heuristic law related to the Poisson ratio. In particular, the growths of the cubic coefficients in the case of shell and 3D elements match perfectly the ratios:
\begin{equation}
\rho_1=\frac{1}{1-\nu^2},\qquad \rho_2=\frac{2}{ (1+\nu)(1-2\nu)},
\end{equation}
related to 2D and 3D constitutive laws. Indeed, the 3D constitutive law for an isotropic elastic material writes:
\begin{equation}
\bm{\pi} = \frac{E}{(1+\nu)(1-2\nu)}\big[\nu\, \trace(\bm{\varepsilon})  \bm{I}_3 + (1-2\nu) \bm{\varepsilon}\big], \label{eq:ConstLaw3D}
\end{equation}
where $\bm{\pi}$ denotes the second Piola-Kirchhoff stress tensor, $\bm{\varepsilon}$ the Green-Lagrange strain tensor, $\bm{I}_3$ the identity operator in 3D and $(E,\nu)$ the Young's modulus and the Poisson ratio of the material.

Moreover, in the case of usual plate theories, a plane stress state is assumed, for which the transverse component $\pi_{zz}=0$ of the stress tensor is zero ($z$ being the direction normal to the middle plane of the plate). In this case, the in-plane counterpart of~(\ref{eq:ConstLaw3D}) reads:
\begin{equation}
\tilde{\bm{\pi}} = \frac{E}{1-\nu^2}\big[\nu\, \trace(\tilde{\bm{\varepsilon}})  \bm{I}_2 + (1-\nu) \tilde{\bm{\varepsilon}}\big], \label{eq:ConstLaw2D}
\end{equation}
where $\tilde{\bm{\pi}}$ and $\tilde{\bm{\varepsilon}}$ denote the plane parts of $\bm{\pi}$ and $\bm{\varepsilon}$, respectively, and $\bm{I}_2$ the identity operator in 2D.

Equations~(\ref{eq:ConstLaw3D}) and (\ref{eq:ConstLaw2D}) makes directly appear the ratios $\rho_1$ and $\rho_2$, which suggests also a direct relationship between the constitutive law and the results presented in Fig.~\ref{fig:SensPoisson}. As a side note, the reference values, used to compare nondimensional values of cubic coefficients in Fig.~\ref{fig:SensPoisson}, differ from those in Figs.~\ref{fig:CONV} and \ref{fig:nu0}(b). Indeed, in Fig.~\ref{fig:SensPoisson}, the normalising coefficient has been selected as  $\beta_{111}^{1,AN}$, the analytical value obtained from the beam theories (see {\em e.g.} \cite{givois2019}). As shown in the previous sections, this coefficient has to be compared with the corrected coefficient used once the convergnce is obtained from 3D models. On the other hand, in Fig.~\ref{fig:CONV} and \ref{fig:nu0}(b), the reported values are normalized with respect to $\beta_{111}^{1}$, {\em i.e.} the uncorrected cubic coefficient, the one coming from direct application of STEP and shown in Table~\ref{tab:ErrorsCoeffsSTEP}. This explains the different values observed between the two figures (e.g. $\beta_{111}^{1}/\beta_{111}^{1,AN} \approx 3.8$ for $\nu = 0.3$ on Fig.~\ref{fig:SensPoisson}, whereas $\beta_{111}^{1}/\Gamma_{111}^{1} = 5.5$ on Fig.~\ref{fig:CONV}).

\subsection{Geometrical nonlinearities}

\begin{figure*}[h]
\centering
\begin{subfigure}{.49\textwidth}
\centering
\includegraphics[width=7.5cm]{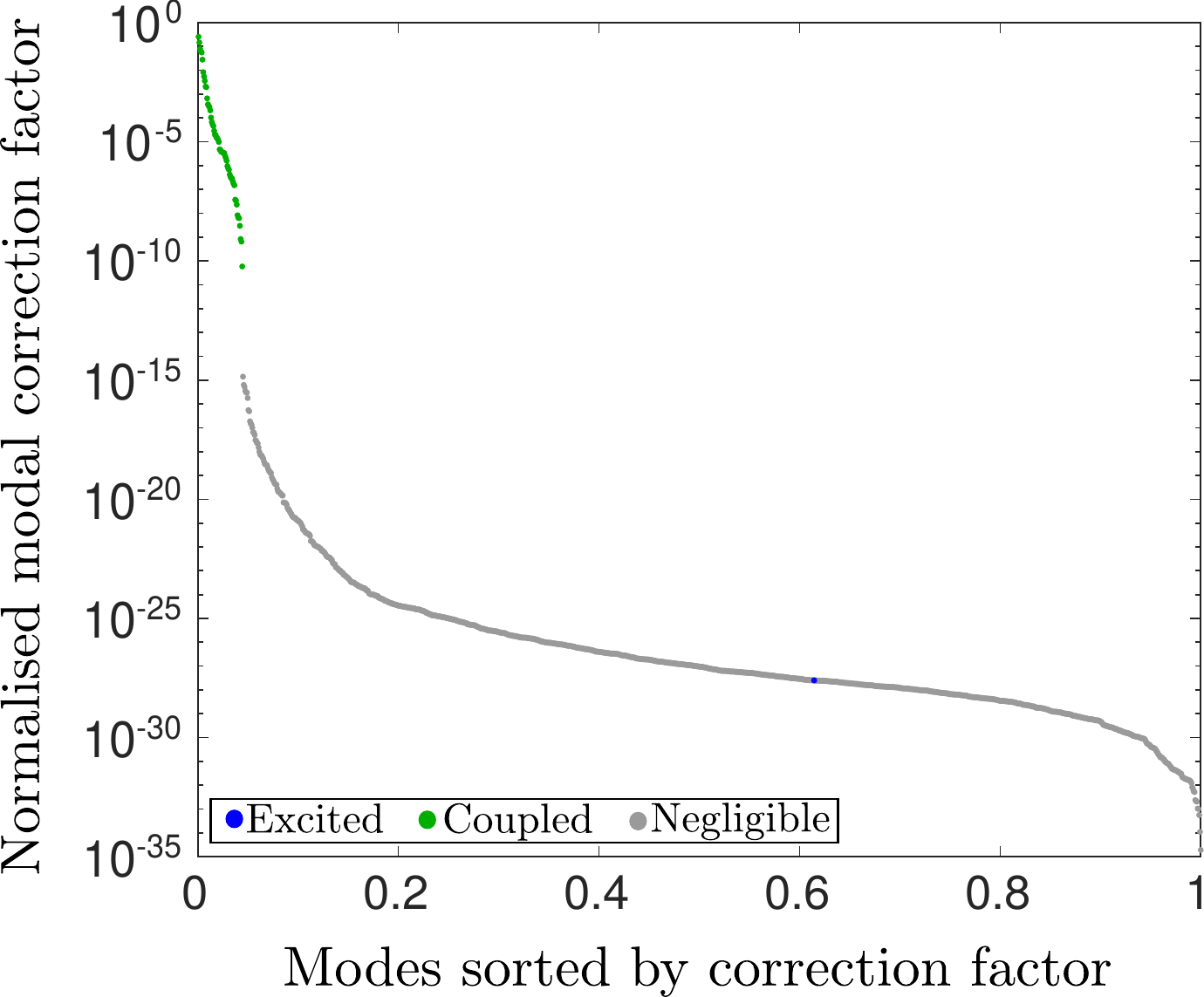}
\caption{}
\end{subfigure}
\begin{subfigure}{.49\textwidth}\centering
\includegraphics[width=7.5cm]{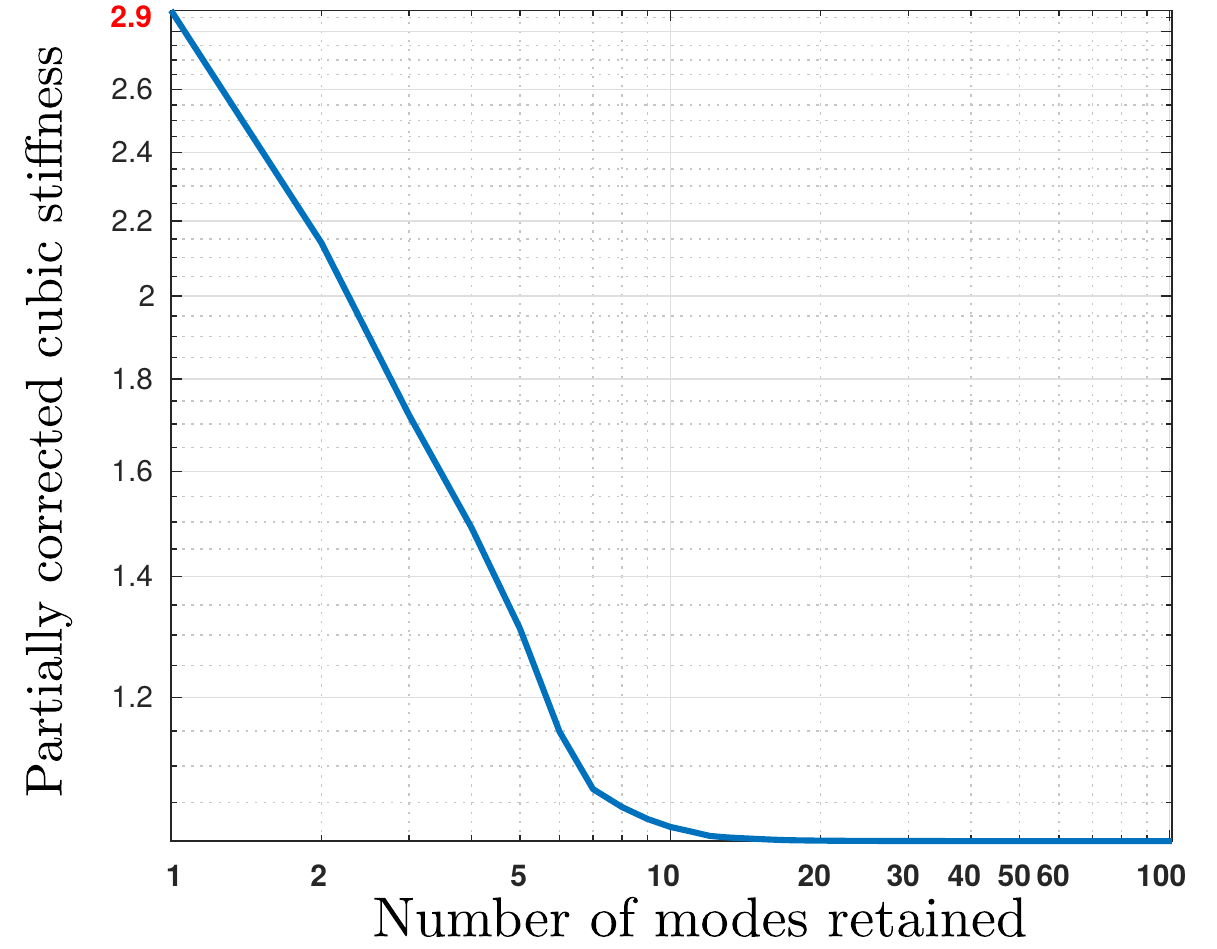}
\caption{}
\end{subfigure}
\begin{subfigure}{\textwidth}
\centering
\begin{tabular}{ccc}
\includegraphics[width=.35\textwidth]{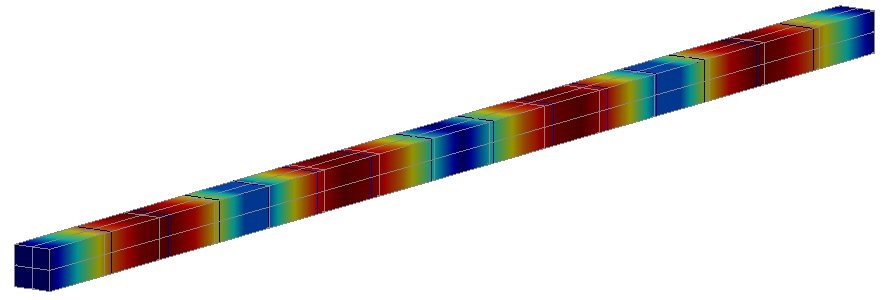} &\hspace{2em} & 
\includegraphics[width=.35\textwidth]{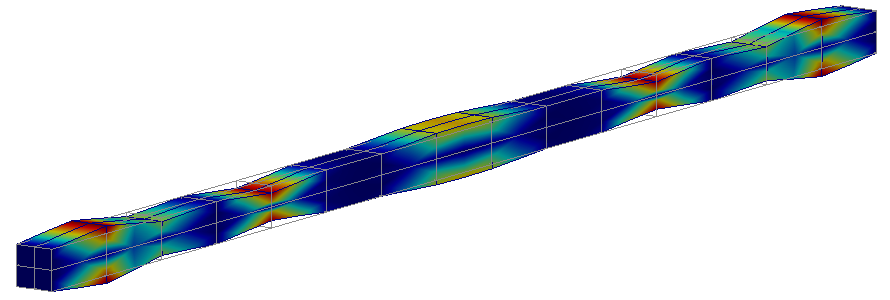} \\[-1em]
{\small \#32} && {\small \#209} \\
\includegraphics[width=.35\textwidth]{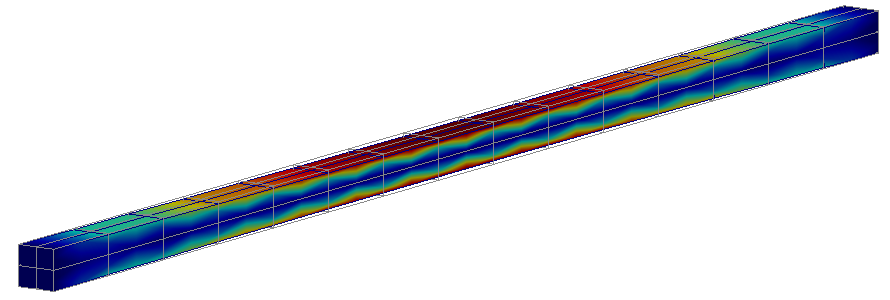} & &
\includegraphics[width=.35\textwidth]{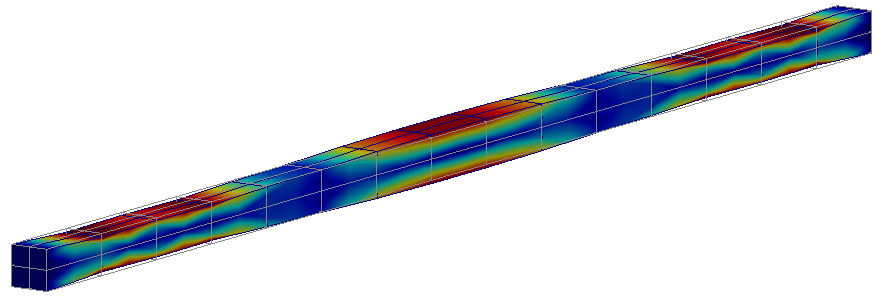}\\[-1em]
{\small \#1036} && {\small \#1043}
\end{tabular}
\caption{}
\end{subfigure}
\caption{Convergence without Poisson effect ($\nu=0$) for the thick beam. (a) Nondimensional correction factor $\mathcal{C}^{1s}_{111}/\beta^1_{111}$ of all the modes over the nondimensional mode number $s/N$. (b) Convergence of the evaluated corrected cubic stiffness coefficient $\Gamma^1_{111}$, defined in Eq.~\eqref{eq:Gammappp}, with increasing number of linear modes kept in the truncation. (c) Mode shapes and order of appearance in the basis (\#) of four of the most relevant modes coupled with the first bending mode.}
\label{fig:nu0}
\end{figure*}

\begin{figure*}[h]
\begin{center}
\includegraphics[width=.45\textwidth]{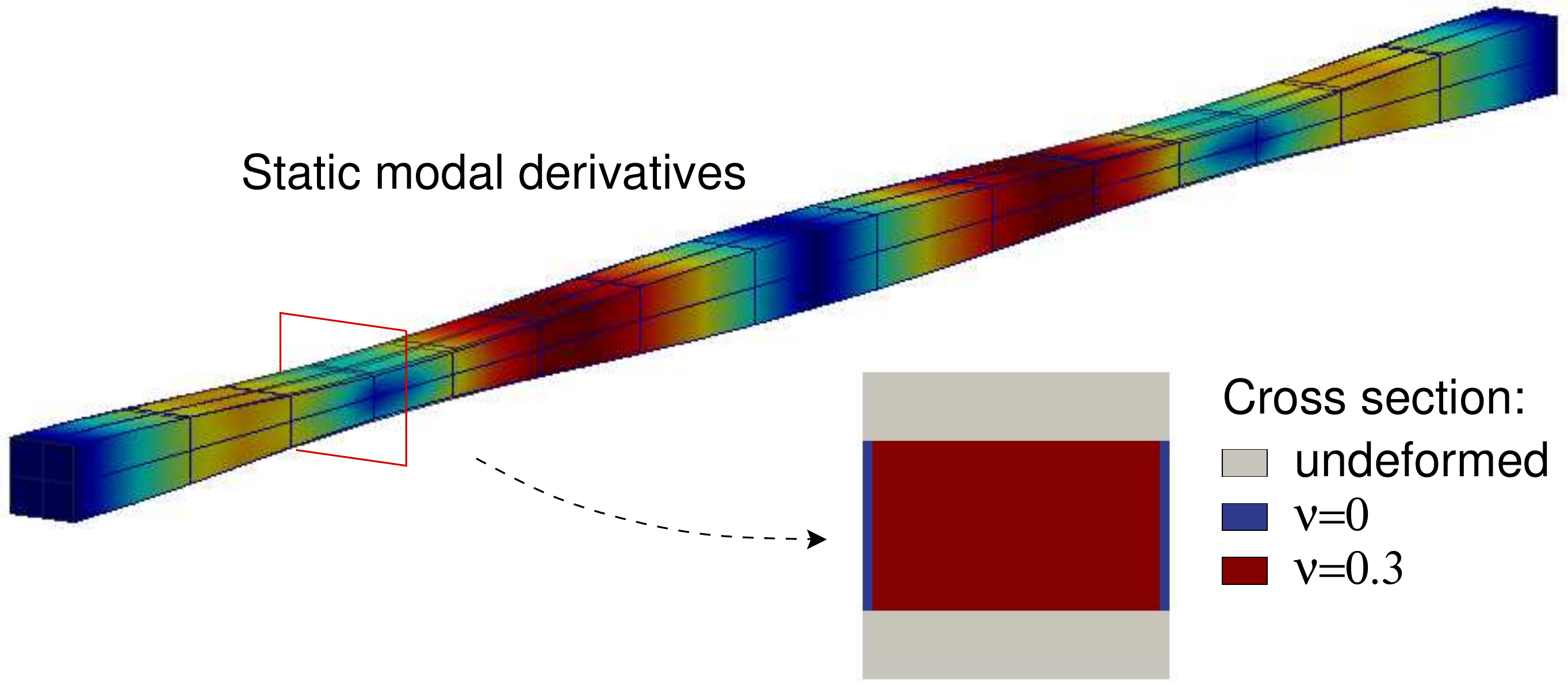}
\end{center}
\caption{Static modal derivative and wiew of the cross-section in the undeformed and deformed configurations, with both $\nu=0$ and $\nu=0.3$.}
\label{fig:SMDcrosssection}
\end{figure*}

In this section, we focus on the physical explanation of the nonlinear couplings with thickness modes, whose origin is the geometrical nonlinearities.  Considering first Fig.~\ref{fig:SensPoisson}, it can be inferred  that the couplings are amplified by the Poisson ratio, but that they are present even without Poisson effect, since there is a factor~2 between the FE value of $\beta_{iii}^{i}$ with respect to its corresponding analytical value in the case $\nu=0$. This leads us to investigate the couplings in this particular case.

Figure~\ref{fig:nu0} is the analog, with $\nu=0$, of Figs.~\ref{fig:sorted}(a), \ref{fig:FRF1}(b) and Tab.~\ref{tab:BeamModes}. Comparing those figures shows that the number of coupled modes is much smaller in the case $\nu=0$ than for $\nu=0.3$: the relevant modes correspond to 5\% of the modal basis if $\nu=0$, whereas it was 20\% for $\nu=0.3$ (see the modes with a correction factor above $10^{-15}$ in Figs~\ref{fig:sorted}(a) and \ref{fig:nu0}(a)). Moreover, the deformations of the cross section in the case $\nu=0$ are purely in the bending transverse-$y$ direction in the case $\nu=0$ ($y$, as defined in Figs.~\ref{fig:meshbeam} and \ref{fig:beamanalytic}, is the deformation direction of the first bending mode considered in all computations of the present article), whereas they were more complex (in 2D) in the case $\nu=0.3$ (compare the mode shapes of Tab.~\ref{tab:BeamModes} and Fig.~\ref{fig:nu0}). Finally, taking a close look at the static modal derivative (SMD), shown in Fig.~\ref{fig:SMD}(a), that gathers all the corrections brought by the NB modes, shows that it has almost the same shape in both cases $\nu=0$ and $\nu=0.3$. In particular, Fig.~\ref{fig:SMDcrosssection} shows that the deformations of the SMD cross section occur without distorsion: initially a square, it is deformed in a rectangle. In the case of no Poisson's ratio, the deformation is purely in the bending $y$-direction, whereas the Poisson's effects adds a slight deformation in the lateral $z$-direction.

Those effects are purely geometrical and come from (i) the particular 3D shape of the eigenvectors and (ii) how these particular shapes, resulting from a linear computation, are modified by the geometrical nonlinearities. A closed form solution in a simple bending case is exposed in Appendix~\ref{sec:purebending}. It shows that the 3D shape of the eigenmodes is the combination of three contributions (see Eq.~(\ref{eq:Ulin})):
\begin{itemize}
\item the deformation of the neutral axis / plane of the structure, described by classical beam / plate theories;
\item the 3D rotation of the cross section around the $z$-axis, that produces an axial deformation with a linear dependence in the thickness coordinate $y$;
\item the 3D Poisson effect, that distorts the cross section in its two ($y$ and $z$) directions.
\end{itemize}
Then, by computing the Green-Lagrange strain tensor with this particular linear deformation, it is shown that the geometrical nonlinearities adds two contributions to the classical von~K\'arm\'an beam / plate model:
\begin{itemize}
\item 3D effects that are {\it independent of the Poisson effect}, that explains a stretching in the transverse $y$ direction, without any deformation in the lateral $z$ direction. Those effects are a direct consequence of {\em the 3D rotation of the cross section} created by the bending;
\item 3D Poisson effect, that involve stretching in both the transverse $y$ and lateral $z$ directions.
\end{itemize}
Those two geometrical effects are purely 3D and are additional to the classical membrane / bending coupling. 

Having in mind those observations, we can now explain those nonlinear thickness coupling. We have first to remark that in both cases of a beam(1D)/plate(2D) von~K\'arm\'an model and the present 3D model, the modal expansion of Eq.~(\ref{eq:eqq}) is exact, provided $N$ is the number of degrees of freedom of the model. 
Moreover, Tab.~\ref{tab:Coeff} and Fig.~\ref{fig:FRF1}(a) show that all models converge to the same solution, which proves that the relevant modes of the basis combine themselves in different ways to give, at the end, the same solution. In fact, as a consequence of the above observations, the thickness modes are here to geometrically compensate (i) the nonlinear deformations in the transverse bending direction due to the 3D rotation of the cross sections, shown in blue on Fig.~\ref{fig:SMDcrosssection} and (ii) the additional deformations of the cross section due to the Poisson effect. Looking again at the deformed cross section shown in red in Fig.~\ref{fig:SMDcrosssection}, one can understand that at the end, the complex 3D distorsions of the cross section due to the Poisson effect (shown in Fig.~\ref{fig:beamanalytic}(c)) must be fully compensated by the NB modes, which then need to be numerous. This explains the bad convergence of the modal expansion, even worse in the case of a non zero Poisson ratio (5\% of the modal basis if $\nu=0$, and 20\% for $\nu=0.3$).



\section{Numerical examples}

\begin{figure*}[h!]\centering
\includegraphics[width=.52\textwidth]{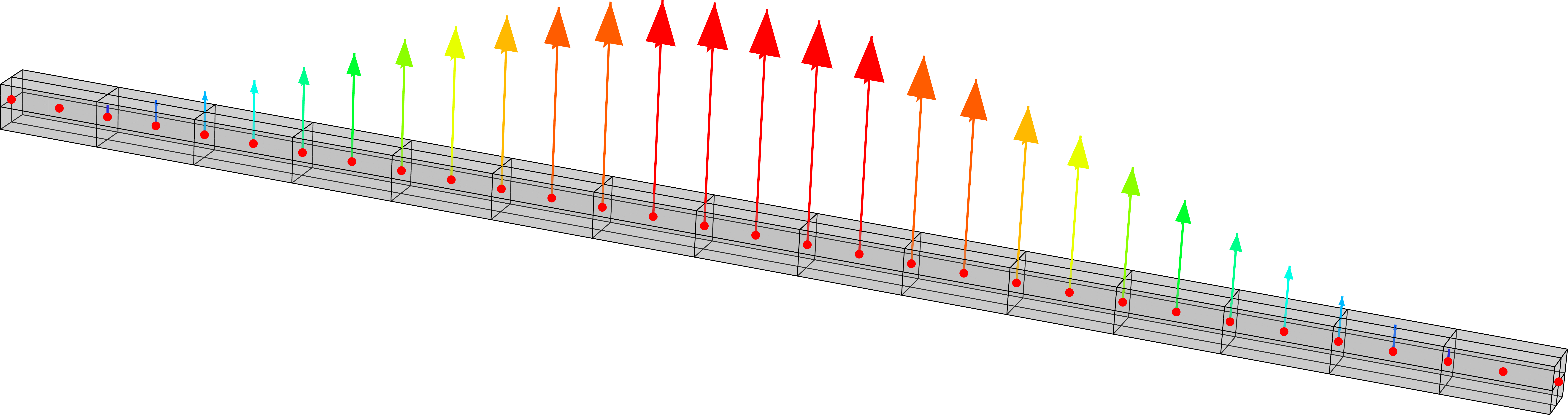}\hfill
\includegraphics[width=.44\textwidth]{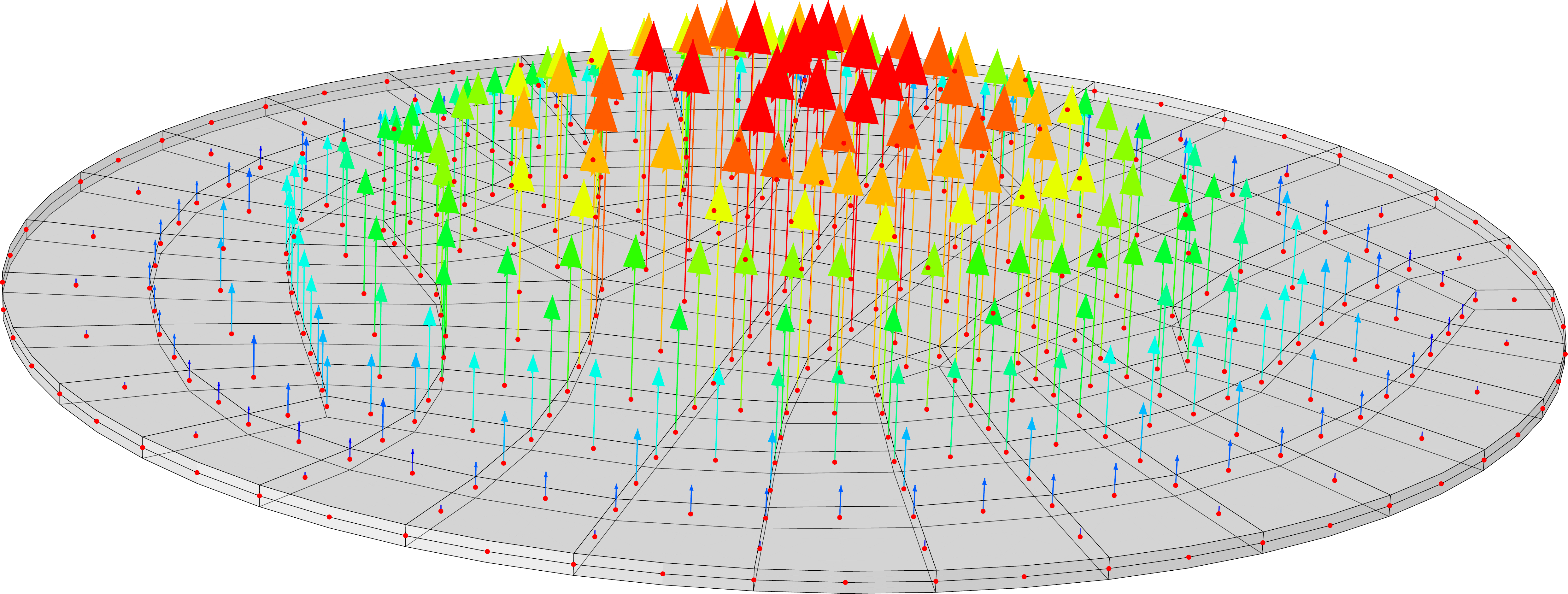}
\caption{First-mode displacements prescribed on the middle line of the clamped beam and the middle surface of the circular plate. For visualisation purpose, the mesh of the plate is less fine than the one use for the computations of the ROM coefficients. }
\label{fig:beam_mstep_arrows}
\end{figure*}

In this section, numerical examples on the three different strategies proposed in order to overcome the bias observed when using 3D elements, are given. In each case, the dominant cubic coefficient of a single mode is compared, using either the M-STEP, the static condensation of all the coupled linear modes, or the static modal derivative. Two test cases are used for the comparisons:  the clamped-clamped beam of Fig. \ref{fig:meshbeamb}, and a clamped circular plate.

\subsection{Application to a clamped-clamped beam}
\label{sec:num}

Computations of the condensed coefficients $\Gamma_{ijk}^r$ are first performed with the three methods. For the M-STEP, the prescribed displacement is depicted in Fig. \ref{fig:beam_mstep_arrows}. The comparison made in Tab. \ref{tab:Coeff} attests that the condensation with all the eigenmodes and the first static modal derivative are quasi equivalent, whereas the M-STEP gives very close values. The relative errors are very small ($<0.5$ \%) in each case.  The analytical reference values present slightly larger errors, between $1$ \% and $5$ \%, probably due to unavoidable differences between the analytical beam theory and the numerical 3D computation. This could be explained by the aspect ratio $h/L=0.03$ of the beam, which is not so small to fully verify Euler-Bernoulli assumptions.

\begin{table*}[h]
\begin{center}
\begin{tabular}{cccccc}
\hline
\multicolumn{6}{c}{Corrected Coefficients}
\\  
& $\Gamma^1_{111}$
& $\Gamma^3_{111}$
& $\Gamma^3_{113}$ 
& $\Gamma^1_{333}$
& $\Gamma^3_{333}$
\\\hline\\
M-StEP
& 2.9790e+08
& -2.3555e+08
& 2.7964e+09
& -1.9044e+09
& 1.9235e+10
\\\hline\\
Static condensation
& 2.9792e+08
& -2.3572e+08
& 2.8003e+09 
& -1.9083e+09 
& 1.9288e+10
\\\hline\\
Static Modal Derivative
& 2.9792e+08
& -2.3573e+08
& 2.8004e+09
& -1.9084e+09
& 1.9288e+10
\\\hline\\
Analytic coefficients
& 2.9150e+08
& -2.3151e+08
& 2.7082e+09
& -1.8310e+09
& 1.8687e+10
\\\hline
\end{tabular}
\end{center}
\caption{Corrected cubic coefficients with the three condensation methods and comparison with anaytical values. Due to the symmetry of the mode shapes, the nonlinear coefficients which couple the first and the third modes have all nonzero values, and are therefore chosen for these computations.}
\label{tab:Coeff}
\end{table*}

\begin{figure*}[h]
\centering
\begin{subfigure}{.49\textwidth}\centering
\includegraphics[width=\textwidth]{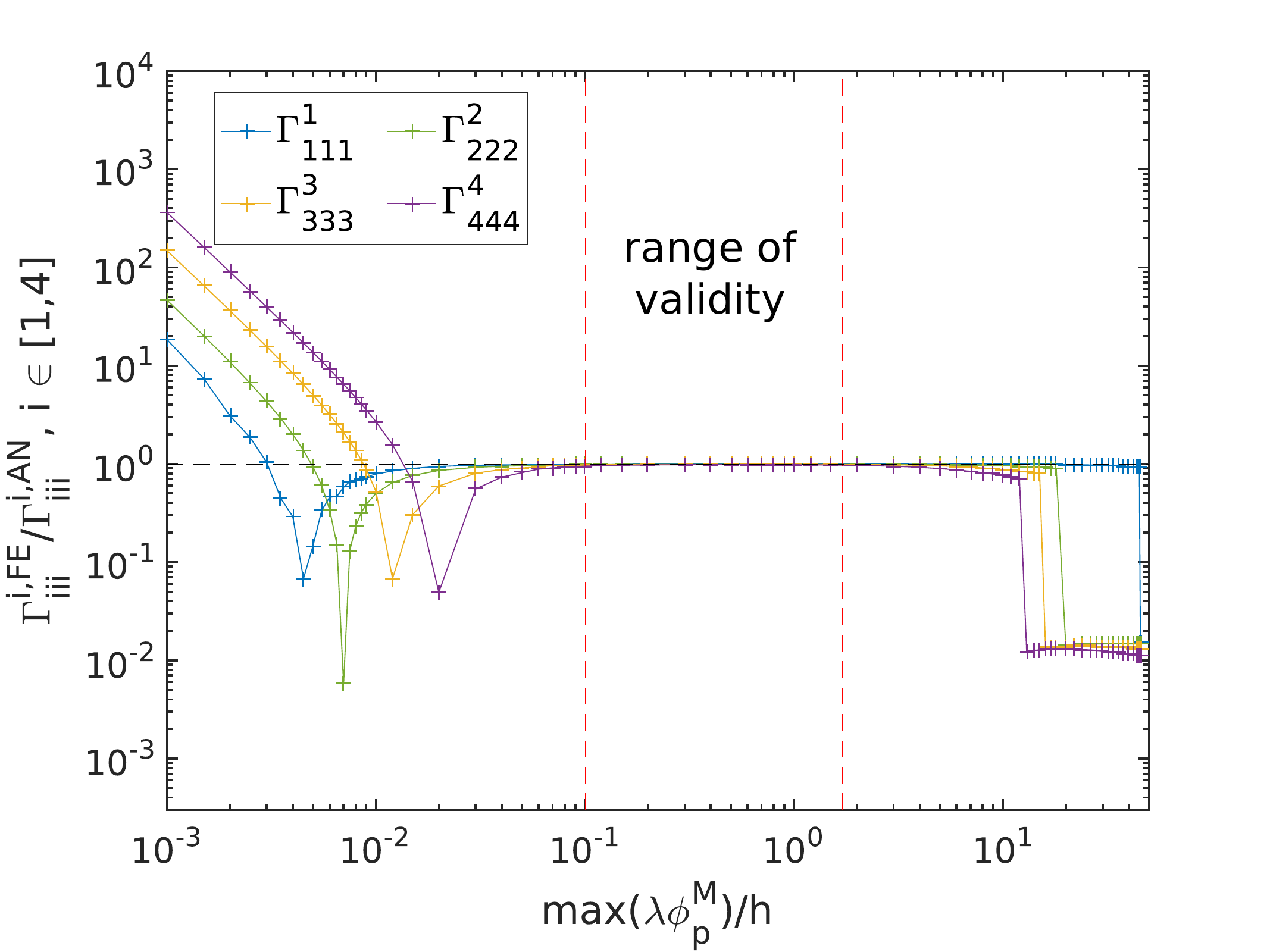}
\caption{}\label{fig:mstep_dependance_ampl_moda}
\end{subfigure}
\begin{subfigure}{.49\textwidth}\centering
\includegraphics[width=\textwidth]{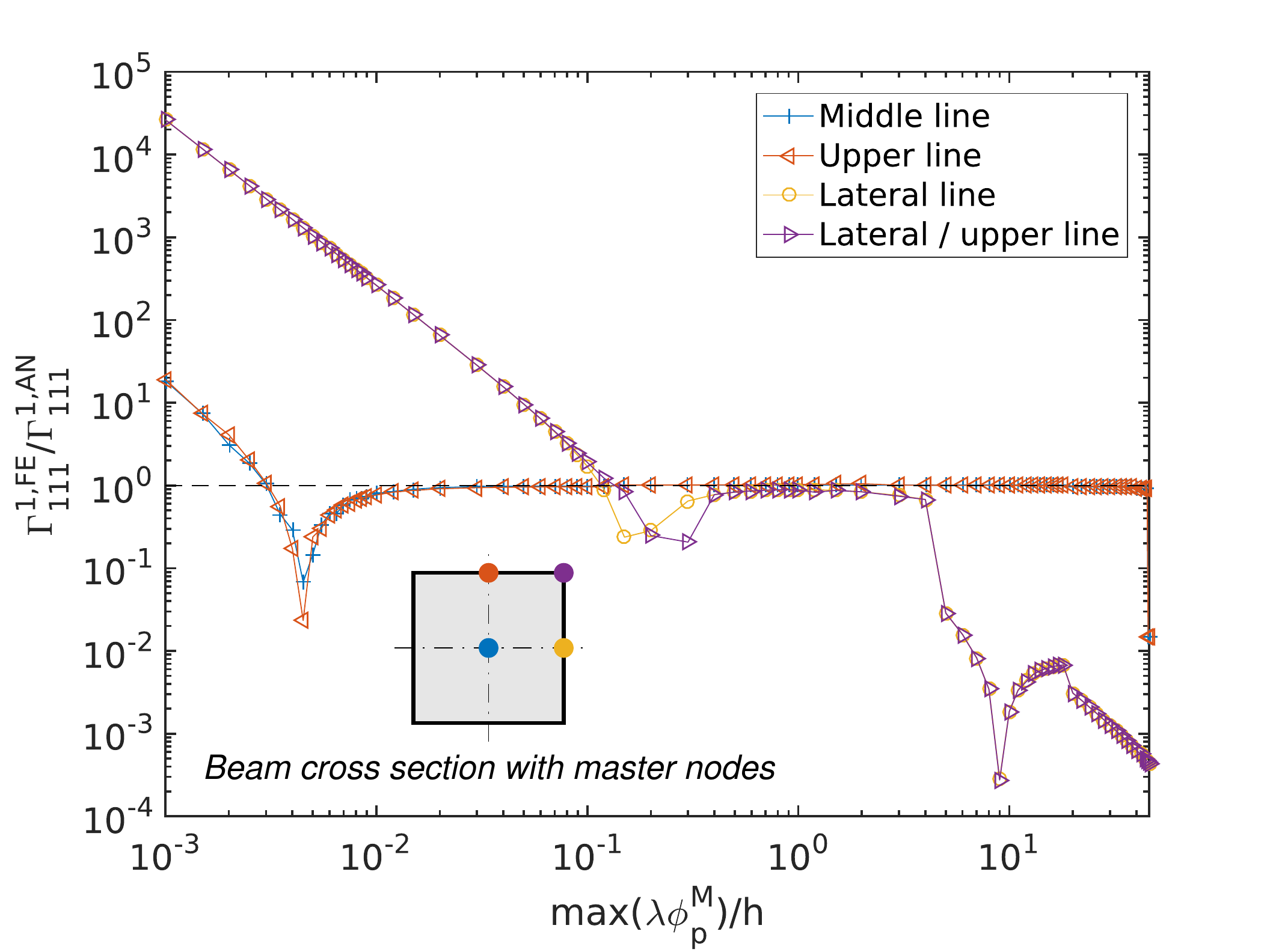}
\caption{}\label{fig:mstep_dependance_ampl_line}
\end{subfigure}
\caption{(a) Dependence of the cubic coefficient $\Gamma_{iii}^i$ with regard to the prescribed displacement amplitude, for different modes $i$, with $i=1,2,3,4$. Black dashed line : reference analytical value ($\Gamma_{iii}^{i,FE} = \Gamma_{iii}^{i,AN}$), red dashed lines: limits of the range of validity. (b) Dependance of the cubic coefficient $\Gamma_{111}^1$ with regard to the prescribed displacement amplitude for different lines where the displacements are prescribed. Black dashed line : reference value ($\Gamma_{111}^{1,FE} = \Gamma_{111}^{1,AN}$). The inset shows the location, in the cross section of the beam, of the lines in which the master displacement is prescribed.}
\end{figure*}

In the case of the M-STEP with the displacement field prescribed on the neutral fiber, Fig.~\ref{fig:mstep_dependance_ampl_moda} gives the sensitivity to the prescribed displacement amplitudes of the corrected cubic coefficients $\Gamma_{iii}^i$ for different modes $i$. It is shown that a range of validity for the displacements amplitude centered around $\text{max}(\lambda \phi_p) \simeq h/2$ can be defined. As it could be expected, the length of this validity range, defined on Fig.~\ref{fig:mstep_dependance_ampl_moda} by a relative error smaller than $3\%$, decreases with the mode order.

Fig. \ref{fig:mstep_dependance_ampl_line} shows what happens if the M-STEP is applied with different selections of the master degrees of freedom. We tried to prescribe the displacement field on three other lines of the beam, parallel to but different than the neutral fiber, as shown in the inset of Fig.~\ref{fig:mstep_dependance_ampl_moda}. We can conclude that the neutral fiber seems the best choice and that the upper line gives very close results. On the other hand, the values obtained when the displacements are prescribed on one of the lateral lines are far from the reference values. This feature will be analyzed in the following considering error estimators.

\begin{figure*}[h]
\centering
\begin{subfigure}{.49\textwidth}\centering
\includegraphics[width=\textwidth]{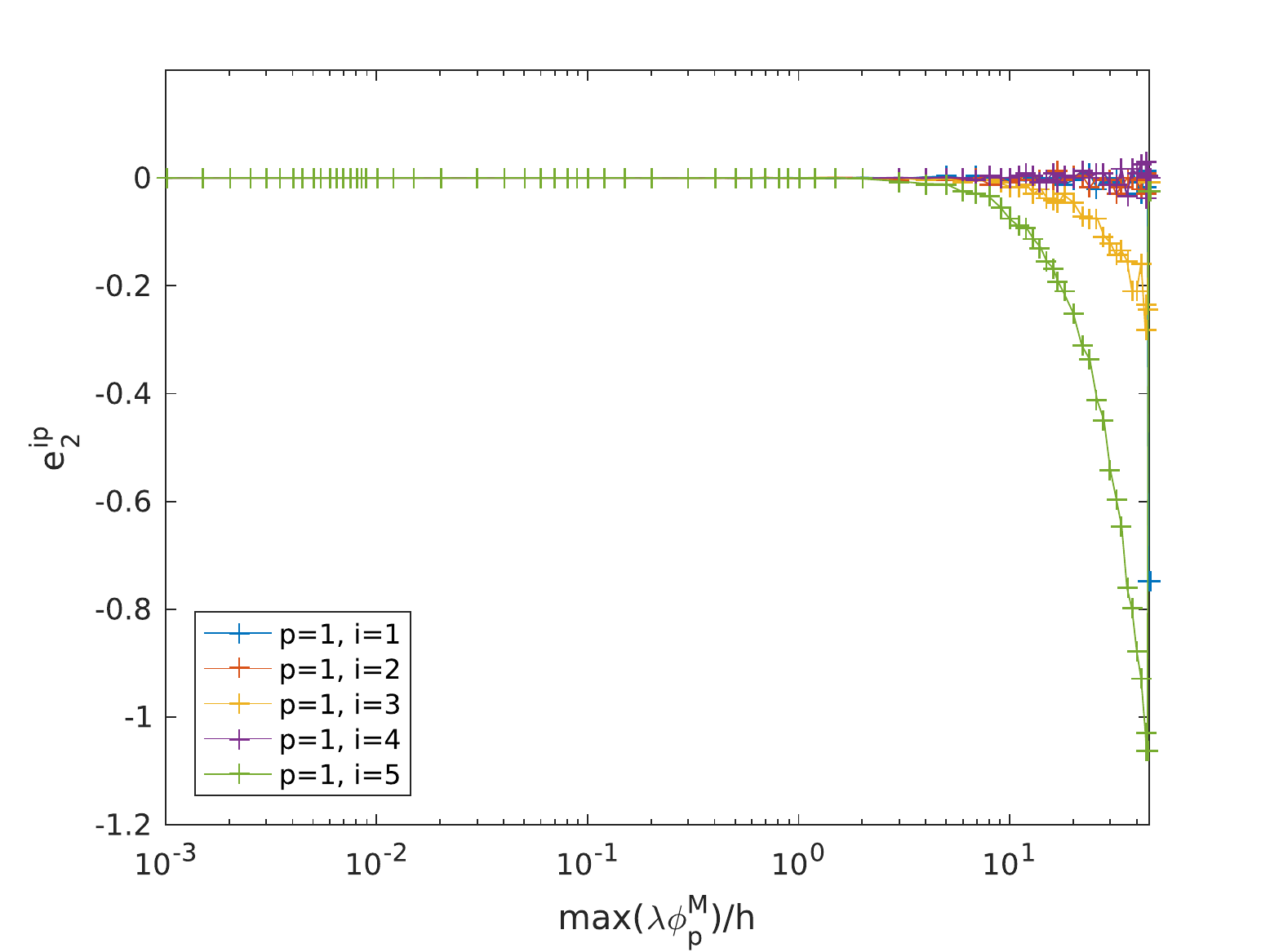}
\caption{}\label{fig:VerifCR2_1}
\end{subfigure}
\begin{subfigure}{.49\textwidth}\centering
\includegraphics[width=\textwidth]{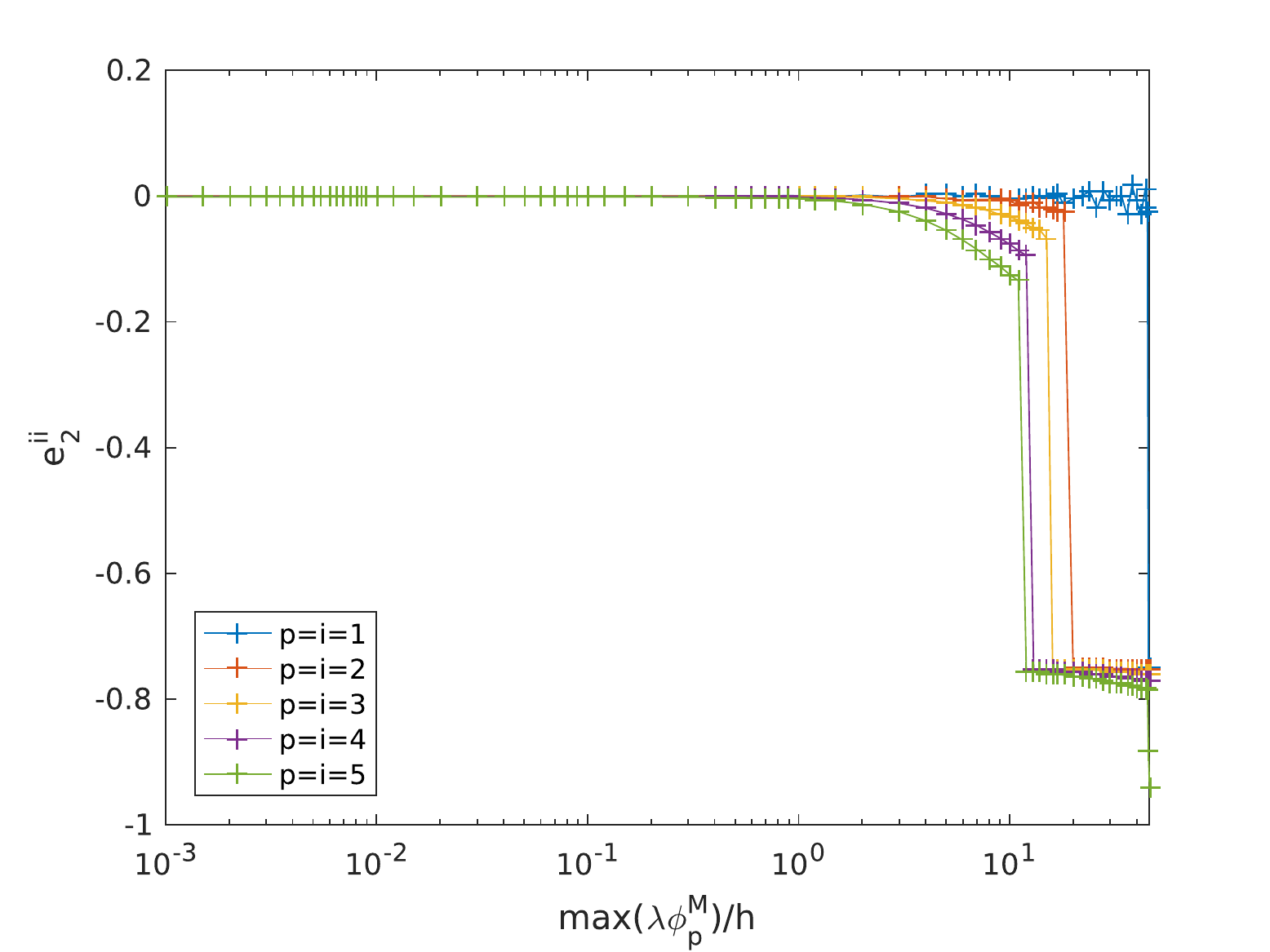}
\caption{}\label{fig:VerifCR2_2}
\end{subfigure}
\caption{Dependence of the criterion $e_1^{pi}$ with regard to the prescribed displacement amplitude when displacements are prescribed on the neutral fiber. The values are presented on (a) for $p=1$ and different modes $i=1,2,3,4,5$. On (b), the values $e_1^{ii}$ are presented, also for $i=1,2,3,4,5$.}\label{fig:mstep_dependance_ampl}
\end{figure*}

\begin{table*}[h]
\begin{center}
\begin{tabular}{cccccc}
\\\hline\\
& Neutral line
& Upper line
& Lateral line 
& Up/Lat line
\\\hline\\
$e_1^{11}$
& -3.496e-04
& 1.234e-04
& -0.0210
& -0.0213
\\\hline\\
$e_1^{12}$
& -6.993e-04
& -2.183e-04
& -8.136e-05
& 2.006e-04
\\\hline\\
$e_1^{13}$
& 5.604e-04
& -3.466e-04
& 0.0577
& 0.0579
\\\hline\\
$e_1^{14}$
& 3.347e-04
& 6.647e-04
& 8.501e-05
& -6.071e-04
\\\hline\\
$e_1^{15}$
& 1.553e-04
& -2.552e-04
& 0.1231
& 0.1228
\\\hline\\
$e_2^1$
& 1.425e-05
& 1.424e-05
& 0.0227
& 0.0227
\\\hline\\
\end{tabular}
\end{center}
\caption{Values of the relative error $e_1^{pi}$ for different $p$, $i$, and master dofs. Nonlinear computations are performed with {$\text{max}(\lambda \vect{\phi}_p^\text{{M}})) = h/2$}}
\label{tab:Cr2}
\end{table*}

The error estimate $e_1^{pi}$, introduced in Section~\ref{subsec:MStepFormulation}, is now analyzed. The criterion $e^{pi}_1$ is first computed with the master transverse dofs prescribed in the neutral fiber, a single master mode $p=1$ and different transverse modes $k\in\{1,...,5\}$. The values presented in Fig.~\ref{fig:VerifCR2_1} show that the orthogonality is well verified, until the upper limit of validity range observed on Fig.~\ref{fig:mstep_dependance_ampl_moda}, after which the values $e_1^{13}$ and $e_1^{15}$ deviate from 0. For $p>1$, the coefficients $e_1^{pp}$ evolve in a similar way as $e_1^{11}$, as depicted on Fig.~\ref{fig:VerifCR2_2}. Tab.~\ref{tab:Cr2} give the numerical values of $e_1^{1i}$ in the plateau of Fig.~\ref{fig:VerifCR2_1}, proving that the error is the order of $10^{-4}$. Consequently, the orthogonality of the prescribed displacement field $\vect{x}$ to the transverse modes $\vect{\phi}_i$ is well verified, thus validating assumptions~(\ref{eq:assume1}a,b).  

Then, the same error estimate $e_1^{pi}$ is computed in the cases of a master displacement prescribed in the other lines of the inset of Fig.~\ref{fig:mstep_dependance_ampl_line}. The obtained values are given in Tab.~\ref{tab:Cr2}. In this cases, we quantitatively confirm the observation linked to Fig.~\ref{fig:mstep_dependance_ampl_line}: the orthogonality of the displacement are not verified when the master dofs are placed on a line of the lateral surfaces of the beam. In particular, the values of the criterion $e_1^{pi}$ presented with $p=1$ and $i=1,2,3,5$ in Table \ref{tab:Cr2} highlight a loss of orthogonality between the first and the odd modes $i=1,3,5$, in the case of master dofs on a lateral line: indeed, the values of $e_1^{11}$, $e_1^{13}$ and $e_1^{15}$ deviate from 0. 

The second error estimate $e_2^1$, also introduced in Section~\ref{subsec:MStepFormulation} and linked to an estimate of the collinearity of the prescribed displacement $\vect{x}_l$ to the master transverse modes $\vect{\phi}_1$, in the case of a linear computation (see Eq.~(\ref{eq:Kxl=f})). This estimate confirms the above results, in particular that the displacements must preferentially be prescribed on the neutral line.

A physical explanation of those effects can be deduced from the 3D displacement field of the modes. Because of the Poisson effect and the rotation of the sections , the displacement field on the nodes at other locations from the neutral line is not purely transverse for a bending eigenvector $\vect{\phi}_p$ and not zero for a NB eigenvector $\vect{\phi}_s$. This explains the losses of orthogonality observed above, as well as the loss of condition~(\ref{eq:assume1}c), since the master part of $\vect{\phi}_s$ is not zero: $\vect{\phi}_s^\text{s}\neq\vect{0}$.

\subsection{Application to a clamped circular plate}

In order to extend the results obtained on the beam test examples, the case of a clamped circular plate is here investigated. The selected plate has a radius $R=0.3~\text{m}$, a thickness $h=0.005\text{m}$, and the material properties are: density $\rho=7800~\text{kg/m}^3$, Young's modulus $E=210~\text{GPa}$ and Poisson ratio $v=0.3$. As for the beam cases, a coarse mesh is chosen so as to compute all the modes and apply the different proposed methodologies. Consequently, 540 HEX20 elements on the face and 2 HEX20 elements in the thickness were used, with a total of 1931 nodes and 4928 degrees of freedom.

The convergence study and appearance of thickness modes are investigated for the fundamental axisymmetric bending mode of the clamped plate, as well as the first asymmetric (1,0) mode, having one nodal line and no nodal circle. The case of the first axisymmetric mode is awaited to share the same complexity as the beam case for symmetry reasons, but the asymmetric mode might be more difficult to achieve convergence.

\begin{figure*}[h]
\centering
\begin{subfigure}{.44\textwidth}\centering
\includegraphics[width=\textwidth]{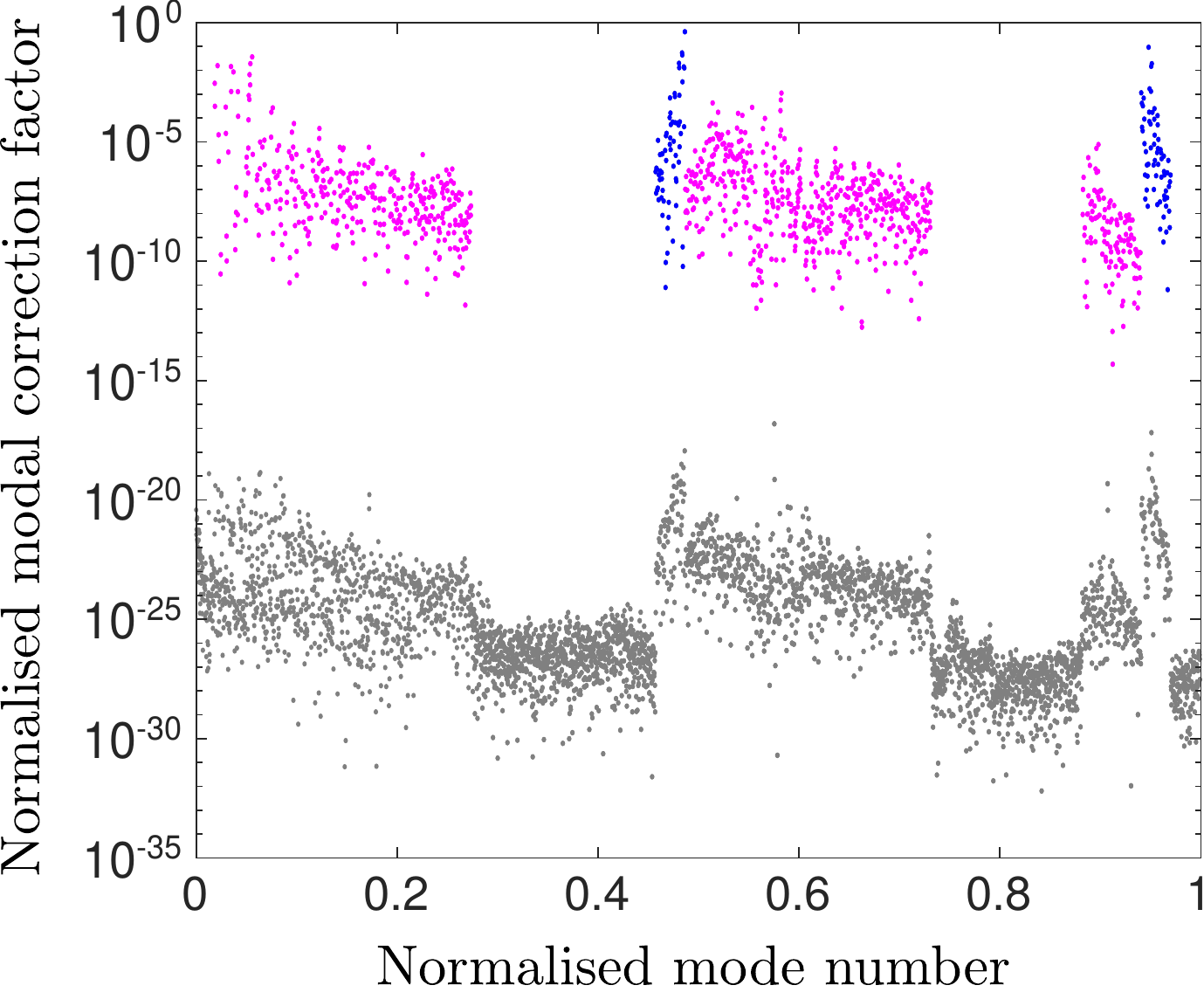}
\caption{}
\end{subfigure}
\hspace{4em}
\begin{subfigure}{.44\textwidth}\centering
\includegraphics[width=\textwidth]{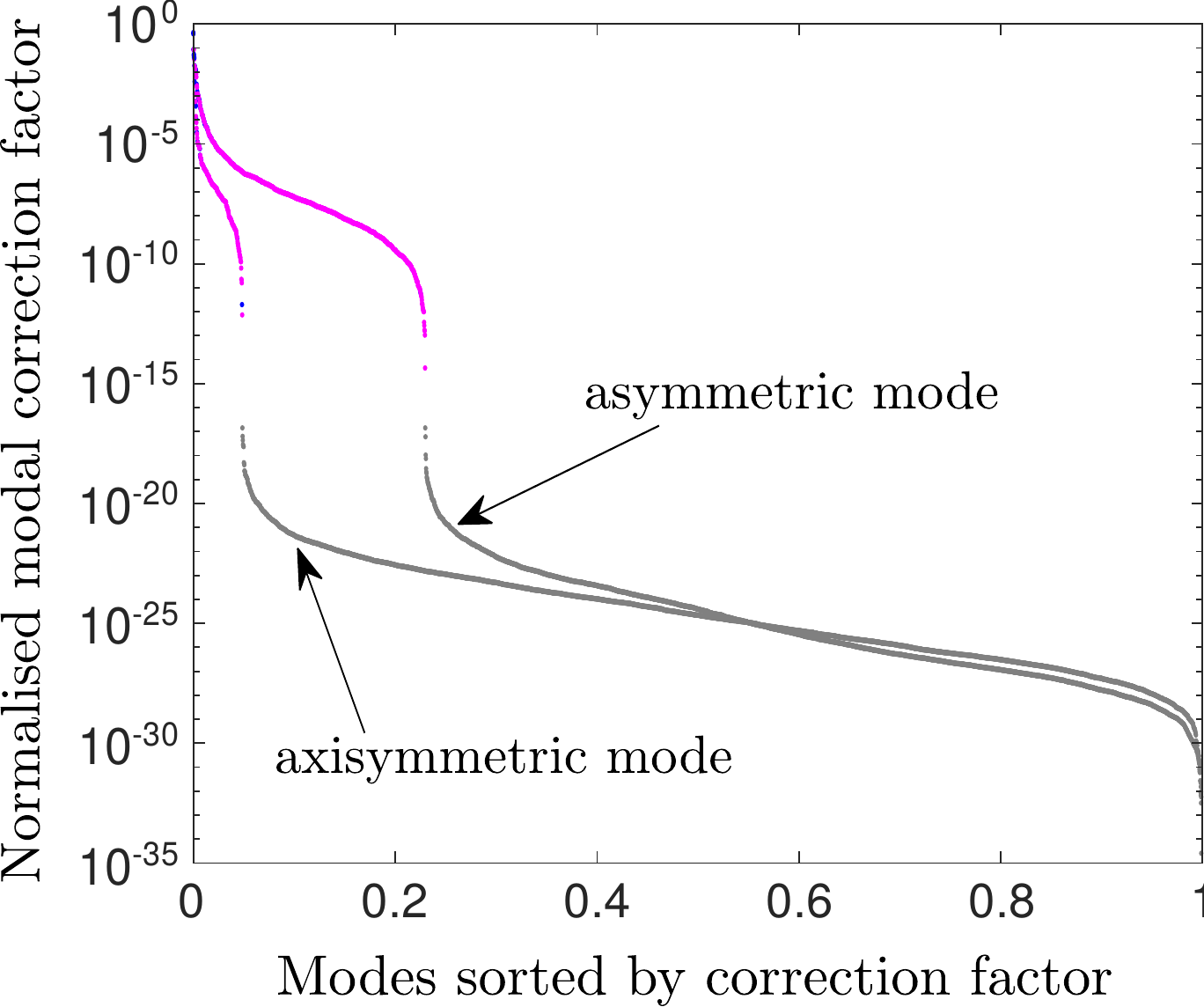}
\caption{}
\end{subfigure}
\caption{(a) Normalised modal correction factor for the clamped circular plate, as a function of the normalised mode number (normalization by the number of dofs), for the first asymmetric (1,0) mode of the plate. (b) The correction factors are now sorted by decreasing values, and two cases are shown : the case of the first asymmetic mode, corresponding to sorting (a), and the case of the first axisymmetric mode, showing a faster convergence. Grey points are negligible modes in terms of coupling, magenta points are the important in-plane coupled modes while blue points are the important thickness modes.}\label{fig:convcircplate}
\end{figure*}

Fig. \ref{fig:convcircplate}(a) shows the behaviour of the normalised modal correction factor $2(\alpha_{pp}^n/\omega_n)^2/\beta_{ppp}^p$ used in the previous sections, where $p$ refers to the master mode (either $p=1$ for the first axisymmetric mode, or $p=2$ for the first asymmetric) and $n\,\in\, \{1,N\}$ with $N$ the number of dofs. In Fig.~\ref{fig:convcircplate}(a) only the case of the first asymmetric mode is shown for the sake of brevity (thus $p=2$), but for $p=1$ the trend was very similar.
As for the beam, a strong coupling with very high frequency modes is also observed. Investigating more precisely which modes are involved in the couplings, it is found that the ones having the most important correction factor are once again thickness modes. 

Table~\ref{tab:thicknessmodescp} shows the deformed shape of the first nine modes, sorted according to their correction factor, which are thus the most important in the coupling with the bending (1,0) mode. Two purely in-plane modes are found, in position 5 and 8, and all others are thickness modes. The deformed shapes can be compared to that obtained for the beam and shown in Table~\ref{tab:BeamModes}. Indeed, the first thickness mode having the most important correction factor shows a similar geometry for both structures. Strong similarities are also observed between the second mode of the beam and mode (c) in Table~\ref{tab:thicknessmodescp}, and the ninth mode in each case.

Fig.~\ref{fig:convcircplate}(b) shows the normalized correction factor now sorted by order of decreasing values, and for the two cases of the axisymmetric fundamental mode and first asymmetric mode. It shows in particular that the convergence on the correct cubic coefficient is more rapidly achieved for the axisymmetric mode, where less than 10\% of the modes are needed. On the other hand, the convergence is more difficult for the first asymmetric mode. Concerning the coupling with high-frequency modes and thickness modes for these two first bending modes, it is interesting to note that the subset of coupled modes is almost exactly the same in the two cases, showing in particular that the coupling with the thickness modes is not very dependent on the selected bending mode. Indeed, more than 90\% of the coupled modes are the same for the two cases investigated.

\begin{table*}[h]
\centering
\begin{tabular}{cccccc}
\toprule
\includegraphics[height=2.5cm,width=2.5cm]{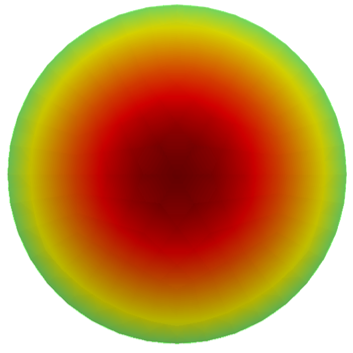}  \vline&
\includegraphics[height=2.5cm,width=2.5cm]{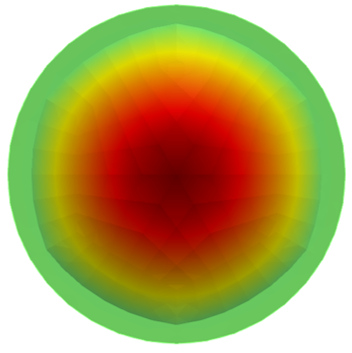} & 
\includegraphics[height=2.5cm,width=2.5cm]{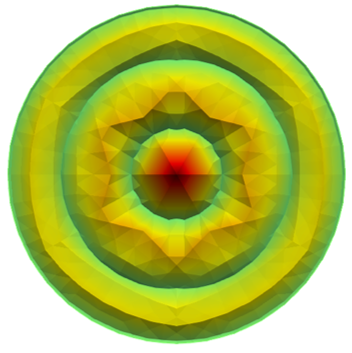}& 
\includegraphics[height=2.5cm,width=2.5cm]{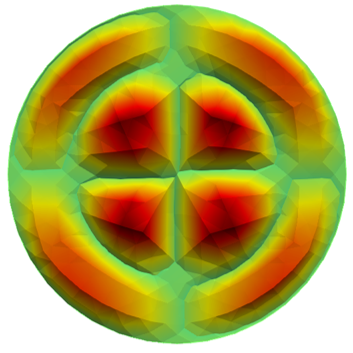}&
\includegraphics[height=2.5cm,width=2.5cm]{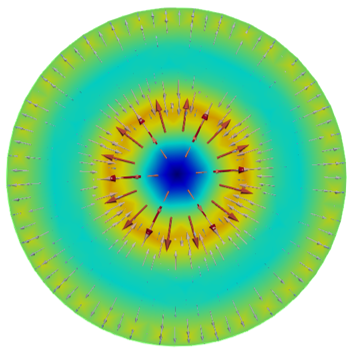}\\
\midrule
(a1)&(b)&(c)&(d)&(e)\\
\midrule
\includegraphics[height=2.5cm,width=2.5cm]{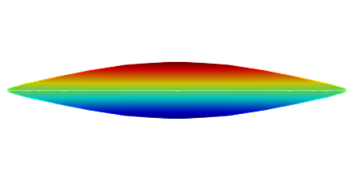} \vline&
\includegraphics[height=2.5cm,width=2.5cm]{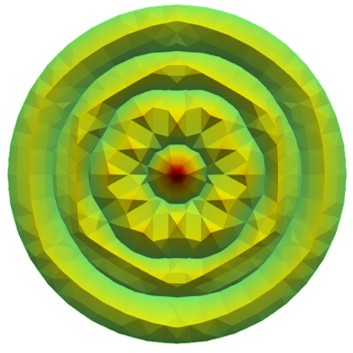} & 
\includegraphics[height=2.5cm,width=2.5cm]{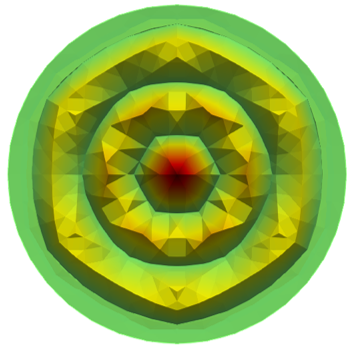}& 
\includegraphics[height=2.5cm,width=2.5cm]{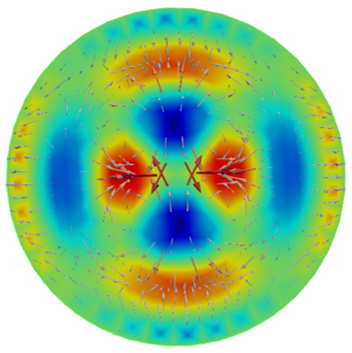}&
\includegraphics[height=2.5cm,width=2.5cm]{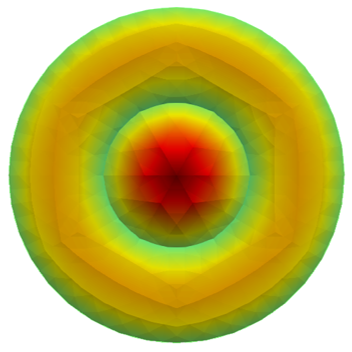}\\
\midrule
(a2)&(f)&(g)&(h)&(i)\\
\midrule
\end{tabular}
\caption{Mode shapes of the 9 most relevant modes coupled with the first flexural asymmetric (1,0) mode. Only two of them are in-plane modes: (e) and (h), while all the others are thickness modes. (a2) is a side view of the top view (a1) of the first thickness mode, in order to show the strong dependence on thickness deformation.}\label{tab:thicknessmodescp}
\end{table*}

\begin{table*}[h]
\begin{center}
\begin{tabular}{cccccc}
\hline
\multicolumn{6}{c}{Corrected cubic Coefficients}
\\  
& $\Gamma^2_{222}$
& $\Gamma^4_{222}$
& $\Gamma^4_{224}$ 
& $\Gamma^2_{444}$
& $\Gamma^4_{444}$
\\\hline\\
M-StEP
& 6.4763e+10
& -2.8539e+09
& 1.5374e+11
& -57.3021
& 3.8776e+11
\\\hline\\
Static condensation
& 6.4762e+10
& -2.8535e+09
& 1.5372e+11
& 5.3298e+03
& 3.8775e+11
\\\hline\\
SMDs condensed
& 6.4762e+10
&-2.8536e+09
& 1.5372e+11
& 5.3080e+03
& 3.8775e+11
\\\hline
\end{tabular}
\end{center}
\caption{Corrected cubic coefficient $\Gamma_{ijk}^p$, for two flexural modes, {\em i.e.} ${i,j,k,p}\in[2,4]$, where 2 refers to the first (1,0) asymmetric mode while 4 refers to the second (2,0) asymmetric mode; anf for three different methods : the modified StEP, the static condensation where all the coupled modes are statically condensed, and the Modal derivative where the added modal derivative is then statically condensed to the master mode.}
\label{tab:cubiccoeffplate}
\end{table*}

Table~\ref{tab:cubiccoeffplate} gathers the numerical results for the corrected cubic coefficient $\Gamma_{ijk}^p$ defined in Eq.~\eqref{eq:Gammadef}, with  $p=2$ for the first asymmetric (1,0) mode and $p=4$ for the (2,0) asymmetric mode (the first bending modes being sorted by increasing frequencies, $p=1$ is the fundamental axisymmetric, $p=2,3$ for the two configurations of the (1,0) asymmetric mode and $p=4,5$ for the two configurations of the (2,0) mode with two nodal lines). This choice has been guided by the fact that these two asymmetric modes are coupled and thus shows important cubic coefficients that are needed if one wants to derive a reduced-order model. The three methods presented in the previous sections: M-STEP, static condensation of all the linear modes and of the modal derivative, give the same results, showing the convergence of the methods also in this case. Only $\Gamma^2_{444}$ shows a different result for the M-STEP, however the value is very small as compared to the other ones so that this coefficient can be compared as negligible.

\section{Conclusion}

In this article, a nonlinear coupling of bending modes with thickness modes of very high frequency has been exhibited, due to geometrical nonlinearities in thin flat structures. This effect adds itself to the classical longitudinal / bending coupling and is the cause of a very slow convergence of a reduced order model (ROM) blindly built on a modal expansion of the nonlinear problem. It has been shown that if all eigenmodes are computed, it is possible to embed the effect of the non-bending modes into a master bending one, thus obtaining a  reduced order model composed of only one nonlinear Duffing oscillator. This procedure can be done either by static condensation or by a normal form reduction, equivalent to the reduction on a single nonlinear mode. Finally, two alternative methods have been proposed to overcome the problem: the use of a static modal derivative or the direct computation of the cubic coefficients by an original method, the M-STEP, inspired by the standard STEP. Those methods have been successfully verified on dedicated examples, showing equivalent results.

Most of the results presented in this paper are restricted to the case of flat structures. Indeed, the specific shape of the equations of motion (see Eqs.~(\ref{eq:qsimple}),(\ref{eq:psimple})) has been used to obtain exact equivalences between different methods. One can await that the obtained results should extend to shallow curved structures. However, for more generic shells with all the nonlinear couplings, most of the equivalences found here won't probably hold anymore.

We focused on the case of a 3D model discretized by finite elements. In section \ref{subsec:test}, we have shown on an example that some convergence problems might also appear for thin structures meshed with plate or shell elements, and having at least one long edge free. Our experience on thin ribbon have shown that the same kind of phenomenon appears when blindly using the STEP with the modal basis, and are again due to the loss of invariance of the modal eigenspaces. Indeed, high-frequency modes involving lateral deformations of the two free edges appeared. We also made computation on a circular plate with a free edge, and found circumferential modes appearing. Consequently, our finding is not restricted to 3D elements, and is completely linked to the use of the modal basis. Note that STEP calculations can also be realized with other input functions, and this has be done in this paper {\em e.g.} in Eq.~\eqref{eq:Gamma_smd}. The complete investigation of the analogy between this contribution, focused on 3D elements, and the problems related  to plate and shell finite elements,  is postponed to a future work.

\begin{acknowledgements}
The author A. Vizzaccaro is thankful to Rolls-Royce plc for the financial support. The author A. Givois is grateful to the French Ministry of Research and Arts et Metiers Institute of Technology for their financial support through the Ph.D. Grant of the author. The author Y. Shen wishes to thank China Scholarship Council (No.201806230253) for the funding of a three-year doctoral position at IMSIA, ENSTA Paris and EDF Lab. The author L. Salles is thankful to Rolls-Royce plc and the EPSRC for the support under the Prosperity Partnership Grant "Cornerstone: Mechanical Engineering Science to Enable Aero Propulsion Futures", Grant Ref: EP/R004951/1. The authors thanks J. Bla\v{s}os for the computation of the frequency response of the full finite element model used as reference.
\end{acknowledgements}

%
\section*{Conflict of interest}
The authors declare that they have no conflict of interest.

\bibliographystyle{spmpsci}
\bibliography{biblio}

\appendix
\section{Definition of cubic nonlinear terms}
\label{app:A}

A particular feature of the manipulations of nonlinear coupling coefficients lies in the fact that they are related to monoms having symmetry relationships. For example, a cubic coefficient $\beta^p_{ijj}$ is related to the monom $q_i q_j^2$, which is also the case of $\beta^p_{jij}$ and $\beta^p_{jji}$. Consequently many formulations  use upper triangular forms for quadratic and cubic coefficients $\alpha^p_{ij}$ (assuming $j \geq i$) and $\beta^p_{ijk}$ (with $k \geq j \geq i$). However, direct calculations produce coefficients that have not this ordering property built-in. The formula given here allows one to get from to another formulation.

The corrected cubic coefficients in Eq.~(\ref{eq:qcubic}) can be derived by replacing the value of $p_s$ from Eq.~\eqref{eq:pstatic} into the quadratic term of Eq.~\eqref{eq:qsimple}:
\begin{equation*}
\sum_{i=1}^{N_B}\sum_{s=N_B+1}^{N} \alpha_{is}^r q_ip_s=
\sum_{i=1}^{N_B}\sum_{j=1}^{N_B}\sum_{k=j}^{N_B}\sum_{s=N_B+1}^{N}  -\dfrac{\alpha_{is}^r\alpha_{jk}^s}{\omega_s^2}q_iq_jq_k
\end{equation*}
Manipulating the right hand side to make the sums over $i$, $j$, and $k$ consistent with the ones of the cubic term of Eq.~\eqref{eq:qsimple}, i.e. having the sum over $j$ from $i$ to $N_B$, reads:
\begin{equation}
\begin{split}
\sum_{i=1}^{N_B}\sum_{j=1}^{N_B}\sum_{k=j}^{N_B}
\sum_{s=N_B+1}^{N}  -\dfrac{\alpha_{is}^r\alpha_{jk}^s}{\omega_s^2}q_iq_jq_k=
\sum_{i=1}^{N_B}\sum_{j=i+1}^{N_B}\sum_{k=j+1}^{N_B}&\sum_{s=N_B+1}^{N} \left(
-\dfrac{{\alpha}^r_{is}\alpha^s_{jk}}{\omega_s^2}
-\dfrac{{\alpha}^r_{js}\alpha^s_{ik}}{\omega_s^2}
-\dfrac{{\alpha}^r_{ks}\alpha^s_{ij}}{\omega_s^2}\right)q_iq_jq_k+\\
\sum_{i=1}^{N_B}\sum_{k=i+1}^{N_B}&
\sum_{s=N_B+1}^{N} \left(
-\dfrac{{\alpha}^r_{is}\alpha^s_{ik}}{\omega_s^2}
-\dfrac{{\alpha}^r_{ks}\alpha^s_{ii}}{\omega_s^2}\right)q_i^2q_k+\\
\sum_{i=1}^{N_B}\sum_{k=i+1}^{N_B}&
\sum_{s=N_B+1}^{N} \left(
-\dfrac{{\alpha}^r_{is}\alpha^s_{kk}}{\omega_s^2}
-\dfrac{{\alpha}^r_{ks}\alpha^s_{ik}}{\omega_s^2}\right)q_iq_k^2+\\
\sum_{i=1}^{N_B}&
\sum_{s=N_B+1}^{N} \left(
-\dfrac{{\alpha}^r_{is}\alpha^s_{ii}}{\omega_s^2}
\right)q_i^3.
\end{split}
\label{eq:corr_term}
\end{equation}
This term can be rewritten in a more compact form by defining the correction factor:
\begin{equation*}
\mathcal{C}^{rs}_{ijk}=
\begin{cases}
\rule[0pt]{0pt}{21pt}
\dfrac{{\alpha}^r_{is}\alpha^s_{jk}}{\omega_s^2}+
\dfrac{{\alpha}^r_{js}\alpha^s_{ik}}{\omega_s^2}+
\dfrac{{\alpha}^r_{ks}\alpha^s_{ij}}{\omega_s^2}\qquad i<j<k\\
\rule{0pt}{21pt}
\dfrac{{\alpha}^r_{is}\alpha^s_{ik}}{\omega_s^2}+
\dfrac{{\alpha}^r_{ks}\alpha^s_{ii}}{\omega_s^2}\hfill i=j<k\\
\rule{0pt}{21pt}
\dfrac{{\alpha}^r_{is}\alpha^s_{kk}}{\omega_s^2}+
\dfrac{{\alpha}^r_{ks}\alpha^s_{ik}}{\omega_s^2}\hfill i<j=k\\
\rule{0pt}{21pt}
\dfrac{{\alpha}^r_{is}\alpha^s_{ii}}{\omega_s^2}\hfill i=j=k
\end{cases}
\end{equation*}
leading to:
\begin{equation*}
\sum_{i=1}^{N_B}\sum_{j=1}^{N_B}\sum_{k=j}^{N_B}
\sum_{s=N_B+1}^{N}  -\dfrac{\alpha_{is}^r\alpha_{jk}^s}{\omega_s^2} q_iq_jq_k=
\sum_{i=1}^{N_B}\sum_{j=i}^{N_B}\sum_{k=j}^{N_B}\sum_{s=N_B+1}^{N} 
-\mathcal{C}^{rs}_{ijk}q_iq_jq_k
\end{equation*}
now with summation indexes consistent with the ones of Eq.~(\ref{eq:qsimple}).\\
Finally, the quadratic and cubic nonlinear terms in Eq.~(\ref{eq:qsimple}) read:
\begin{equation*}
\sum_{i=1}^{N_B}\sum_{s=N_B+1}^{N} \alpha_{is}^r q_ip_s+
\sum_{i=1}^{N_B}\sum_{j=i}^{N_B}\sum_{k=j}^{N_B}
\beta^r_{ijk} q_iq_jq_k=
\sum_{i=1}^{N_B}\sum_{j=i}^{N_B}\sum_{k=j}^{N_B}
\left(
\beta^r_{ijk}-\sum_{s=N_B+1}^{N} 
\mathcal{C}^{rs}_{ijk}
\right)
q_iq_jq_k.
\end{equation*}
and the corrected cubic coefficient in Eq.~\eqref{eq:qcubic}:
\begin{equation}\label{eq:Gammadef}
\Gamma^r_{ijk}=\beta^r_{ijk}-\sum_{s=N_B+1}^{N} 
\mathcal{C}^{rs}_{ijk}
\end{equation}
\section{Expression of static modal derivatives in terms of quadratic coupling coefficients}\label{app:B}
Given the general equation of a system with quadratic and cubic nonlinearities in physical coordinates:
\begin{equation}
\vect{M}\ddot{\bm{x}}+\vect{C}\dot{\bm{x}}+\vect{K}\bm{x}+\vect{f}_\text{nl}(\bm{x})=\vect{f}_\text{e},
\end{equation}
the nonlinear force can be written in terms of nonlinear tensors as:
\begin{equation}
\vect{f}_\text{nl}(\bm{x})=\vect{A}\bm{xx}+\vect{B}\bm{xxx},\label{eq:nonlinear_force}
\end{equation}
where we used the compact tensor notation also employed in~\cite{Jain2017,Rutzmoser}. In order to explicit the notation, the products $\vect{A}\bm{xx}$ and $\vect{B}\bm{xxx}$ are here given with explicit indicial notation:
\begin{subequations}
\begin{align*}
\vect{A}\bm{xx}
=&
\sum_{i=1}^N \sum_{j=1}^N
\vect{A}_{ij}{x}_{j}{x}_{i} \, ,\\
\vect{B}\bm{xxx}=&
\sum_{i=1}^N \sum_{j=1}^N \sum_{k=1}^N
\vect{B}_{ijk}{x}_{i}{x}_{j}{x}_{k}.
\end{align*}
\end{subequations}
The most inner  product defined above coincides with a matrix product performed on the last index of the tensors.

In order to derive Eq.~\eqref{eq:SMD_pp}, we first express the $i$-th column of nonlinear stiffness matrix as:
\begin{equation}
\left(\dfrac{\partial\vect{f}_\text{nl}}{\partial\vect{x}}\right)_i=\sum_{j=1}^N
\vect{A}_{ij}{x}_{j}+\sum_{j=1}^N
\vect{A}_{ji}{x}_{j}+
\sum_{j=1}^N \sum_{k=1}^N
\vect{B}_{ijk}{x}_{j}{x}_{k}
+
\sum_{j=1}^N \sum_{k=1}^N
\vect{B}_{jik}{x}_{j}{x}_{k}
+
\sum_{j=1}^N \sum_{k=1}^N
\vect{B}_{jki}{x}_{j}{x}_{k}.
\label{eq:nonlinear_stiffness_ext}
\end{equation}
Exploiting the symmetry of the tensors $\vect{A}$ and $\vect{B}$ which implies that $\vect{A}_{ij}=\vect{A}_{ji}$ and similarly $\vect{B}_{ijk}=\vect{B}_{jik}=\vect{B}_{jki}$, that stems from the fact that geometric nonlinear forces can be derived from a potential (see \cite{muravyov}), the nonlinear stiffness matrix can be written in compact form as:
\begin{equation}
\dfrac{\partial\vect{f}_\text{nl}}{\partial\vect{x}}=
2\vect{A}\bm{x}+3\vect{B}\bm{xx}.
\label{eq:nonlinear_stiffness}
\end{equation}
The nonlinear stiffness matrix evaluated along mode $p$ reads:
\begin{equation}
\dfrac{\partial\vect{f}_\text{nl}}{\partial{\vect{x}}}(\boldsymbol{\Phi}_p q_p)=
2q_p \vect{A}\vect{\Phi}_p +3q_p^2 \vect{B}
\boldsymbol{\Phi}_p\boldsymbol{\Phi}_p \, ,
\label{eq:nonlinear_stiffness_along_mode_p}
\end{equation}
and its derivatives with respect to $q_p$ evaluated at $q_p=0$ reads:
\begin{equation}
\left.\left(\dfrac{\partial}{\partial q_p}
\dfrac{\partial\vect{f}_\text{nl}}{\partial{\vect{x}}}(\boldsymbol{\Phi}_p q_p)
\right)
\right|_{q_p=0}=2\vect{A}\vect{\Phi}_p.
\label{eq:nonlinear_stiffness_along_mode_p_derived}
\end{equation}
Hence the static modal derivatives defined in Eq.~\eqref{eq:Jain} leads to:
\begin{equation}\label{eq:SMD_apr}
\vect{\theta}_{pr}=-2\vect{K}^{-1}\vect{A}\vect{\Phi}_p\vect{\Phi}_r\, ,
\end{equation}
and, when $p=r$, to:
\begin{equation}\label{eq:SMD_app}
\vect{\theta}_{pp}=-2\vect{K}^{-1}\vect{A}\vect{\Phi}_p\vect{\Phi}_p.
\end{equation}

We now want to relate these expressions to the quadratic modal coupling coefficients $\alpha^s_{pr}$ obtained from the STEP method, as well as expliciting directly how to compute the modal derivative from specific static evaluation of the internal force vector. When $p=r$, Eq.~\eqref{eq:zicoefSTEPq} reads:
\begin{equation}
\label{eq:alpha_coefficients}
\alpha^s_{pp}=\tp{\vect{\Phi}_s}\left(\dfrac{ \bm{f}_\text{nl}(q_p\bm{\Phi}_p)+ \bm{f}_\text{nl}(-q_p\bm{\Phi}_p)}{2q_p^2}\right)/m_s.
\end{equation}
When $p \neq r$, the STEP method needs to call two static evaluations with $q_p\bm{\Phi}_p$ and $q_s\bm{\Phi}_s$ so that the expression writes:
\begin{equation}
\label{eq:alpha_coefficientspr}
\alpha^s_{pr}=\tp{\vect{\Phi}_s}\left(\dfrac{ \bm{f}_\text{nl}(q_p\bm{\Phi}_p+q_r\bm{\Phi}_r)+ \bm{f}_\text{nl}(-q_p\bm{\Phi}_p-q_r\bm{\Phi}_r)-\bm{f}_\text{nl}(q_p\bm{\Phi}_p)-\bm{f}_\text{nl}(-q_p\bm{\Phi}_p)-\bm{f}_\text{nl}(q_r\bm{\Phi}_r)-\bm{f}_\text{nl}(-q_r\bm{\Phi}_r)}{2q_p q_r}\right)/m_s.
\end{equation}
Let us now assume that the eigenvectors are mass normalised so that $m_s=1,\;\forall s$, using Eq.~\eqref{eq:nonlinear_force} we can find the relation between $\alpha^s_{pp}$ and $\vect{A}$ as:
\begin{equation}
\label{eq:alpha_coefficients_pp}
\alpha^s_{pp}=\tp{\vect{\Phi}_s}\;
\vect{A}\bm{\Phi}_p\bm{\Phi}_p \, .
\end{equation}
Similarly for the coefficients $\alpha^s_{pr}$ with $p<r$:
\begin{equation}
\label{eq:alpha_coefficients_pr}
\alpha^s_{pr}=2\tp{\vect{\Phi}_s}\;
\vect{A}\bm{\Phi}_p\bm{\Phi}_r \, ,
\end{equation}
where the factor 2 appears for symmetry reasons and is related to the usual problem of representing the polynomial monoms by counting them separately or not. Indeed, in the usual polynomial representation, 
 the coefficients $\alpha^s_{pr}$ with $p<r$ contains both values of $\tp{\vect{\Phi}_s}\;\vect{A}\bm{\Phi}_p\bm{\Phi}_r$ and  $\tp{\vect{\Phi}_s}\;\vect{A}\bm{\Phi}_r\bm{\Phi}_p$ whereas the coefficients $\alpha^s_{rp}$ are set to zero. We can now observe that each coefficient $\alpha^s_{pr}$ can be seen as the row of a vector $\bm{\alpha}_{pr}$ whose compact expression would be:
\begin{equation}
\label{eq:alpha_vector}
\vect{\alpha}_{pr}=2\tp{\vect{V}}\;
\vect{A}\bm{\Phi}_p\bm{\Phi}_r \, ,
\end{equation}
where the full matrix of eigenvector $\vect{V}$ has been introduced. In the same line, the expression for the vector $\bm{\alpha}_{pp}$ when $p=r$ has the same shape but without the factor 2:
\begin{equation}
\label{eq:alphapp_vector}
\vect{\alpha}_{pp}=\tp{\vect{V}}\;
\vect{A}\bm{\Phi}_p\bm{\Phi}_p \, .
\end{equation}
Introducing Eqs.~\eqref{eq:alpha_vector}-\eqref{eq:alphapp_vector} respectively in Eqs.~\eqref{eq:SMD_apr}-\eqref{eq:SMD_app}, and using the relationships of the quadratic coupling coefficients from the STEP method, Eqs.~\eqref{eq:alpha_coefficientspr}-\eqref{eq:alpha_coefficients}, one obtains the important formulas allowing one to compute directly the modal derivative from static FE calculations. The expression for $\vect{\theta}_{pp}$ is given in the main text as Eq.~\eqref{eq:SMD_fnl} while the formula for $\vect{\theta}_{pr}$ reads:
\begin{equation}\label{eq:thetapr_fnl}
\vect{\theta}_{pr}=-\vect{K}^{-1}\left(\dfrac{ \bm{f}_\text{nl}(\lambda(\bm{\Phi}_p+\bm{\Phi}_r))+ \bm{f}_\text{nl}(-\lambda(\bm{\Phi}_p+\bm{\Phi}_r))-\bm{f}_\text{nl}(\lambda\bm{\Phi}_p)-\bm{f}_\text{nl}(-\lambda\bm{\Phi}_p)-\bm{f}_\text{nl}(\lambda\bm{\Phi}_r)-\bm{f}_\text{nl}(-\lambda\bm{\Phi}_r)}{2\lambda^2}\right)
\end{equation}

We now want to relate more closely the modal derivative to the modal representation and derive an explicit expression showing that the SMD gathers the contributions of all coupled modes. For that purpose, one needs to express the usual  orthonormality conditions shared by the matrix of eigenvectors $\vect{V}$, assumed to be mass normalized:
\begin{equation*}
\tp{\vect{V}}\vect{M}\vect{V}=\vect{I},\qquad\qquad
\tp{\vect{V}}\vect{K}\vect{V}=\vect{\Omega}^2,
\end{equation*}
where $\vect{\Omega}^2$ is the diagonal matrix containing the squared eigenfrequencies.

Recalling eq. \eqref{eq:SMD_apr} we can now express the static modal derivatives in terms of $\bm{\alpha}_{pr}$:
\begin{equation}
\bm{\theta}_{pr}=-\vect{K}^{-1}\vect{V}^{-\text{T}}\;\bm{\alpha}_{pr},
\end{equation}
and for $p=r$:
\begin{equation}
\bm{\theta}_{pp}=-2\vect{K}^{-1}\vect{V}^{-\text{T}}\;\bm{\alpha}_{pp}.
\end{equation}

Moreover, using the orthogonality conditions:
\begin{equation}
\vect{K}^{-1}\vect{V}^{-\text{T}}=\vect{V}\vect{\Omega}^{-2},
\end{equation}
we can express the static modal derivatives in the general case as:
\begin{align}
\label{eq:link_0}
&\bm{\theta}_{pr}=
-\vect{V}\vect{\Omega}^{-2}\;\bm{\alpha}_{pr},\\
&\bm{\theta}_{pp}=
-2\vect{V}\vect{\Omega}^{-2}\,\bm{\alpha}_{pp}.
\end{align}
In the particular case of a flat structure, the equations of motions have a simple structure recalled in Eqs.~\eqref{eq:qsimple}-\eqref{eq:psimple}, so that $\alpha^s_{pr}$ is nonzero only for the non-bending modes. Thanks to this simplification, one can express the static modal derivatives as a linear combination of the non-bending modes only as:
\begin{align}
\label{eq:link}
&\bm{\theta}_{pr}=
-\sum^{N}_{s=N_B+1}\bm{\Phi}_s \dfrac{\alpha^s_{pr}}{\omega^2_s},\\
&\bm{\theta}_{pp}=
-\sum^{N}_{s=N_B+1}2\,\bm{\Phi}_s \dfrac{\alpha^s_{pp}}{\omega^2_s}.
\end{align}

\section{Corrected cubic coefficient obtained from static modal derivatives}
\label{app:C}

This appendix aims at demonstrating that coefficient $\tilde{\Gamma}_{ppp}^p$ introduced in Eq.~\eqref{eq:romSMDsc}, {\em i.e.} by using an imposed displacement composed of the master mode plus a quadratic part containing the SMD,  is exactly equal to that given in Eq.~\eqref{eq:Gammappp}, and obtained thanks to the static condensation. For that purpose, let us recall that the imposed displacement in the first case reads:
\begin{equation}\label{eq:x_qp}
\bm{x}(q_p)=q_p\;\bm{\Phi}_p+\frac{1}{2}\;q_p^2\;\bm{\theta}_{pp}.
\end{equation}
The cubic coefficient $\tilde{\Gamma}_{ppp}^p$ can be computed by using the general formula from the STEP method, Eq.~\eqref{eq:zicoefSTEPq}, by replacing the imposed displacement by the one given in Eq.~\eqref{eq:x_qp}. Consequently, one arrives at:
\begin{equation}
\tilde{\Gamma}_{ppp}^p=\tp{\bm{\Phi}_p}\left(
\vect{f}_\text{nl}(\lambda\bm{\Phi}_p+\frac{1}{2}\lambda^2\bm{\theta}_{pp})-\vect{f}_\text{nl}(-\lambda\bm{\Phi}_p+\frac{1}{2}\lambda^2\bm{\theta}_{pp})\right)/2\lambda^3.
\end{equation}
Using the explicit expression of the nonlinear forces from Eq.~\eqref{eq:nonlinear_force}:
\begin{align}
&\vect{f}_\text{nl}(\lambda\bm{\Phi}_p+\frac{1}{2}\lambda^2\bm{\theta}_{pp})=
\lambda^2\vect{A}\bm{\Phi}_p\bm{\Phi}_p\;+\;
\lambda^3\left(\frac{1}{2}
\vect{A}\bm{\Phi}_p\bm{\theta}_{pp}+\frac{1}{2}
\vect{A}\bm{\theta}_{pp}\bm{\Phi}_p+
\vect{B}\bm{\Phi}_p\bm{\Phi}_p\bm{\Phi}_p
\right)\;+\;\mathcal{O}(\lambda^4)\\
&\vect{f}_\text{nl}(-\lambda\bm{\Phi}_p+\frac{1}{2}\lambda^2\bm{\theta}_{pp})=
\lambda^2\vect{A}\bm{\Phi}_p\bm{\Phi}_p\;-\;
\lambda^3\left(\frac{1}{2}
\vect{A}\bm{\Phi}_p\bm{\theta}_{pp}+\frac{1}{2}
\vect{A}\bm{\theta}_{pp}\bm{\Phi}_p+
\vect{B}\bm{\Phi}_p\bm{\Phi}_p\bm{\Phi}_p
\right)\;+\;\mathcal{O}(\lambda^4).
\end{align}
Thus the difference between the above nonlinear forces reads:
\begin{equation}
\vect{f}_\text{nl}(\lambda\bm{\Phi}_p+\frac{1}{2}\lambda^2\bm{\theta}_{pp})-\vect{f}_\text{nl}(-\lambda\bm{\Phi}_p+\frac{1}{2}\lambda^2\bm{\theta}_{pp})=2\lambda^3\left(\frac{1}{2}
\vect{A}\bm{\Phi}_p\bm{\theta}_{pp}+\frac{1}{2}
\vect{A}\bm{\theta}_{pp}\bm{\Phi}_p+
\vect{B}\bm{\Phi}_p\bm{\Phi}_p\bm{\Phi}_p
\right)\;+\;\mathcal{O}(\lambda^5)
\end{equation}
and the coefficient $\tilde{\Gamma}_{ppp}^p$, neglecting high order terms, reads:
\begin{equation}\label{eq:Gamma_AB}
\tilde{\Gamma}_{ppp}^p=\frac{1}{2}\tp{\bm{\Phi}_p}
\vect{A}\bm{\Phi}_p\bm{\theta}_{pp}+
\frac{1}{2}\tp{\bm{\Phi}_p}
\vect{A}\bm{\theta}_{pp}\bm{\Phi}_p+
\tp{\bm{\Phi}_p}
\vect{B}\bm{\Phi}_p\bm{\Phi}_p\bm{\Phi}_p
\end{equation}
Following a similar argument as the one of Eq.~\eqref{eq:alpha_coefficients} it is possible to show that the last term on the right hand side of Eq.~\eqref{eq:Gamma_AB} is the uncorrected cubic coefficient:
\begin{equation}
\beta^p_{ppp}=\tp{\bm{\Phi}_p}\vect{B}\bm{\Phi}_p\bm{\Phi}_p\bm{\Phi}_p
\end{equation}
As regards to the first two terms on the right hand side of Eq.~\eqref{eq:Gamma_AB}, using the link between SMD and non-bending modes of Eq.~\eqref{eq:SMD_pp} valid in case of a flat structure, they can be written as:
\begin{align}
&\frac{1}{2}\tp{\bm{\Phi}_p}
\vect{A}\bm{\Phi}_p\bm{\theta}_{pp}
=
-\frac{1}{2}\sum^{N}_{s=N_B+1}
\tp{\bm{\Phi}_p}
\vect{A}\bm{\Phi}_p\vect{\Phi}_s
\;\dfrac{2\alpha^s_{pp}}{\omega^2_s}
\\
&\frac{1}{2}\tp{\bm{\Phi}_p}
\vect{A}\bm{\theta}_{pp}\bm{\Phi}_p
=
-\frac{1}{2}\sum^{N}_{s=N_B+1}
\tp{\bm{\Phi}_p}
\vect{A}\vect{\Phi}_s\bm{\Phi}_p
\;\dfrac{2\alpha^s_{pp}}{\omega^2_s}
\end{align}
where the index $s$ spans over all the non-bending modes. Using the symmetry of the tensor $\vect{A}$ once again:
\begin{equation}
\tp{\bm{\Phi}_p}
\vect{A}\vect{\Phi}_s\bm{\Phi}_p+\tp{\bm{\Phi}_p}
\vect{A}\bm{\Phi}_p\vect{\Phi}_s=2\;\tp{\bm{\Phi}_p}
\vect{A}\bm{\Phi}_p\vect{\Phi}_s
\end{equation}
and in light of Eq.~\eqref{eq:alpha_coefficients_pr}:
\begin{equation}
2\;\tp{\bm{\Phi}_p}
\vect{A}\bm{\Phi}_p\vect{\Phi}_s
=
\alpha^p_{ps}
\end{equation}
because $p>s$ for each non-bending mode.

It is now possible to express the corrected cubic coefficient as:
\begin{equation}
\tilde{\Gamma}_{ppp}^p=\beta^p_{ppp}-\sum^{N}_{s=N_B+1}
\alpha^p_{ps}
\;\dfrac{\alpha^s_{pp}}{\omega^2_s}=\Gamma_{ppp}^p
\end{equation}
and recover the same expression of the corrected coefficient given in Eq.~\eqref{eq:Gammappp} obtained with static condensation of all non-bending modes.

It is worth mentioning that the demonstration could have been done with another pathway. Indeed, the displacement introduced in Eq.~\eqref{eq:x_qp} follows the general strategy proposed in~\cite{Jain2017,Rutzmoser}, consisting of using SMD to build a quadratic manifold approach in order to define a reduced-order models thanks to a nonlinear mapping between physical and reduced coordinates. However, the SMD can also be used more simply, by considering $\bm{\theta}_{pp}$ as an enrichment of 
the modal basis, composed here of the single master eigenvector $\bm{\Phi}_p$. One could then derive a two-dofs reduced-order model, by projecting the general equations of motion onto these two vectors. Then, since the motions associated to the SMD are linked to NB modes having high frequencies, one can neglect their inertia and proceed to the static condensation of the part coming from the static modal derivative. By doing so, one would show again that the corrected cubic coefficient is still equal to $\Gamma_{ppp}^p$. Hence the static condensation of the SMD (a single vector) is thus strictly equivalent, in our simplified case of a flat structure (and considering only one master mode), to the static condensation of all coupled NB modes (including thickness modes).

\section{Analytical solution for the pure bending of a beam}
\label{sec:purebending}

\begin{figure*}[ht]
\begin{center}
\raisebox{6em}{
\parbox{.5\textwidth}{
\centering
\begin{picture}(0,0)%
\includegraphics{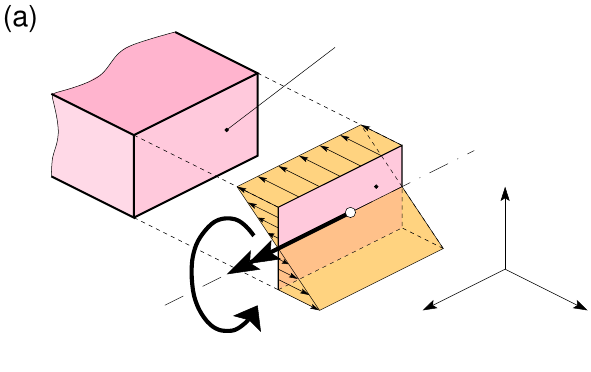}%
\end{picture}%
\setlength{\unitlength}{1302sp}%
\begin{picture}(8562,5376)(8251,-10549)
\put(13126,-5686) {\makebox(0,0)[b]{\smash{$\mathcal{S}$}}}
\put(11043,-10413){\makebox(0,0)[b]{\smash{$\vect{M}_b$ }}}
\put(14251,-6736) {\makebox(0,0)[b]{\smash{$\sigma_{xx}$}}}
\put(16651,-9361) {\makebox(0,0)[b]{\smash{$\vect{e}_x$ }}}
\put(14476,-10036){\makebox(0,0)[b]{\smash{$\vect{e}_z$ }}}
\put(15976,-8236) {\makebox(0,0)[b]{\smash{$\vect{e}_y$ }}}
\end{picture}%
\\
\includegraphics[scale=.8]{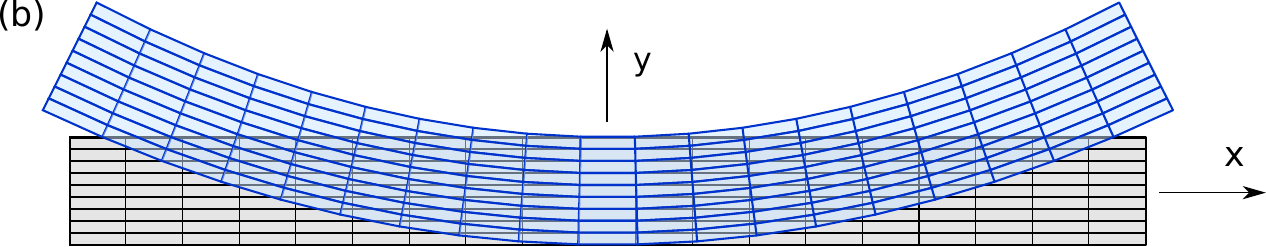}
}}
\hspace*{2em}\includegraphics[scale=.5]{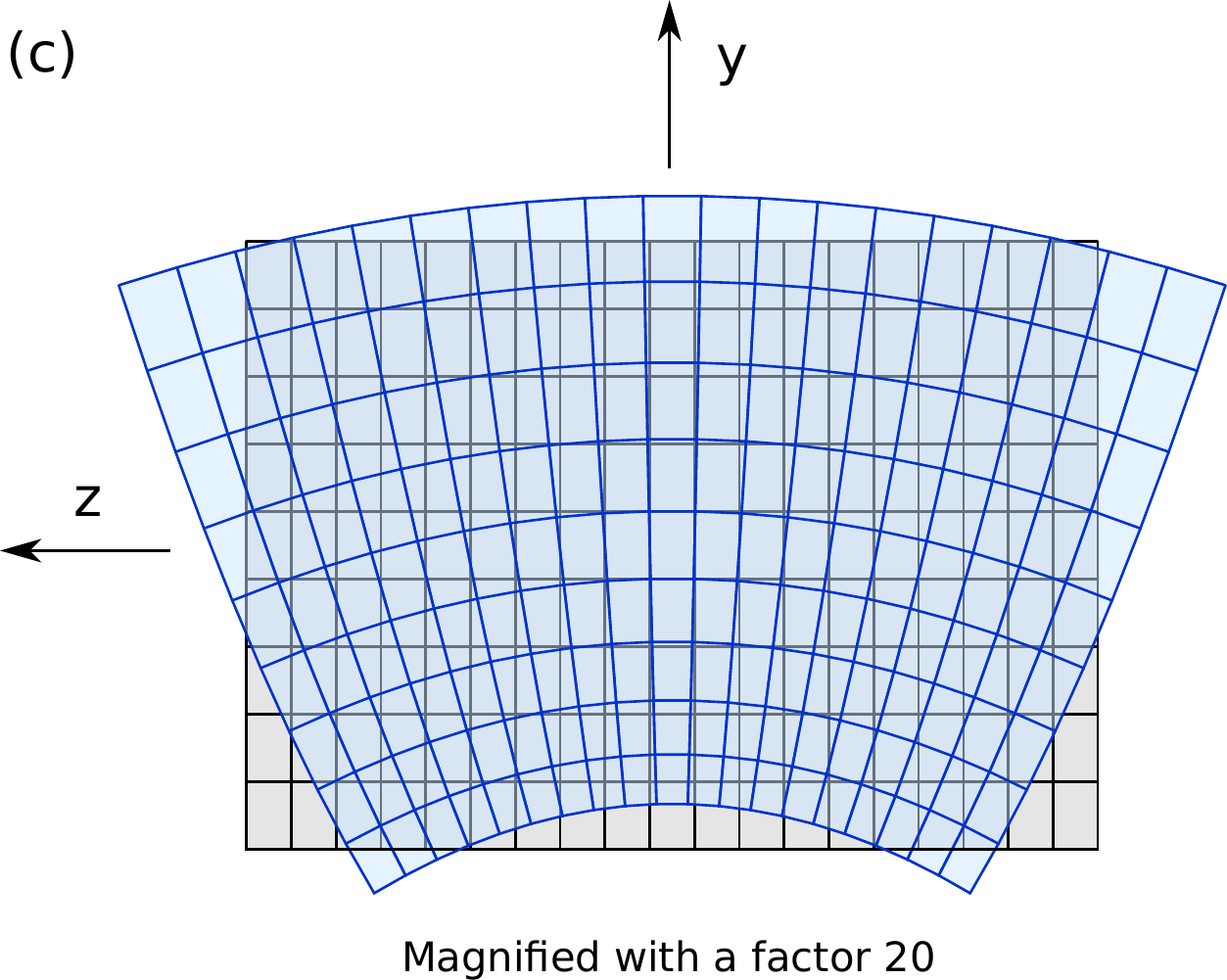}
\end{center}
\caption{Analytical solution of a beam in pure bending. (a) Stress state $\sigma_{xx}$ in a cross section $\mathcal{S}$; (b,c) reference (in grey) and deformed (in blue) configurations of the beam. The Poisson effect has been magnified with a factor 20 in the cross section view (c).}
\label{fig:beamanalytic}
\end{figure*}
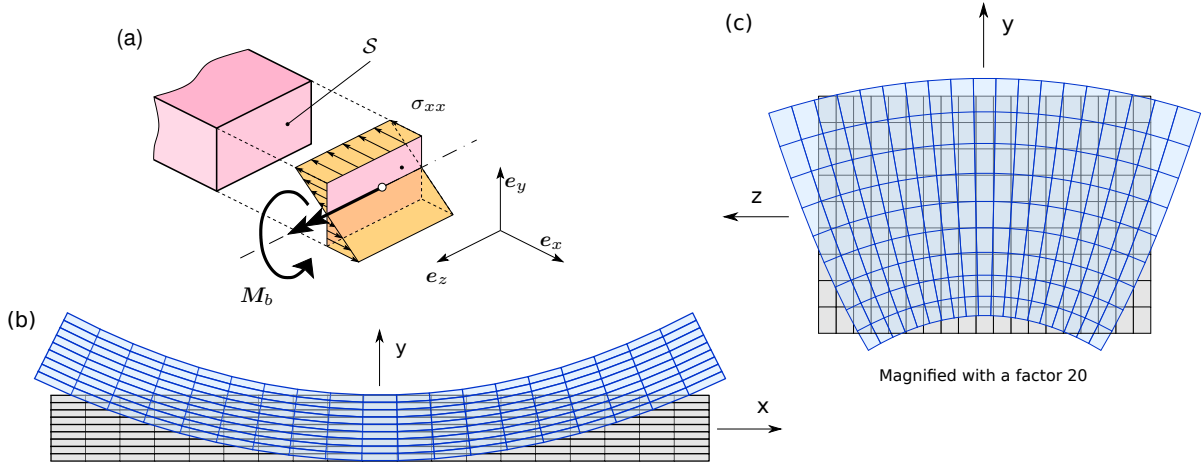

We consider the linear elastic solution of a beam of rectangular cross section under pure bending (see Fig.~\ref{fig:beamanalytic}). The beam is thus subjected to a uniform bending moment $\vect{M}_b=M\vect{e}_z$. The material of the beam is assumed linear elastic with Young's modulus $E$ and Poisson's ratio $\nu$. The orthonormal frame $(\vect{e}_x,\vect{e}_y,\vect{e}_z)$ is used, with $\vect{e}_x$ colinear to the middle axis of the beam, $\vect{e}_y$ the direction of bending and $\vect{e}_z=\vect{e}_x\wedge\vect{e}_y$ with $\wedge$ the vector product. The local equilibrium of the beam is exactly verified by the following axial stress state, linear through the thickness of the beam:
\begin{equation}
\vdiver\tens{\sigma}=\vect{0}\quad\Rightarrow\quad
\tens{\sigma}=\begin{pmatrix}
\sigma_{xx} & 0 & 0 \\
0 & 0 & 0 \\
0 & 0 & 0
\end{pmatrix}
,\quad\sigma_{xx}=-\alpha y,
\end{equation}
where  $\vdiver$ is the divergence operator, $\tens{\sigma}$ is the stress tensor and $\alpha$ is a constant. The bending moment writes $M=\alpha I$ where $I$ is the second moment of inertia of the beam. The linear strain tensor $\tens{\varepsilon}$ verifies:
\begin{equation}
\tens{\varepsilon}=\frac{1}{2}\left(\tgrad\vect{U} + \tptgrad\vect{U}\right) = \frac{1+\nu}{E}\tens{\sigma}-\frac{\nu}{E}\trace\tens{\sigma}\tens{I}_3\quad\Rightarrow\quad\tens{\epsilon}=-\alpha\begin{pmatrix}
 y /E & 0 & 0 \\
0 & -\nu y/E & 0 \\
0 & 0 & -\nu y/E
\end{pmatrix}
\end{equation}
As exposed in \cite{salencon2001}, the following displacement field verifies exactly the above equations:
\begin{equation}
\label{eq:Ulin}
\vect{U}=\alpha xy\,\vect{e}_x-\frac{\alpha}{2}\left[x^2+\nu(y^2-z^2)\right]\,\vect{e}_y-\nu \alpha y z \,\vect{e}_z.
\end{equation}
This displacement is composed of three parts:
\begin{itemize}
\item the transverse ($\vect{e}_y$) component, proportional to $x^2$, which is the standard transverse displacement of the neutral line due to the bending. This term is the one directly computed in a beam theory;
\item the axial ($\vect{e}_x$) component, which is the 3D linearized rotation of the cross section around vector $\vect{e}_z$, also due to bending. In a beam theory, it is the consequence of the Euler-Bernoulli kinematics;
\item two additional terms in the transverse ($\vect{e}_y$) and lateral ($\vect{e}_z$) directions, proportional to $\nu$ and thus directly linked to the Poisson effect. These terms are responsible of the distortion of the cross section, as seen in Fig~\ref{fig:beamanalytic}(c).
\end{itemize}
To see the effect of the geometrical nonlinearities, we use the displacement field~(\ref{eq:Ulin}) to compute the nonlinear Green-Lagrange strain tensor. We obtain:
\begin{align}
\tens{\gamma} & =\frac{1}{2}\left(\tgrad\vect{U} + \tptgrad\vect{U} + \tptgrad\vect{U}\tgrad\vect{U}\right) \\
& = \tens{\varepsilon} + \underbrace{\frac{\alpha^2}{2}
\begin{pmatrix}
x^2 & 0 & 0\\
0 & 0 & 0 \\
0 & 0 & 0
\end{pmatrix}
}_{\tens{\gamma}_1}+\underbrace{\frac{\alpha^2}{2}
\begin{pmatrix}
y^2 & xy & 0 \\
xy & x^2 & 0 \\
0 & 0 & 0
\end{pmatrix}
}_{\tens{\gamma}_2}+\underbrace{\frac{\nu\alpha^2}{2}
\begin{pmatrix}
0 & xy & -xz \\
xy & \nu(y^2+z^2) & 0 \\
- xz & 0 & \nu(y^2+z^2)
\end{pmatrix}
}_{\tens{\gamma}_3}
\end{align}
The above equation shows that, in addition to the linear part $\tens{\varepsilon}$, the nonlinear part has three type of components:
\begin{itemize}
\item the first nonlinear term $\tens{\gamma}_1$ is purely axial, with the $\gamma_{xx}$ term $\alpha^2x^2/2$ being the leading term responsible of the standard axial / bending coupling due to the geometrical nonlinearities. It is the term predicted by the von~K\'arm\'an theory: if $v(x)=\alpha/2x^2$ denotes the transverse displacement of the neutral fiber, the nonlinear terms added by the von~K\'arman theory in $\gamma_{xx}$ is $v'(x)^2/2=\alpha^2x^2/2$ (see \cite{givois2019});
\item the second nonlinear term $\tens{\gamma}_2$ comes from purely 3D effects {\it independent of the Poisson effect}, that add (i) a stretch in the axial $\vect{e}_x$ direction, proportional to $y^2$; (ii) a positive and homogeneous stretch in the transverse $\vect{e}_y$ direction (proportional to $\vect{x}^2$); (iii) a transverse shear;
\item the third nonlinear term $\tens{\gamma}_3$ gathers the effects of the 3D Poisson effect. It involves stretching in both the transverse $\vect{e}_y$ and lateral $\vect{e}_z$ direction as well as shear.
\end{itemize}
Even if the present results are strictly valid for a beam in pure bending, they can be extended and applied to understand qualitatively any bending state.

\end{document}